\documentclass[sigconf]{acmart}
\usepackage{colortbl}
\usepackage{multirow}
\usepackage{graphicx}
\usepackage{float}
\usepackage{placeins}
\usepackage{tabularx}
\usepackage{soul,array}
\usepackage{balance}       
\usepackage{graphics}      
\usepackage[english]{babel}
\usepackage[T1]{fontenc}   
\usepackage[autostyle=true]{csquotes}

\usepackage{color}
\usepackage{booktabs}
\usepackage{textcomp}

\usepackage{algorithm}
\usepackage{algpseudocode}
\usepackage{siunitx}
\usepackage{amsmath}
\usepackage{amsfonts}
\makeatletter
\let\Bbbk\@undefined
\makeatother
\usepackage{amssymb}
\usepackage{gensymb}
\usepackage{array}
\usepackage{multicol}
\usepackage{url}
\usepackage{subcaption}
\usepackage[percent]{overpic}

\AtBeginDocument{%
  }

\copyrightyear{2026}
\acmYear{2026}
\setcopyright{cc}
\setcctype{by}
\acmConference[CHI '26]{Proceedings of the 2026 CHI Conference on Human Factors in Computing Systems}{April 13--17, 2026}{Barcelona, Spain}
\acmBooktitle{Proceedings of the 2026 CHI Conference on Human Factors in Computing Systems (CHI '26), April 13--17, 2026, Barcelona, Spain}
\acmPrice{}
\acmDOI{10.1145/3772318.3791820}
\acmISBN{979-8-4007-2278-3/2026/04}

\begin{document}

\title{SwEYEpinch: Exploring Intuitive, Efficient Text Entry for Extended Reality via Eye and Hand Tracking}


\author{Ziheng `Leo' Li}
\authornote{These authors contributed equally to this work.}
\email{zihengleoli@cs.columbia.edu}
\affiliation{%
  \institution{Department of Computer Science, Columbia University}
  \city{New York}
  \state{NY}
  \country{USA}
}

\author{Xichen He}
\authornotemark[1]
\email{xh2623@columbia.edu}
\affiliation{%
  \institution{Department of Computer Science, Columbia University}
  \city{New York}
  \state{NY}
  \country{USA}
}

\author{Mengyuan `Millie' Wu}
\email{mw3209@columbia.edu}
\affiliation{%
  \institution{Department of Computer Science, Columbia University}
  \city{New York}
  \state{NY}
  \country{USA}
}

\author{Zeyi Tong}
\email{zt2373@columbia.edu}
\affiliation{%
  \institution{Department of Computer Science, Columbia University}
  \city{New York}
  \state{NY}
  \country{USA}
}

\author{Haowen Wei}
\email{hw2892@columbia.edu}
\affiliation{%
  \institution{Department of Computer Science, Columbia University}
  \city{New York}
  \state{NY}
  \country{USA}
}

\author{Benjamin Yang}
\email{by2297@columbia.edu}
\affiliation{%
  \institution{Department of Computer Science, Columbia University}
  \city{New York}
  \state{NY}
  \country{USA}
}

\author{Steven Feiner}
\email{feiner@cs.columbia.edu}
\affiliation{%
  \institution{Department of Computer Science, Columbia University}
  \city{New York}
  \state{NY}
  \country{USA}
}

\author{Paul Sajda}
\email{psajda@columbia.edu}
\affiliation{%
  \institution{Department of Biomedical Engineering, Columbia University}
  \city{New York}
  \state{NY}
  \country{USA}
}

\renewcommand{\shortauthors}{Li and He et al.}
\newif\ifreviewversion
\reviewversionfalse

\newif\iffinalreview
\finalreviewfalse

\definecolor{DarkGreen}{HTML}{5DAC81}
\definecolor{DarkBlue}{HTML}{00008B}
\definecolor{FinalRed}{HTML}{B00020}
\definecolor{FinalPurple}{HTML}{6A1B9A} 

\ifreviewversion
  \newcommand\review[1]{\textcolor{DarkGreen}{#1}}
  \newcommand\minorreview[1]{\textcolor{DarkBlue}{#1}}
\else
  \newcommand\review[1]{#1}
  \newcommand\minorreview[1]{#1}
\fi

\iffinalreview
  \newcommand\finalreview[1]{\textcolor{FinalRed}{#1}}
  \newcommand\finalminorreview[1]{\textcolor{FinalPurple}{#1}}
\else
  \newcommand\finalreview[1]{#1}
  \newcommand\finalminorreview[1]{#1}
\fi

\begin{abstract}
Despite steady progress, text entry in Extended Reality (XR) often remains slower and more effortful than typing on a physical keyboard or touchscreen. We explore a simple idea: use gaze to swipe through a virtual keyboard for the fast, low-effort \emph{where} and a manual pinch held throughout the swipe for the \emph{when}, extending and validating it through a series of user studies. We first show that a basic version including a low-latency decoder with spatiotemporal Dynamic Time Warping and fixation filtering outperforms selecting individual keys sequentially, either by finger tapping each or gazing at each while pinching. We then add mid-swipe prediction and in-gesture cancellation, improving words per minute (WPM) without hurting accuracy. We show that this approach is faster and more preferred than previous gaze-swipe approaches, finger tapping with prediction, or hand swiping with the same additions. Furthermore, a seven-day, 30-session study demonstrates sustained learning, with peak performance reaching 64.7~WPM.
\end{abstract}


\begin{teaserfigure}
    \centering
    \includegraphics[width=\textwidth]{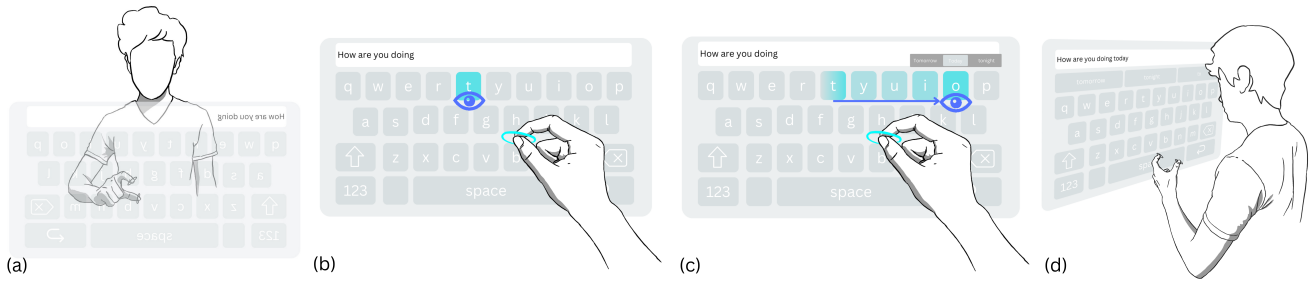}
    \caption{Typing in XR with \textit{SwEYEpinch}. (a) The user (Head-Worn Display not shown) is about to type the word ``today''. (b) They look 
    at the first letter ``t'' and start pinching. (c) Keeping fingers pinched, they swipe their gaze 
    to ``o''. The system predicts likely word candidates based on the current gaze trace and preceding context, ``How are you doing''. (d) The user releases the pinch to confirm the first candidate ``today'', which is placed in the text box automatically. The other candidates are displayed above the keyboard and the user may choose one by looking at it and pinching.}
    \Description{A four-part illustration sequence demonstrating how a user enters the word “today” using the SwEYEpinch technique: (a) Shows a sketched, third-person view of a user preparing to input text using a virtual keyboard, with one hand raised for a pinch gesture (the headset is not rendered). The text input box already contains “How are you doing” (b) The user fixates on the letter “t” and begins the pinch gesture to start a swipe. (c) While continuing the pinch, the user’s gaze moves across the keyboard from “t” to “o”; the system predicts the top three most likely word candidates based on this gaze trace and the prior sentence context (“How are you doing”). The top three candidates are tomorrow, today, tonight. (d) The user releases the pinch to confirm the top candidate “today.” Three additional predictions appear above the keyboard, allowing the user to select other candidates if needed.}
    \label{fig:teaser}
\end{teaserfigure}


\maketitle

\section{Introduction}

Efficient text entry in Extended Reality (XR) remains a significant challenge, impeding the widespread adoption of Head-Worn Displays (HWDs). \review
{While visual fidelity in XR has advanced rapidly, input methods have generally struggled to balance the trade-off between text entry rate, physical effort, and mobility.} 

Controller–raycast keyboards demand precise pointing at arm’s length and yield modest \review{typing speed} (e.g., $\sim$15–17 WPM with $\sim$11\% total error rate), with fatigue from sustained shoulder elevation and depth/occlusion ambiguity near key boundaries \cite{boletsis2019controller, akhoroz2024poke}. Hand-based mid-air tapping \review{avoids} controllers but still lacks tactile feedback, amplifying “fat-finger” ambiguity on closely spaced keys \cite{akhoroz2024poke}. 
External peripherals (physical keyboards) restore speed but break mobility and situational use, effectively tying users to a surface, or encumbering a hand if hand-worn \cite{grubert2018text, knierim2018physical, rosenberg2002chording}. Voice dictation can be fast, yet it fails in noisy or shared spaces and raises privacy concerns. Gaze, on the other hand, has clear appeal as a \emph{targeting} signal in this landscape---eyes move quickly with very low muscular effort and \review{eye tracking is} increasingly available on HWDs \cite{andersson2017one, cui2023glancewriter}---making it well-suited for text entry. Prior work on gaze-swipe typing has shown that users can naturally trace word-shaped paths with their eyes and that such “eye-gesture” input can outperform dwell-based gaze typing in entry speed \cite{kurauchi2016eyeswipe}. \review{However, gaze-only interaction typically relies on dwell-based selection for activation, which imposes an inherent speed cap and is susceptible to false activations—or on additional instrumentation (e.g., external touch surfaces or controllers \cite{kumar2020tagswipe, yu2017tap}), while hand-based activation is reliable but incurs a movement cost \cite{luong2023controllers, buckingham2021hand, akhoroz2024poke}.} This motivates \review{the desirability of} a hybrid that preserves continuous, word-gesture \emph{swipes} with gaze for rapid, low-effort \emph{where}, while offloading the precise \emph{when} to a \review{manual} pinch delimiter that marks word start and end. \review{Our goal is an XR typing method that (1) approaches practical, everyday text entry rate, (2) reduces physical demand compared to mid-air tapping, and (3) avoids gaze-only pitfalls such as dwell latency and Midas touch~\cite{Jacob-TOIS-1991}.}

\review{We propose a hybrid interaction that bridges this gap: \textit{SwEYEpinch}. We operationalize a simple principle: decouple targeting from commitment. The eyes are used exclusively to \textit{trace} the word-gesture (fast, low effort), while a manual mid-air pinch is used to \textit{delimit} the word (explicit, reliable). This preserves the speed of gaze and the reliability of manual input, without the fatigue of reaching for keys or the immobility of physical surfaces. To rigorously evaluate this approach against the spectrum of XR text entry, we present the design and evaluation of \textit{SwEYEpinch} through four progressive user studies.}


In an initial \textit{User Study 1} (\textbf{US1}), we compare \textit{SwEYEpinch-Basic}—a minimal gaze-swipe and pinch-delimiter variant that outputs word candidates upon pinch release—with two letter-by-letter baselines used in commercial XR headsets, \review{focusing specifically on their interaction mechanics}: \textit{Finger-Tap} \cite{speicher2018selection, meier2021tapid, streli2022taptype, kern2023text} (\review{Microsoft} HoloLens, Meta Quest) and \textit{Gaze\&Pinch}, a well-studied selection technique in XR \cite{pfeuffer2017gaze, mutasim2021pinch, kim2025pinchcatcher} (Apple Vision Pro).

Motivated by \textit{US1}, we next ask whether \emph{live, mid-swipe feedback} could further reduce effort and increase speed without compromising accuracy. The design must also mitigate eye fatigue from large in-view keyboards \emph{and} reduce off-keyboard verification glances. To this end, we introduce \textit{SwEYEpinch}, which continuously decodes words during the swipe, allowing users to see candidate words while tracing the letters rather than when they finish a swipe. \review{We also add} two safeguards to manage errors: \emph{mid-swipe deletion} (allowing the user to cancel mid-swipe) and a \emph{deletion peek window} (letting the user see what will be removed when they gaze at the delete button). These additions raise a natural question: \emph{What impacts the user experience the most: live mid-swipe feedback, the pinch delimiter, or their combination?}

To answer this, \textit{User Study 2} (\textbf{US2}) disentangles the effects of mid-swipe prediction and the delimiter (how the user signals the start/end of a word) by comparing \textit{SwEYEpinch} against \textit{SwEYEpinch-Basic} and two gaze-only \review{swipe-based approaches} with different delimiters: \textit{SkiMR}~\cite{hu2024skimr}, delimited by gazing at the space key (dwell-free), and \textit{GlanceWriter XR}, adapted \review{for} XR from its  screen-based version~\cite{cui2023glancewriter}, where the user starts and ends a word by moving their gaze in and out of the letter-key area. This isolates the contribution of \emph{mid-swipe prediction} and of the \emph{delimiter} itself. Our results show \textit{SwEYEpinch} stands at the Pareto frontier in terms of typing speed and user preference while maintaining accuracy.

To test whether the gains observed in US2 hold up against strong baselines that users would encounter in commercial products,
\review{\textit{User Study 3} (\textbf{US3}) evaluates \textit{SwEYEpinch} against: \textit{Finger-Tap with Prediction\&Completion} and \textit{Hand-Swipe} (mid-air word-gesture typing with hand ray and pinch delimiter \cite{markussen2014vulture, dudley2019performance, Chen2019exploring, henderson2020stat}; also available in Meta Quest) to the same level of word prediction and completion as \textit{SwEYEpinch}.}

Finally, \textit{User Study 4} (\textbf{US4}) examines whether, with sustained use, users can gain proficiency fit for everyday use—how fast their performance improves, where they plateau, and what failure modes appear over time. We followed \review{nine} participants; each used \textit{SwEYEpinch} for 30 sessions over seven consecutive days, allowing us to observe learning trends akin to what users would have had with daily exposure to the technique. By the end, all users achieved on average \textit{above 40 WPM}, and three achieved \textit{above 60 WPM}.

Our results support a simple design principle: \emph{decouple targeting from commitment}. Use gaze only to \emph{trace} the word (fast, low effort) and a \review{manual} pinch to \emph{commit} (the delimiter); give the user immediate feedback with mid-swipe candidates, and reduce correction effort \review{through} mid-swipe deletion and \review{a} peek window. Together, these cut verification glances, remove dwell delays, and preserve explicit control.

\review{We thus make the following contributions}:
\begin{itemize}
    \item We introduce \textit{SwEYEpinch-Basic}, a gaze–gesture hybrid for XR text entry, and \textit{SwEYEpinch}, an enhanced variant offering mid-swipe prediction, mid-swipe deletion, and a deletion peek window. \review{These provide an optimized decoding pipeline (filtering by fixation detection combined with density-based clustering) that overcomes the high-frequency noise of eye-tracking data, enabling real-time mid-swipe feedback that was previously computationally prohibitive for gaze.}
    \item Through three comparative studies, we benchmark \review{\textit{SwEYEpinch-Basic} and \textit{SwEYEpinch}} with other prominent \review{XR} text input methods. In \textbf{US1}, we establish that users prefer \textit{SwEYEpinch-Basic} compared to traditional XR baselines, and gather feedback \review{enabling} us to design \textit{SwEYEpinch}. In \textbf{US2}, we map the design space by isolating the roles of mid-swipe prediction and delimiter mechanics (\textit{SwEYEpinch} vs.\ \textit{SwEYEpinch-Basic} vs.\ gaze-only \textit{SkiMR}/\textit{GlanceWriter XR}). In \textbf{US3}, we compare \textit{SwEYEpinch} against production-realistic contenders (\textit{Finger-Tap w/ Prediction}, \textit{Hand-Swipe}). For all user studies, we report on participants' speed, accuracy, and preference.
    \item Finally, we show that with repeated daily practice (\textbf{US4}), users of \textit{SwEYEpinch} can achieve typing speeds comparable to a conventional keyboard (up to 64.7~WPM), highlighting its potential for everyday XR text-entry.
\end{itemize}

\textbf{Code and Data Availability.} Our decoding algorithm (Gaze2Word), Unity demo, and swipe trace data compiled from user studies are publicly available. See Section \ref{sec:code and data availability} for details.

\section{Related Work}
We review prior work, from well-studied conventional methods used in commercial products to more recent developments. For a broader overview of XR text entry as a whole, see the comprehensive survey by Bhatia et al. \cite{bhatia2025text}.


\subsection{Text-entry in HWDs---Conventional Methods and Alternatives}

\textbf{Controller and Hand-tracking.}
Many XR text-entry approaches rely on handheld controllers with virtual keyboards (point-and-click via raycasting) or mid-air typing with hand tracking \cite{wan2024design}. For instance, one study reported $16.65$ WPM with $11.05\%$ total error rate (TER) using controller-based raycasting
\cite{boletsis2019controller}.
A more recent study \cite{akhoroz2024poke} found $15.3$ WPM with controllers, slightly outperforming hand tracking (13.8 WPM). Using 3D protruding virtual keys (keys with depth) can also improve typing speed (15.6 \review{WPM} with 3D keys vs $13.4$ WPM with flat keys) and user experience \cite{akhoroz2024poke}. Overall, controllers provide reliable selection and tactile feedback (via trigger clicks) but still yield low text entry rate compared to desktop typing; users retain $60\%$ of their desktop performance  \cite{grubert2018text}, or equal it when typing with a conventional keyboard in XR \cite{knierim2018physical}.

Holding controllers can be cumbersome for on-the-go scenarios. Xu et al. note that while one can pair a physical keyboard with an AR display, it effectively locks the user to a stationary setup, whereas virtual keyboards (operated via hand gestures) better support on-the-go usage \cite{xu2019pointing}. Hand-tracking and mid-air keyboards improve immersion by allowing direct interaction. Luong et al. report that while controllers were faster for distant pointing tasks, participants actually favored freehand interaction for direct mid-air selection tasks \cite{luong2023controllers}. This implies that for text input styles where users “touch” keys in the air (a near-field task), hand tracking can be more natural and even preferred, but it can suffer from a lack of tactile feedback, leading to increased arm fatigue as reported in a few studies \cite{akhoroz2024poke, buckingham2021hand}. 

\textbf{Alternative Approaches.} Alternative approaches integrate other bodily signals,  encompassing a diverse range of interaction techniques that eliminate the need for physical controllers or hands, improving accessibility and reducing physical strain. Voice dictation in XR can be extremely fast in WPM and is already used in practice (e.g., voice-to-text on HoloLens). In a controlled study \cite{adhikary2021text}, speaking sentences aloud and then making corrections with hand-tracking averaged about 28 WPM (up to 36 WPM with perfect recognition) with low error rates ($0.5\%$), far surpassing the $11$ WPM of purely hand-based mid-air typing. However, voice is impractical in noisy or shared spaces. 

Some hands-free typing experiments have tried using facial muscle movements to select letters, but suffer from very slow speeds. One comparison found that using head-gaze plus a “mouth-open” gesture to confirm each character yielded only about $3.07$ WPM, and using an “eyebrow raise” yielded ~$2.85$ WPM \cite{gizatdinova2012comparison}. Other techniques integrate other body movements, such as NeckType, which allows users to select characters via subtle head gestures \cite{lu2020handsfree}. \review{On the other hand, \textit{Hummer}~\cite{hedeshy2021hummer} uses vocal humming as a delimiter, offering a hands-free alternative, yet vocal input faces social acceptability challenges in public or quiet environments.}
Despite the variety of alternative hands-free input methods, gaze emerges as the most technically suitable modality for XR text entry. Eye movements are extremely fast and require minimal effort, involving very small, well-conditioned muscles, enabling rapid targeting of virtual keys or UI elements.

\subsection{Gaze-based text entry}

\textbf{DwellType.} Gaze-based text entry has been widely studied in XR as a hands-free alternative to traditional input. Among these methods, DwellType—where users select letters by fixating for a set duration—is one of the most common baselines due to its simplicity and accessibility \cite{lu2020handsfree}. However, its reliance on delayed activation fundamentally limits speed. MacKenzie et al. estimated a theoretical maximum of around 22 WPM (with a 0.5 s dwell and rapid 40 ms saccades to move between keys), yet novice users achieve only ~8--10 WPM, and experienced users plateau around $20$ WPM \cite{mackenzie2010text}. Comparative studies show DwellType ($10.59$ WPM) is significantly slower than TapType ($15.58$ WPM) and GestureType ($19.04$ WPM) \cite{yu2017tap}, and offers the lowest throughput ($1.12$ bits/s) compared to pinch gestures ($1.60$ bits/s) or button click ($1.88$ bits/s)  \cite{mutasim2021pinch}. Although Yu et al. \cite{yu2017tap} evaluate head-gaze–based TapType and GestureType rather than eye-gaze methods, these techniques still outperform DwellType, reinforcing that the dwell-timer mechanism, rather than gaze modality, is the main limiting factor \cite{mott2017improving}. Furthermore, designers face a speed--accuracy trade-off: longer dwell durations are slow but reliable, while shorter ones risk many errors and unintentional activations \cite{gizatdinova2012comparison}.
Beyond performance, DwellType imposes high cognitive load and visual strain, with studies reporting increased blink frequency, altered saccade patterns, and session limits ($\leq15$ minutes) to mitigate fatigue \cite{bafna2021mental,liu2024role}.

\textbf{Eye-swipe Typing.} To overcome dwell-time limitations, researchers have proposed dwell-free gaze typing. Kurauchi et al. \cite{kurauchi2016eyeswipe} introduced \textit{EyeSwipe}, adapting smartphone word-gesture keyboards \cite{Zhai-Kristensson-CACM2012} with reverse crossing for selection, achieving 11.7 WPM compared to 9.5 WPM with dwell-based typing. Similarly, Cui et al. \cite{cui2023glancewriter} developed \textit{GlanceWriter}, leveraging a probabilistic decoding algorithm to dynamically predict user intent, achieving a 67\% speed improvement over EyeSwipe. In addition, Liu et al. combined gaze-tracking with the Moving Window String Matching algorithm for more accurate word entries \cite{liu2015gazetry}. These approaches demonstrate that eliminating dwell-time delays can significantly enhance gaze-based text entry. The adaptation of the swipe in an HWD by Yu et al. \cite{yu2017tap} uses head movement to activate the swipe instead of gaze, reaching 24.7 WPM after one hour of practice, while Lu et al. found that, in head-gesture typing, eye blinks were the most effective selection mechanism compared to dwell or swipe \cite{lu2021iText}. Other XR adaptations have explored using the \emph{space key} as a delimiter to mitigate Midas-touch ambiguity \cite{hu2024skimr} and relax spatial matching constraints for more ergonomic gaze-based typing \cite{amswipe}. Another study focusing on head vs. gaze pointing showed that users
generally prefer gaze over head-based input, describing it as faster, more pleasant, and efficient \cite{gizatdinova2012comparison}. A rich body of prior work has therefore \review{investigated} ways to improve the accuracy of gaze input more generally \cite{kumar2008improving}, including probabilistic correction and snap-to-target techniques.


Pure gaze-based techniques leave ample room for improvement in both WPM and accuracy due to their reliance on precise gaze trace and potential high visual strain \cite{feng2021hgaze, kurauchi2016eyeswipe}. Studies indicate that hybrid approaches, combining gaze with additional input modalities such as hand gestures, head movements, or predictive models,
can increase speed and usability \cite{lystbaek2022gaze}.
\review{ While multimodal gaze-typing has been explored on desktop and mobile surfaces, its application in mobile XR presents unique constraints. \textit{TagSwipe}~\cite{kumar2020tagswipe} demonstrated that anchoring gaze swipes with a touch-surface delimiter significantly improves performance. However, TagSwipe relies on physical contact (touchscreen or trackpad), which limits user mobility in everyday XR scenarios. }

\subsection{Combining Hand Tracking and Gaze}
Combining gaze with manual input has long been proposed as a natural division of labor: the eyes rapidly target, while the hand provides confirmation or fine-grained control. Early work such as Gaze-touch \cite{pfeuffer2014gaze} demonstrated this principle on touchscreens, showing that gaze can efficiently pre-select targets for subsequent touch interaction. More recent XR studies extend this idea by using the hand as a pointing device while leveraging gaze implicitly to accelerate cursor movement \cite{zhao2023gaze}, or by combining gaze selection with controller confirmation \cite{rajanna2018gaze}. Extending this principle, Wagner et al. \cite{wagner2024eye} analyzed eye–hand coordination for object manipulation in near-space XR, finding systematic gaze-leading-hand dynamics that can be harnessed for input design. These insights motivate text entry methods that combine gaze and hand actions, as in our SwEYEpinch approach.

Another promising direction is to integrate gaze with body-based gestures. HGaze exemplifies this by combining subtle head gestures with gaze-based swipe typing 
\cite{feng2021hgaze}. Pfeuffer et al. \cite{pfeuffer2024design} outlined core design principles and challenges for gaze–pinch interaction in XR, emphasizing simplicity, immediacy, and clear feedback as critical factors for usable systems. Empirical findings echo these principles: Park et al. \cite{park2024impact} showed that increasing gesture complexity (e.g., multi-step or compound gestures) significantly degrades performance compared to simpler gaze–pinch combinations. This suggests that, while gaze–pinch can be powerful, designers should keep gestures lightweight to maximize usability.

Freehand selection refined by gaze tracking has also been explored in XR. Lystbæk et al. \cite{lystbaek2022gaze} proposed S-Gaze\&Finger and S-Gaze\&Hand, where gaze assists freehand input to enhance accuracy and reduce hand movement. Their study found that S-Gaze\&Finger reduced hand movement by over 50\% compared to Finger-Tap (as in HoloLens 2), though the study showed no improvement in speed ($10.7$ WPM for S-Gaze\&Finger and 11.4 for Finger-Tap). This reveals an interesting observation: the baseline techniques in commercial headsets if included in a study, have repeatedly been found hard to beat.

While gaze-only methods such as \textit{SkiMR} \cite{hu2024skimr} have been compared against other gaze-only techniques in XR, prior work has not examined gaze-based swipe techniques alongside commercial XR baselines such as Finger-Tap and Hand-Swipe, or hybrid approaches such as Gaze\&Pinch. This leaves an open question of how gaze-based typing compares to the input techniques users already encounter in practice. Our work addresses this question by conducting a comparative analysis of these input techniques to identify the most effective methods for text entry in XR.

Moreover, most recent XR studies on gaze-based swipe input have focused on short-term use and have not assessed performance at the expert level. Earlier work shows that text-entry performance typically plateaus after extended training \cite{matias1996one}, with expert typists on physical keyboards eventually matching their real-world typing speeds in XR \cite{knierim2018physical}. To address this critical gap, we conducted an extended longitudinal user study (US4), wherein nine participants performed 30 sessions (around 30 min per session, including setup) with the best-performing technique identified from US1, US2, and US3. This extensive practice allows us to examine expert-level typing performance, providing additional insight into the potential of gaze-based swipe techniques for text input.

\subsection*{Scope and Baseline Selection}
\review{
Our experimental scope is constrained by two design goals: (1) use only sensors that are already integrated into many modern HWDs (eye tracking and hand tracking), and (2) study \emph{silent}, everyday text entry that does not rely on external surfaces or voice. This excludes influential techniques such as TagSwipe \cite{kumar2020tagswipe} and GestureType \cite{yu2017tap}, which require an external device for input, and Hummer \cite{hedeshy2021hummer}, which uses humming as the primary activation channel. We instead treat these as \emph{external reference points}, comparing reported WPM and learning trends, while reserving in-study baselines for techniques that can be implemented with HWD-only sensing under our apparatus.}

\section{Design of SwEYEpinch: Tiny-Explicit Gesture-Delimited Gaze \review{Swipe} Typing}


Unlike existing XR text-entry methods that rely on dwell-based gaze selection (slow and error-prone) or swipe-only gaze gestures (ambiguous activation) \cite{kurauchi2016eyeswipe, feng2021hgaze, kumar2020tagswipe, cui2023glancewriter,hu2024skimr,amswipe}, \textit{SwEYEpinch} offers 
two key innovations: a gaze–explicit gesture-combined input method and an optimized word-completion algorithm tailored for XR text entry. Prior work shows that mid-air gesture typing in XR imposes substantial visuomotor demands, requiring users to allocate their visual attention to the keyboard area and maintain tight eye–hand synchronization to compensate for the lack of physical boundaries~\cite{hu2025seeingandtouchingtheair}. \textit{SwEYEpinch} mitigates these burdens by leveraging a simple pinch delimiter to explicitly confirm word input, which minimizes unintentional selections and reduces both cognitive and visual fatigue compared to prolonged dwell or gaze-only activation. Moreover, its backend prediction algorithm is optimized to decode words on the fly during a swipe. This combination of gesture-based control and predictive completion enables users to input text more fluidly, with reduced physical strain and fewer errors.

\subsection{How SwEYEpinch Works}\label{sec:sweyepe}

To understand the interaction loop of \textit{SwEYEpinch}, we can begin by considering \textit{Gaze\&Pinch}, a popular XR text-entry technique that  \review{uses eye tracking to select individual keys, each confirmed with a pinch gesture}. \textit{SwEYEpinch} \review{also} leverages
\review{eye tracking and gestures}. In smartphones, “swipe typing” \review{\cite{Zhai-Kristensson-CACM2012}} extends the traditional “tap key to type,” and \textit{SwEYEpinch}, together with other gaze-based swipe techniques, follows the same logic for gaze input: Instead of tapping each key individually, you can swipe across multiple letters while pinching your fingers. As noted by Yu et al.~\cite{yu2017tap}, swipe-enabled keyboards are advantageous because they let users quickly tap out a single letter or swipe an entire word, thus addressing the out-of-vocabulary problem. 
We refer to this baseline version as  \textit{SwEYEpinch-Basic}, to distinguish it from the \textit{SwEYEpinch} variant introduced later in Section \ref{sec: partial_sweyepe}.

The \textit{SwEYEpinch-Basic} interaction loop is straightforward. As illustrated in \autoref{fig:teaser}, to type a word, the user first looks at the word's initial letter, then pinches their fingers and swipes \review{their gaze} through all intended letters. After traversing the letters, they release the pinch to end the swipe. A decoding algorithm (detailed in the next section) then suggests the top four most likely words. Note that \autoref{fig:teaser} shows \textit{SwEYEpinch}, the improved version \review{(Section \ref{sec: partial_sweyepe})}, where predictions appear on the fly, rather than only after releasing the pinch.

Once the user finishes swiping, the system automatically inserts the first candidate into the text box and shows the remaining three candidates right below it. Selecting any of these alternate candidates is done by gazing at \review{it} and pinching, just as if tapping a single key. If the user presses the delete key immediately following a swipe, it removes the entire last word rather than a single space. Once another input key (e.g., a letter, number, or space) is pressed, the delete key reverts to its normal behavior and removes only one character per press. 

\review{
\textbf{Seamless Mode Switching.} Crucially, \textit{SwEYEpinch-Basic} inherently supports character-level entry too \review{\cite{Zhai-Kristensson-CACM2012}}. If the user performs a \textit{pinch-and-release} on a key without generating a swipe trajectory (i.e., without looking at other keys), the system registers a standard single-letter tap. This allows users to fluidly mix word-level swiping with character-level tapping (e.g., for out-of-vocabulary words or passwords) using the same modality, effectively combining the benefits of \textit{SwEYEpinch} and \textit{Gaze\&Pinch} in a single interface.
}

\begin{algorithm*}[t]
\caption{\textbf{Gaze2Word}: spatiotemporal Dynamic Time Warping (DTW) on pruned swipe trace with $n$-gram fusion}
\label{alg:g2w}
\begin{algorithmic}[1]
\Require Raw gaze samples $\mathbf{g}_{raw}=\{(x_i,y_i,t_i)\}_{i=1}^{N}$, language context $C$, lexicon $\mathcal{V}$, keyboard geometry $\mathcal{K}$, top-$k$, weights $\alpha\!\in\![0,1]$, $\epsilon\!>\!0$, I\!-VT params $\theta_{\mathrm{ivt}}$, DBSCAN params $(\varepsilon,\mathrm{minPts})$
\Ensure Ranked list $\mathrm{TopK}$
\State $\mathbf{g}\leftarrow \textsc{I-VT}(\mathbf{g}_{raw};\theta_{\mathrm{ivt}})$ \Comment{Fixation detection; returns fixation points with timestamps}
\State $\mathbf{g}\leftarrow \textsc{DBSCAN\_Reduce}(\mathbf{g};\varepsilon,\mathrm{minPts})$ \Comment{Cluster and keep time-ordered centroids}
\State $\mathcal{V}_{\mathrm{filtered}}\leftarrow \textsc{FilterCandidates}(\mathbf{g}[1],\mathcal{V},\mathcal{K})$ \Comment{Gate by first-fixation key proximity / length}
\If{$\mathcal{V}_{\mathrm{filtered}}=\emptyset$}
  \State $\mathcal{V}_{\mathrm{filtered}}\leftarrow$ LM prefix-trie shortlist from $C$
\EndIf
\ForAll{$w\in \mathcal{V}_{\mathrm{filtered}}$}
  \State $\mathbf{q}_w\leftarrow \textsc{TemplateTrace}(w,\mathcal{K})$ \Comment{Key-center path; set $t'_j\!\leftarrow\!j$}
  \State $\delta_w \leftarrow \textsc{DTW}_{xyz}(\mathbf{g}, \mathbf{q}_w)$ \Comment{DP on distance $d((x,y,t),(x',y',t'))$}
\EndFor
\State $d_{\min}\leftarrow \min_{w}\delta_w$, \quad $d_{\max}\leftarrow \max_{w}\delta_w$
\ForAll{$w\in \mathcal{V}_{\mathrm{filtered}}$}
  \State $p_{\mathrm{dist}}(w)\leftarrow 1 - \dfrac{\delta_w - d_{\min}}{(d_{\max}-d_{\min})+\num{1e-12}}$ \Comment{Distance$\!\to\!$[0,1]}
  \State $p_{\mathrm{ng}}(w)\leftarrow p_{\mathrm{ngram}}(w\mid C)$
  \State $s(w)\leftarrow \bigl(p_{\mathrm{ng}}(w)+\epsilon\bigr)^{\alpha}\cdot \bigl(p_{\mathrm{dist}}(w)\bigr)^{1-\alpha}$ \Comment{Fusion (Sec.~\ref{sec: sweyepe decoding algorithm})}
\EndFor
\State \Return top-$k$ words by $s(w)$ (break ties by higher $p_{\mathrm{dist}}$)
\end{algorithmic}
\end{algorithm*}

\subsection{SwEYEpinch Decoding Algorithm and Optimization}
\label{sec: sweyepe decoding algorithm}

The \textit{SwEYEpinch} decoding algorithm, which we call \texttt{Gaze2Word}, builds on previous work~\cite{kurauchi2016eyeswipe,yu2017tap,cui2023glancewriter} and is described in Algorithm \ref{alg:g2w}. In essence, \textit{Gaze2Word} takes a user’s gaze trace as input and outputs the top-$k$ candidate words by matching this gaze trace to template traces stored in a dictionary. The resulting similarity scores are then combined with an $n$-gram language model to generate a final ranked list of candidate words.

\review{
\noindent\textbf{The Challenge of Real-Time Gaze Decoding.}
While swipe prediction is standard on mobile touchscreens, porting this to gaze presents a specific computational challenge: signal density. Touch events are discrete and relatively sparse. In contrast, eye tracking generates high-frequency data (200\,Hz) laden with jitter and micro-saccades. A raw two-second gaze swipe contains $\sim$400 points; running \review{Dynamic Time Warping (DTW) on this raw stream for every frame causes latency that breaks the feedback loop.
}

\review{
To enable the \textit{mid-swipe predictions} introduced in this work, we implement a specific optimization pipeline before the DTW step. We chain \textit{Velocity-Threshold Identification} (I-VT) \cite{salvucci2000identifying} to remove saccades with \textit{Density-Based Spatial Clustering} (DBSCAN) \cite{ester1996density} to reduce the gaze trace cardinality by over 90\% (from $\sim$320 to $\sim$15 points per word, see \autoref{appendix: Optimization in SwEYEpinch Decoding}).
}

Below, we describe the core components of Algorithm~\ref{alg:g2w} that utilize this pruned data, emphasizing the key optimizations that make on-the-fly decoding feasible (Section \ref{sec: partial_sweyepe})—that is, displaying candidate words while the user is still swiping, rather than only at the end. In addition, we added a temporal axis to the DTW distance measure, improving decoding accuracy at the cost of a slight increase in latency. A post-hoc analysis using US1 data shows that, with the proposed optimization and spatial-temporal DTW, the proposed algorithm is more accurate and has lower latency when compared with decoders in prior work (Table \ref{tab:decoder_comp}). For more algorithm details, see the supplementary material.


\noindent
\paragraph{Spatiotemporal Dynamic Time Warping}
After we reduce the swipe trace, the next step is to match it with word template traces. We align the preprocessed gaze trace 
$\mathbf{g} = \{(x_i, y_i, t_i)\}_{i=1}^n$ with template word paths 
$\mathbf{q} = \{(x'_j, y'_j, t'_j)\}_{j=1}^m$ from $\mathcal{V}_{\text{filtered}}$, where $q$ consists of the keyboard centers of the letters that make up this word. We represent each gaze point as $(x_i,y_i,t_i)$, where $(x_i,y_i)$ is the 2D position where the gaze hit the keyboard and $t_i$ is its timestamp. Template points are defined by letter-center positions, with $t'_j = j$ serving as a positional index to enforce a weak forward-time prior while remaining robust to speed variation. Then we have the pointwise distance as 
\[
dist\bigl(\mathbf{g}_i, \mathbf{q}_j\bigr)
= \sqrt{\,(x_i - x'_j)^2 
        + (y_i - y'_j)^2 
        + (t_i - t'_j)^2}.
\]

\noindent
\paragraph{Combining with Language Model.}
Lastly, we multiply $p_{\mathrm{dist}}(w; \mathbf{g})$ with an $n$-gram language-model probability $p_{\mathrm{ngram}}(w \mid \text{context})$. Let $\alpha \in [0,1]$ be a weighting parameter and $\epsilon$ a small smoothing term (we used $\epsilon=1e-8$). Then,
\[
\begin{aligned}
& P(w \mid \mathbf{g}, \text{context})
= \Bigl(p_{\mathrm{ngram}}(w \mid \text{context}) + \epsilon \Bigr)^{\alpha}
\times \Bigl(p_{\mathrm{dist}}(w; \mathbf{g})\Bigr)^{1 - \alpha}.
\end{aligned}
 \]

\noindent
Here, $\alpha$ controls how \review{much} the $n$-gram model factors into the combined probability. Through pilot testing, we found $\alpha=0.05$ to be most effective. We then rank the words in descending order of $P(w \mid \mathbf{g}, \text{context})$, retaining the top $k$ as the final output.

\subsubsection{Open-Source Algorithm, Demo, and Multi-Study Swipe Dataset}
\label{sec:code and data availability}
We provide an open-source Python implementation of the SwEYEpinch (\texttt{Gaze2Word}) algorithm as a PyPI package\footnote{\url{https://pypi.org/project/gaze2word/}, \url{https://github.com/SwEYEpinch/Gaze2Word}}, to help researchers and developers build on our work. Computational bottlenecks such as DTW are wrapped in optimized, compiled extensions to meet real-time latency, enabling \textit{SwEYEpinch} to refresh candidates whenever the user saccades to a new key. We additionally release a Unity demo package illustrating \texttt{Gaze2Word} integration in Unity applications, using the cursor as a proxy for gaze\footnote{\url{https://github.com/SwEYEpinch/SwEYEpinch-Unity-Demo}}.

\noindent\textbf{Dataset.} We release a \emph{headset-captured, word-level} gaze-swipe dataset\footnote{\url{https://huggingface.co/datasets/SwEYEpinch/SwEYEpinch}} aggregated from all our user studies covering all swipe conditions (\textit{SwEYEpinch-Basic}, \textit{SwEYEpinch}, \textit{Hand-Swipe}, \textit{SkiMR}, and \textit{GlanceWriter XR}) with recordings from $71$ participants and $9248$ word attempts. Each sample corresponds to one intended word (from delimiter onset to confirmation) and includes raw 200\,Hz gaze, swipe hit points on the keyboard used for decoding, and the intended word. To our knowledge, this is the first public XR \emph{word-level} gaze-swipe dataset of this scope.

\section{User Study Metrics and Apparatus}
\label{sec:performance_metrics}
To compare text-entry methods, we report words per minute (WPM) and total error rate (TER).
Following common practice, WPM normalizes by five-character words:
\begin{equation}
\mathrm{WPM}=\frac{|S|/5}{t},
\end{equation}
where $|S|$ is the transcribed-string length (characters) and $t$ is entry time (minutes).
We compute TER following Soukoreff and MacKenzie~\cite{soukoreff2003metrics}:
\begin{equation}
\mathrm{TER}=\frac{IF+INF}{C+IF+INF},
\end{equation}
where $C$ is correct keystrokes, $IF$ incorrect-but-fixed, and $INF$ incorrect-and-not-fixed.
\review{For learning across sessions, we report Learning Rate as the regression slope $\beta$ in
$\mathrm{WPM}_s=\alpha+\beta s+\epsilon$, where $s$ is session index.}

\noindent\textbf{Statistical analysis and counterbalancing.} To mitigate longitudinal learning and fatigue effects, we employed a Staggered Latin Square to rotate condition orders systematically across sessions. Given the variation in participant counts across studies (unbalanced block sizes), we utilized \textbf{Linear Mixed Models (LMMs)} for all performance analyses rather than repeated-measures ANOVAs. We modeled \textit{Technique} and \textit{Presentation Order} as fixed effects, with \textit{Participant} as a random intercept. \finalreview{Although perfect counterbalancing was not achievable due to sample-size constraints (e.g., uneven multiples of orderings), this approach allows us to statistically isolate intrinsic technique performance from the significant learning effects such as the ordering bias observed over the multi-day study.}

\review{\noindent\textit{Multiple comparisons.} All pairwise comparisons derived from the LMMs were adjusted using the Benjamini–Hochberg false discovery rate (FDR) method to control for expected false discoveries while maintaining power. All $p$-values reported in the results are adjusted values.}

All studies use a Varjo XR-3 (200\,Hz eye tracking) with integrated Ultraleap hand tracking in Unity 2022.3.10f1 on Windows 11
with an Intel® Core™ i9-10900K CPU @ 3.70GHz and an Nvidia GeForce
RTX 3090 graphics card. PhysioLabXR 1.1 \cite{li2024physiolabxr} is used to collect data and serve the decoder algorithm. The same QWERTY layout was used across studies (see the supplementary material). In the experiment scene, the keys were rectangular (width 51.0\,mm, height 56.7\,mm) with a horizontal gap of 11\,mm and a vertical gap of 9.30\,mm. \review{The keyboard was positioned 70\,cm from the user, corresponding to visual angles of roughly $4.18\degree$ horizontally and $4.64\degree$ vertically for each key and the entire keyboard subtends $47.9\degree$ in width and $35.6\degree$ in height, consistent with established parameters for gaze-based text-entry research. During practice, the keyboard distance was adjusted by up to $\pm$15\,cm for comfort. The 200\,Hz eye tracker provided sufficient temporal resolution to capture rapid eye movements between keys without dropouts. Throughout all sessions, the application sustained rendering rates above 60 fps, ensuring stable visual performance and eliminating latency that could influence typing behavior.}
This apparatus and keyboard layout are identical across all user studies (US1–US4).

\section{User Study 1: From Simple Baselines to SwEYEpinch-Basic}
\label{sec:user_study_1}
To test the design intuition \emph{decouple targeting from commitment}, \review{we began with a minimal hybrid, \textit{SwEYEpinch-Basic} (\S\ref{sec:sweyepe}) that decodes at the word-level: gaze swipe for fast targeting, and a pinch gesture to confirm the predicted word. In contrast, current commercial XR systems (e.g., Meta Quest \textit{Finger-Tap}, Apple Vision Pro \textit{Gaze\&Pinch}) expose mainly non-predictive, character-level keyboards. We therefore use these widely deployed character-level techniques—without adding word prediction—as status-quo baselines for US1:}

\begin{figure}[h]
    \centering 
    \includegraphics[width=\columnwidth]{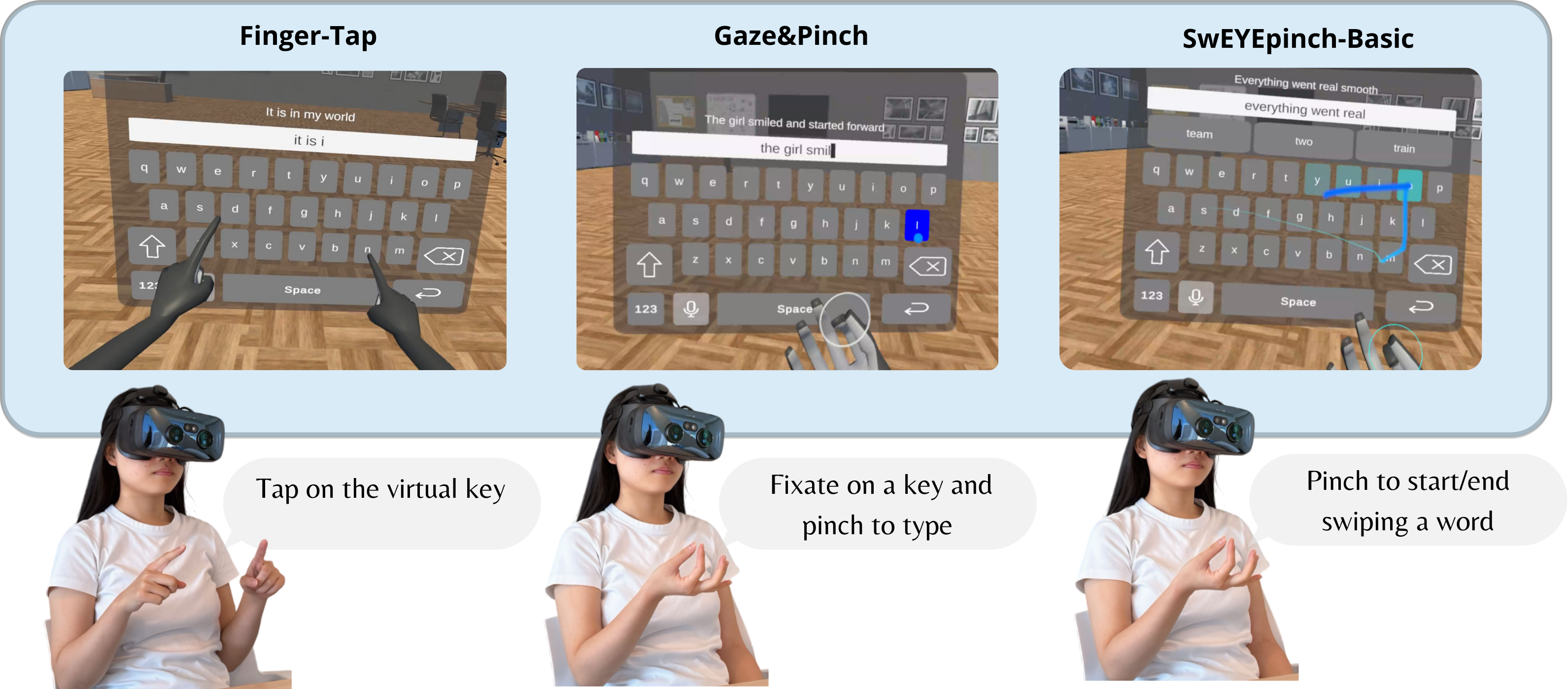} 
    \caption{Techniques evaluated in US1: two simple XR baselines--\textit{Finger-Tap} (letter-by-letter mid-air tapping on a virtual keyboard) and \textit{Gaze\&Pinch} (gaze-targeted key selection with a pinch delimiter)--compared against our proposed \textit{SwEYEpinch-Basic}, our pinch-delimited gaze-swipe technique.} 
    \Description{Three panel images compare Finger-Tap, Gaze\&Pinch, and SwEYEpinch-Basic. The top row shows VR screenshots: Finger-Tap depicts two virtual hands tapping keys; Gaze\&Pinch shows a highlighted key selected while a virtual hand makes a pinch; SwEYEpinch-Basic displays a swipe trace across keys during a pinch gesture. The bottom row shows a person wearing a VR headset demonstrating each gesture: tapping forward with both index fingers, fixating while pinching, and pinching to begin or end a swipe, with short labels under each photo.}
    \label{fig:us1_methods} 
\end{figure}

\paragraph{Finger-Tap.} Users aim their tracked finger at the desired key and tap (Figure \ref{fig:us1_methods} left). This method relies on direct, physical mapping but requires significant arm movement.

\paragraph{Gaze\&Pinch.} This method proceeds similarly to tapping, except the user fixates their gaze on a target key and performs a “pinch and release” gesture (Figure \ref{fig:us1_methods} middle). 

However, both the tracked fingertip and the gaze can easily land near key boundaries, introducing uncertainty over the user’s intended letter. Because letter-by-letter typing is subject to this ``fat-finger'' problem, for both baselines, we compute the most probable letter by combining a 2D Gaussian centered at the tap/gaze location and a character-level $n$-gram (see \autoref{appendix: Fat-Finger Avoidance via 2D Gaussian and Character-Level n-gram} for details). In addition, recognizing the late-trigger problem often found with gaze-based inputs, we include an analysis of this issue in the supplementary material.

\review{While this compares our predictive swipe technique against character-level tapping, this experimental design aligns with established prior work~\cite{kurauchi2016eyeswipe,yu2017tap,kumar2020tagswipe}, which benchmarks novel predictive methods against standard non-predictive defaults to determine the practical performance delta over the status quo.} Prior gaze-only or hand-only methods either incur dwell latency/Midas touch or high physical demand \cite{luong2023controllers, buckingham2021hand, akhoroz2024poke}. \review{ Although SwEYEpinch-Basic employs word prediction as part of its design, \textit{Finger-Tap} and \textit{Gaze\&Pinch} do not natively include predictive assistance. In US1, we intentionally evaluate methods without adding additional prediction to the baselines, so that we could isolate the effect of interaction mechanics alone without confounding these results with differing levels of predictive support.} Thus, US1 compares \textit{SwEYEpinch-Basic} to these production-style XR baselines to test the following hypotheses:


\begin{itemize}
  \item \textbf{H1.1.}
  After limited practice (Session~5), \textit{SwEYEpinch-Basic} achieves higher text-entry speed (WPM) than \textit{Finger-Tap} and \textit{Gaze\&Pinch}.

  \item \textbf{H1.2.}
  By Session~5, \textit{SwEYEpinch-Basic} maintains non-inferior TER relative to \textit{Finger-Tap} and \textit{Gaze\&Pinch}.

  \item \textbf{H1.3.} 
  \textit{SwEYEpinch-Basic} lies on or above the speed–preference Pareto frontier relative to both baselines; at similar speeds it is preferred more, and at similar preference it is faster.
\end{itemize}

\subsection{Study Design}
We used a within-subjects, five-session design (max one session/day). Each session began with two warm-up phrases per technique, then a blocked transcription task with three techniques: \textit{Finger-Tap}, \textit{Gaze\&Pinch}, and \textit{SwEYEpinch-Basic}. Participants transcribed 12 unique MacKenzie–Soukoreff phrases \cite{mackenzie2003phrase} per technique (36 per session), with technique order counterbalanced across participants and sessions. Phrases were not repeated for a given participant across sessions. After each session, participants completed a raw NASA Task Load Index (TLX) and provided a technique preference ranking with a brief justification.

\subsection{Participants}
Forty participants (ages 14 to 34, \finalminorreview{$\bar{x}=24.5$}; \review{23 male, 16 female, 1 non-binary}) completed US1. All reported normal or corrected-to-normal vision and proficient English. Previous exposure to XR was common (62.5\%), but only 10\% had used XR keyboards before.

\subsection{Results and Discussion}
\label{sec: user_study_1_results_and_discussion}

\begin{figure*}[!t]
    \centering
    \begin{subfigure}[t]{0.25\textwidth}
        \centering
        \includegraphics[width=\linewidth]{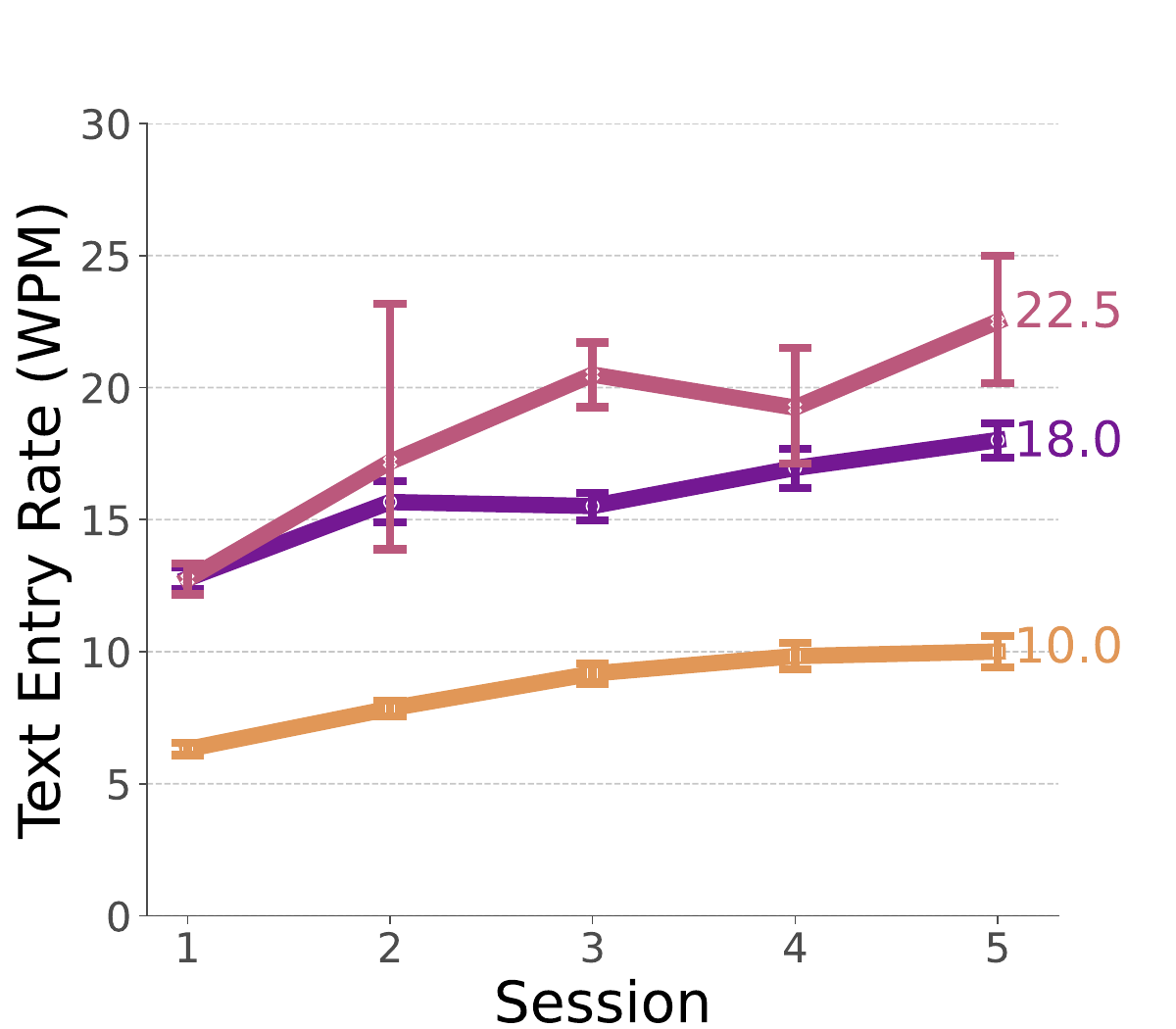}
        \phantomsubcaption\label{fig:us1_wpm}
    \end{subfigure}
    \hfill
    \begin{subfigure}[t]{0.25\textwidth}
        \centering
        \includegraphics[width=\linewidth]{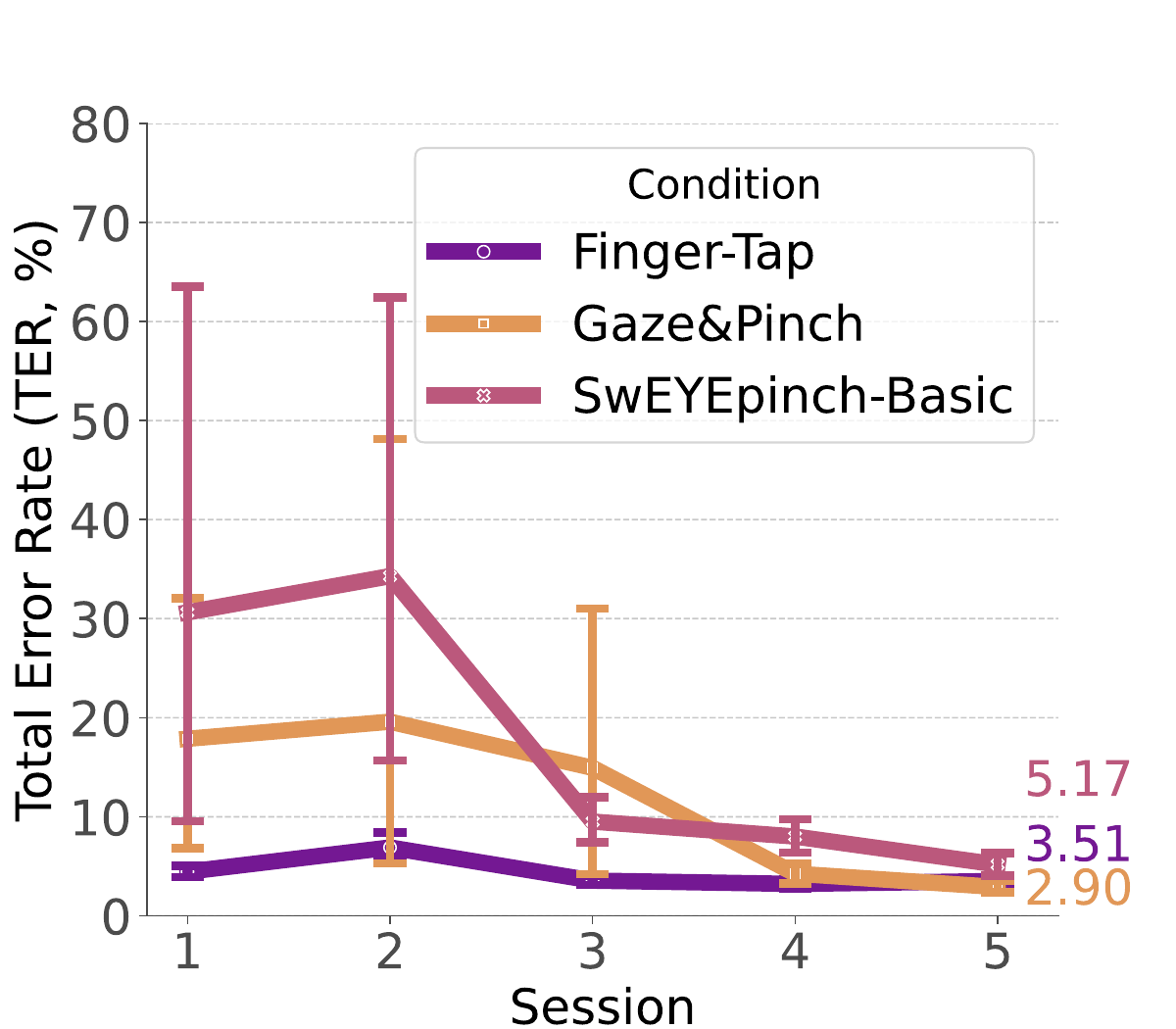}
        \phantomsubcaption\label{fig:us1_total_error_rate}
    \end{subfigure}
    \hfill
    \begin{subfigure}[t]{0.48\textwidth}
        \centering
        \includegraphics[width=\linewidth]{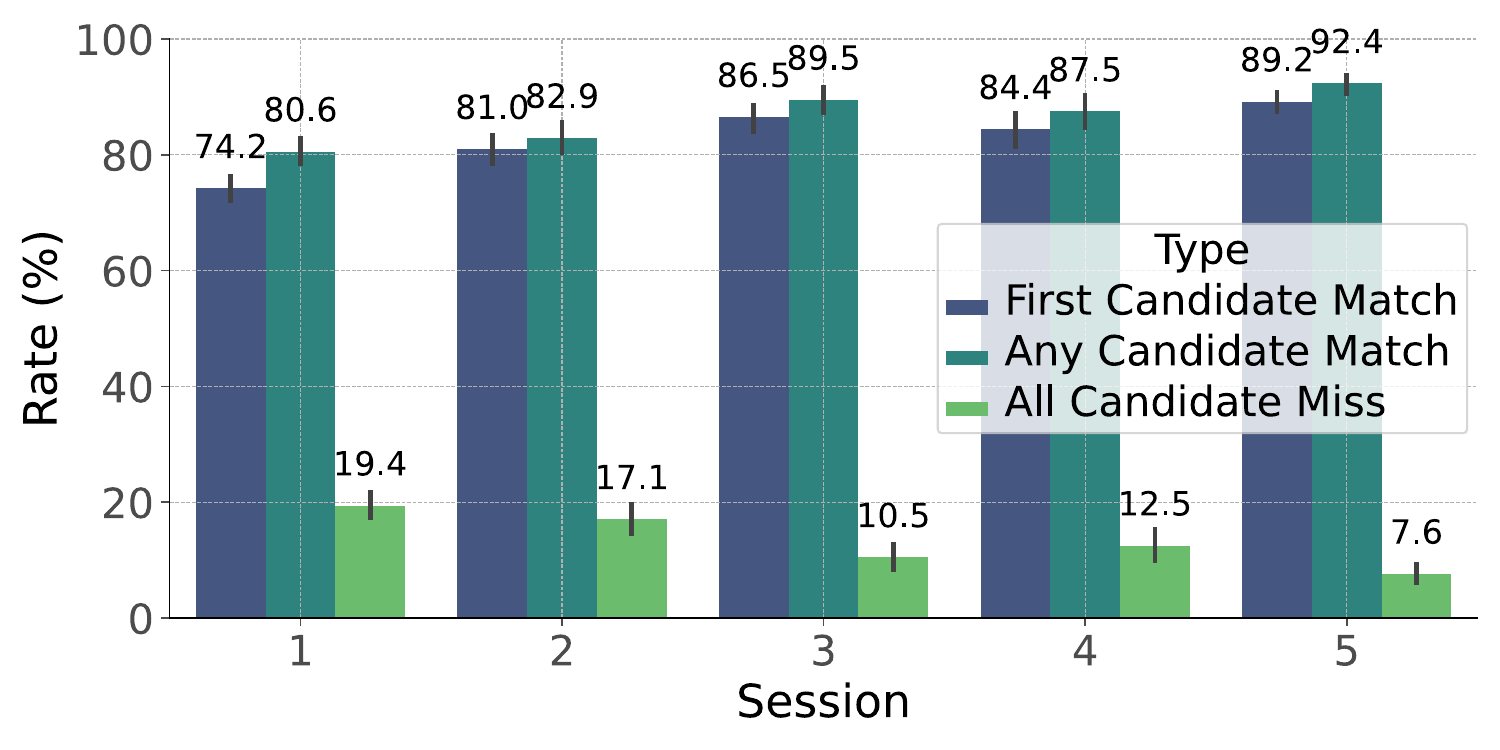}
        \phantomsubcaption\label{fig:us1_sweyepe_miss_rate}
    \end{subfigure}
    
    \caption{Performance results from US1 across five sessions. 
    From left to right, WPM for the three techniques in US1, TER by the conditions in US1. \textit{SwEYEpinch-Basic} error and match rates across sessions. Each bar represents the mean percentage of text-entry outcomes for three categories: 
    (1) \textit{First Candidate Match}: the first suggestion was correct; most desirable because the user can move on to the next word if the first candidate matches,
    (2) \textit{Any Candidate Match}: at least one suggestion was correct,
    and (3) \textit{All Candidate Miss}: no suggested candidates matched.     Overall, the figure reveals how SwEYEpinch-Basic’s suggestion accuracy evolves with practice over multiple sessions. Error bars show standard errors.}
    \Description{Three graphs show performance results from US1 across five sessions for three techniques: Finger-Tap, Gaze\&Pinch, and SwEYEpinch-Basic. The left graph shows text entry rate in words per minute as a line chart, with performance increasing over sessions for all three techniques. Only Session~5 endpoints are labeled: SwEYEpinch-Basic reaches 22.5~WPM, Finger-Tap reaches 18.0~WPM, and Gaze\&Pinch reaches 10.0~WPM. Error bars are shown at each session. The middle graph shows total error rate as a line chart, with error percentages decreasing over sessions for all three techniques. Larger error bars appear in early sessions and shrink over time. At Session~5, SwEYEpinch-Basic has a total error rate of 5.17\%, Finger-Tap has 3.51\%, and Gaze\&Pinch has 2.90\%. The right graph shows match and miss rates as a grouped bar chart with three categories: First Candidate Match, Any Candidate Match, and All Candidate Miss, plotted across Sessions~1 to~5. Each bar includes a numeric label and an error bar. In Session~1, First Candidate Match is 74.2\%, Any Candidate Match is 80.6\%, and All Candidate Miss is 19.4\%. In Session~2, the values are 80.6\%, 82.9\%, and 17.1\%, respectively. In Session~3, the values are 86.5\%, 89.5\%, and 10.5\%. In Session~4, the values are 84.4\%, 87.5\%, and 12.5\%. In Session~5, the values are 89.2\%, 92.4\%, and 7.6\%. Overall, the right graph shows increasing match rates and decreasing miss rates across the five sessions.}
    \label{fig:us1_results}
\end{figure*}

\begin{table}[ht]
\centering
\begin{tabular}{l c}
\hline
\textbf{Condition} & \textbf{Learning Rate (WPM/Session)} \\
\hline
SwEYEpinch-Basic & 2.10 \\
Gaze\&Pinch & 1.07 \\
Finger-Tap  & 1.42 \\
\hline
\end{tabular}
\caption{Average learning rate (WPM per session).}
\Description{A table listing the average learning rate, measured in words per minute per session, for three input conditions. SwEYEpinch-Basic has a learning rate of 2.10 WPM per session, Gaze\&Pinch has 1.07 WPM per session, and Finger-Tap has 1.42 WPM per session.}
\label{tab:us1_wpm_learning_rate}
\end{table}


\begin{figure*}[h]
    \centering
    \begin{subfigure}[t]{\textwidth}
        \centering
        \includegraphics[width=\linewidth]{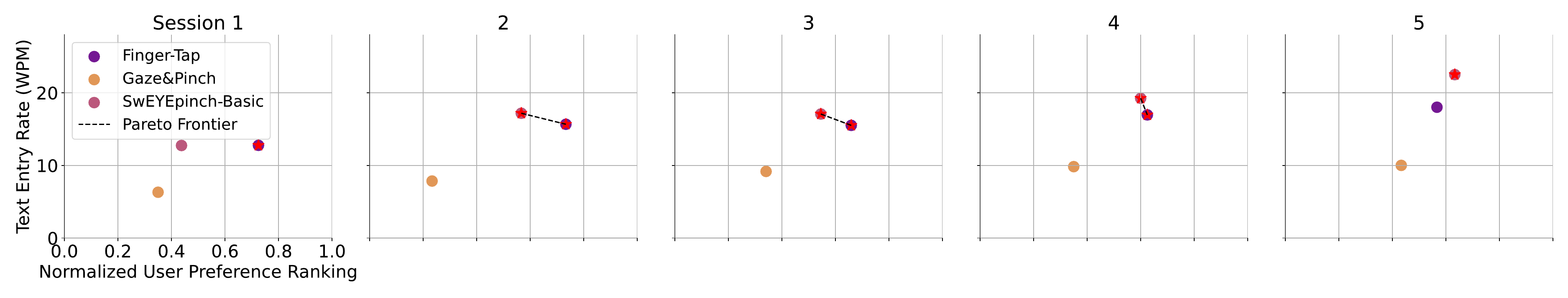}
        \phantomsubcaption\label{fig:us1_pareto}
    \end{subfigure}
    \begin{subfigure}[t]{\textwidth}
        \centering
        \includegraphics[width=\linewidth]{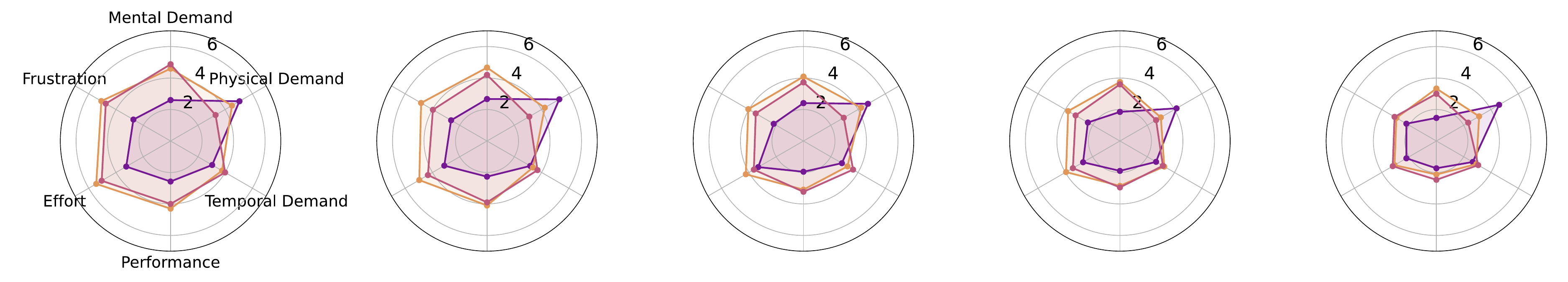}
        \phantomsubcaption\label{fig:us1_tlx}
    \end{subfigure}
    \caption{Top: Pareto frontiers showing normalized user preference vs.\ WPM. Bottom: Raw NASA TLX scores.}
    \Description{A ten-panel figure summarizes speed--preference trade-offs and workload over five sessions for three techniques: Finger-Tap (purple), Gaze\&Pinch (orange), and SwEYEpinch-Basic (red). The top row shows results for Sessions~1 to~5 as scatter plots with the x-axis representing normalized preference ranking, where 0 indicates least preferred and 1 indicates most preferred, and the y-axis representing text entry rate in words per minute. A dashed line indicates the Pareto frontier in each session. Across sessions, SwEYEpinch-Basic moves steadily upward and to the right and lies on the Pareto frontier; by Session~5, it achieves the highest text entry rate of approximately 23~WPM with high preference. Finger-Tap shows mid-to-high text entry rate of approximately 18~WPM with moderate preference, while Gaze\&Pinch remains slower at approximately 10~WPM and less preferred. The bottom row shows results for Sessions~1 to~5 as radar charts of NASA--TLX workload subscales, including Mental Demand, Physical Demand, Temporal Demand, Performance, Effort, and Frustration, measured on a 1--7 scale. Across sessions, overall perceived workload decreases. SwEYEpinch-Basic exhibits comparable or lower physical demand, effort, and frustration than the other techniques, and shows equal or better perceived performance in later sessions.}
    \label{fig:us1_session15_pareto_and_tlx}
\end{figure*}

\noindent\textbf{SwEYEpinch-Basic is faster than both baselines after five sessions, supporting H1.1.} By Session~5, \textit{SwEYEpinch-Basic} was faster than both baselines (Fig.~\ref{fig:us1_results}\subref{fig:us1_wpm}): $22.50$~WPM vs.\ \textit{Finger-Tap} $18.01$~WPM (\review{$p=2.55\times 10^{-5}$}) and \textit{Gaze\&Pinch} $10.01$~WPM (\review{$p<10^{-32}$}). Participants linked the gain to forgiving decoding and reduced letter-by-letter micromanagement: \texttt{P104} “\emph{Swipe was easy to understand and allowed for errors with the suggestion box},” \texttt{P118} “\emph{Swiping did incredibly well at guessing the words I was trying to type}.”

\noindent\textbf{SwEYEpinch-Basic does not meet accuracy non-inferiority at Session~5, not supporting H1.2.} TER was higher than both baselines at Session~5: $5.17\%$ vs.\ \textit{Finger-Tap} $3.51\%$ (\review{$p=0.004$}) and vs.\ \textit{Gaze\&Pinch} $2.90\%$ (\review{ $p=1.65\times10^{-4}$}); \textit{Finger-Tap} and \textit{Gaze\&Pinch} did not differ (\review{$p=0.088$}). Without knowing what word will be predicted during a swipe, users point out they have to frequently gaze upward and check: \texttt{P134}: “\emph{I kept having to look at the input field and it would ruin the word I was typing}.” Several participants found a workable tactic—anchoring the first letter—to stabilize recognition (\texttt{P117}: “\emph{Focusing on getting the first letter correct really helped.}”). Because the decoder conditions on the pinch-onset location, it prunes the lexicon to candidates whose first letter lies near that start point. These observations directly motivate \textit{SwEYEpinch}’s live mid-swipe candidates and low-effort correction (mid-swipe deletion, deletion peek window) to reduce TER.

\noindent\textbf{SwEYEpinch-Basic sits on the speed–preference Pareto frontier and dominates Gaze\&Pinch, supporting H1.3.}  The Session~5 Pareto plot (WPM vs.\ preference) places \textit{SwEYEpinch-Basic} on the frontier and dominating \textit{Gaze\&Pinch}; \textit{Finger-Tap} trades familiarity for higher physical demand (Fig.~\ref{fig:us1_session15_pareto_and_tlx}). Comments reflect this trade-off: \texttt{P108} “\emph{FingerTap is reliable but very physically taxing; swipe was fun and easy to fix when it went wrong},” \texttt{P112} “\emph{Once I mastered the coordination, swipe was the simplest to use}.”

\begin{table}[t]
\centering
\caption{Decoder comparison on US1 (19{,}328 traces; $n{=}40$). Running time (RT) or latency is the average time it takes to decode the candidates from the swipe trace. Our method runs DTW on I-VT{+}DBSCAN–pruned traces. $^{\dagger}$ST-DTW: spatiotemporal DTW. $^{\dagger\dagger}$ w/o td: without time dimension}
\Description{Table 2 compares decoder performance on US1 (19{,}328 swipe traces, $n=40$). Columns report Method, Top-1 accuracy (\%), Top-4 accuracy (\%), and average running time (RT, ms). SwEYEPinch with spatio-temporal DTW (ST-DTW) achieves the highest Top-1 accuracy (77.9\%) and Top-4 accuracy (87.5\%) with low latency (2.38 ms). SwEYEPinch without the time dimension (w/o td) shows reduced accuracy (Top-1: 72.1\%, Top-4: 85.9\%) and similar runtime (2.32 ms). SwipeType reports 75.7\% Top-1 and 86.1\% Top-4 accuracy with substantially higher latency (18.4 ms). HGaze achieves 71.9\% Top-1 and 81.3\% Top-4 accuracy with 2.44 ms runtime.}
\begin{tabular}{lccc}
\toprule
\textbf{Method} & \textbf{Top-1 (\%)} & \textbf{Top-4 (\%)} & \textbf{RT (ms)} \\
\midrule
SwEYEpinch (ST-DTW$^{\dagger}$) & 77.9 & 87.5 & 2.38 \\
SwEYEpinch (w/o td$^{\dagger\dagger}$)      & 72.1 & 85.9 & 2.32 \\
EyeSwipe~\cite{kurauchi2016eyeswipe}             & 75.7 & 86.1 & 18.4 \\
HGaze~\cite{feng2021hgaze}                        & 71.9 & 81.3 & 2.44 \\
\bottomrule
\end{tabular}
\vspace{2pt}
\small 
\label{tab:decoder_comp}
\end{table}

\noindent
\paragraph{Comparison to decoders in prior works.}
To contextualize our algorithmic design, we conducted a post-hoc evaluation on the gaze traces collected from US1's \textit{SwEYEpinch-Basic} condition. Alongside our decoder with filtered gaze trace and spatiotemporal DTW, we re-implemented the two most relevant gaze-trace baselines: (i) \emph{ EyeSwipe}—DTW on (near-)raw gaze with simple outlier removal, and (ii) \emph{HGaze}—Fréchet distance on a temporally averaged (smoothed) trace. As Table~\ref{tab:decoder_comp} shows, SwEYEpinch yields the best Top-1/Top-4 while remaining in the same latency class as DTW variants; EyeSwipe is slower because it aligns largely raw traces, whereas we prune with I-VT{+}DBSCAN. Removing timestamps in our DTW drops Top-1 by 5.8 points (77.9$\rightarrow$72.1), highlighting the benefit of the weak forward-time prior; HGaze underperforms due to Fréchet’s sensitivity to single noisy fixations. We did not include \emph{GlanceWriter} in this offline comparison because its delimiter (crossing a line above the keyboard) changes the gaze-trace geometry, so replaying \textit{SwEYEpinch} traces would be invalid. Instead, we compare against \textit{GlanceWriter XR} directly in US2.

\subsubsection{Hyperparameter sensitivity}
Using the same US1 corpus as the decoder-baseline comparison, we swept the core preprocessing parameters: I-VT velocity threshold in $[50,150]$\,deg/s, DBSCAN $\varepsilon\!\in\![0.05,0.15]$ (normalized keyboard units), and $\mathit{minPts}\!\in\![2,6]$. Across this grid, mean per-trace latency was essentially unchanged (2.32–2.47\,ms; mean 2.39\,ms). Accuracy improved when lowering the I-VT threshold from 150 to 100\,deg/s and then saturated (\,<1\% additional gain). Varying DBSCAN $\varepsilon$ or $\mathit{minPts}$ had a negligible effect on accuracy. These results support the defaults used in our main experiments and indicate that the decoder is robust to reasonable parameter choices.

\noindent\textit{Design insight.} These results reinforce our principle to \emph{decouple targeting from commitment}: gaze for fast, low-effort tracing and a tiny pinch to commit yields substantial speed gains, with a small accuracy cost that, as later studies would reveal, can be mitigated via mid-swipe feedback and low-effort correction tools.

\section{User Study 2: SwEYEpinch and other Gaze-Swipe Baselines}
\label{sec:user_study_2}

US1 confirmed our design intuition ``gaze for targeting, pinch to confirm'' is advantageous: pinch-delimited gaze swipe (\textit{SwEYEpinch-Basic}) can beat production-style, letter-by-letter input on speed and preference. However, it also exposed a key limitation: candidates appear only at commit, prompting off-keyboard verification glances that contribute to higher error rates. We therefore developed an improved version, which we describe next.

\subsection{Improving SwEYEpinch-Basic: SwEYEpinch}
\label{sec: partial_sweyepe}

Participants in US1 found SwEYEpinch-Basic to be the most preferred text-entry method overall, even though the speed advantage emerged later as the sessions progressed. To further enhance both the efficiency and usability of SwEYEpinch-Basic, we introduce a refined variant called \textit{SwEYEpinch}.

\textit{SwEYEpinch} introduces two key improvements over SwEYEpinch-Basic: \textbf{(1)~Mid-Swipe Prediction}, which displays word candidates as the user swipes, and \textbf{(2)~Mid-Swipe Deletion}, which allows users to cancel an in-progress swipe if the predictions are incorrect. We describe each in detail below.

\paragraph{Mid-Swipe Prediction}
\review{Inspired by the Swype keyboard,} we display candidate words above the current key whenever the user saccades from one letter to another \textit{within} the same swipe. This preview allows the user to see potential completions without looking at every letter of the intended word. \review{However, unlike touch-anchored Swype, when a user has just started an eye swipe, and relatively few gaze points have been captured, the gaze-based distance matching can be unreliable. It often proves more effective to rely on the language model.}

To accommodate this, we multiply the fusion parameter $\alpha$ (see Section \ref{sec: sweyepe decoding algorithm}) with an adaptive weight that decreases linearly from $3$ to $1$ as the number of gaze points increases from $1$ to $3$. By doing so, we triple the importance of the $n$-gram probabilities when there is only one gaze point available, then smoothly decrease to the original $\alpha$ when there are three or more gaze points. This increases the influence of the language model when evidence is sparse and smoothly returns to the base $\alpha$ once the trace stabilizes. \review{Our choices for these values were informed by pilot studies.}

While mid-swipe prediction is standard on mobile touchscreens (e.g., Android), porting this interaction to gaze is non-trivial due to signal disparity. Touch input is relatively sparse and precise (high signal-to-noise ratio). In contrast, gaze data is high-frequency (200\,Hz in our setup), voluminous, and prone to jitter and micro-saccades. Naively applying standard decoding (e.g., frame-by-frame DTW on raw gaze) introduces latency that breaks the feedback loop. Prior work on gaze-swipe systems~\cite{kurauchi2016eyeswipe,yu2017tap,cui2023glancewriter, kumar2020tagswipe,feng2021hgaze} delays prediction until after a swipe is complete.
Therefore, our primary technical contribution is the \textbf{optimization pipeline} that makes continuous gaze-decoding feasible. By chaining velocity-thresholding (I-VT) with density-based clustering (DBSCAN), we reduce the swipe trace cardinality by over 90\% (see \autoref{appendix: Optimization in SwEYEpinch Decoding}), allowing our Spatiotemporal DTW to run within the required frame budget for fluid UI updates.}

\noindent
\paragraph{Mid-Swipe Deletion}
In early testing, we observed that users sometimes recognized \textit{during} a swipe that the emerging candidate words were incorrect, and they could not make a correction mid-gesture. This is especially true when the starting letter is wrong because we filter the candidates by gaze position at the start of a swipe. With \textit{SwEYEpinch-Basic}, they would have to abort the swipe by releasing \minorreview{the} pinch, then fixate and pinch again on the delete key. Instead, \review{\textit{SwEYEpinch} allows} the user to \textit{cancel} a swipe mid-gesture: if none of the displayed mid-swipe predictions are correct, the user can continue swiping to the delete key and release pinch there, discarding the swipe entirely. This “Mid-Swipe Deletion” streamlines error correction by eliminating a separate “exit + delete” action.

\noindent
\paragraph{Deletion Peek Window}
During our early testing, we observed that the user, when needing to delete longer text, often had to repeatedly shift their gaze up and down between the delete key to pinch at it and the text input field to check their inputs. To mitigate this disruptive behavior, we added a small contextual window that appears above the deletion button when the user gazes at it; this peek window displays the content of the input box. This way, the user does not need to shift their gaze back to the actual input box to check its content while deleting.

\subsection{Other XR-native, Gaze-Swipe baselines}
\review{Moreover, to isolate specific contributions of our design against comparable \textit{XR-native} techniques (i.e., those not requiring a physical surface like TagSwipe~\cite{kumar2020tagswipe}), we benchmark against} two representative gaze-swipe approaches: \textit{SkiMR} and \textit{GlanceWriter XR}.

\begin{figure*}[h] 
    \centering 
    \includegraphics[width=\textwidth]{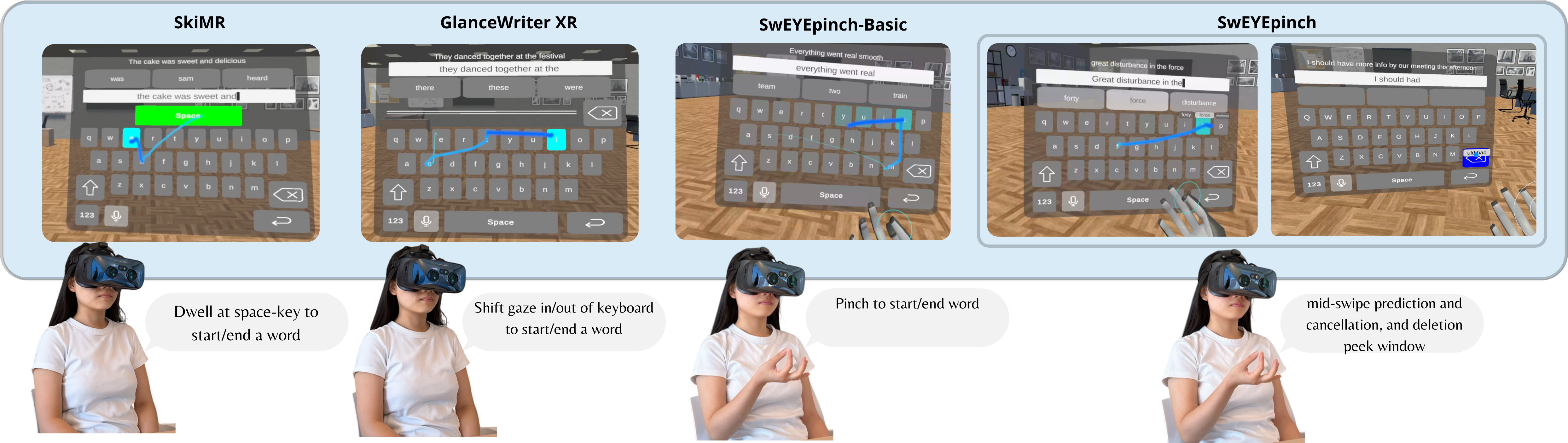} 
    \caption{Techniques evaluated in US2 are gaze-swipe baselines with different delimiters: \textit{SkiMR} (the delimiting space key is placed on top of the keyboard following \cite{hu2024skimr}), \textit{GlanceWriter XR}, and our proposed \textit{SwEYEpinch-Basic} and its improved version: \textit{SwEYEpinch}.}
    \Description{The figure compares four gaze-swipe text-entry techniques using VR screenshots and photos of a user demonstrating each delimiter gesture. The top row shows keyboard screenshots with colored swipe traces. Under SkiMR, a blue trace ends on a highlighted green space bar. Under GlanceWriter XR, a blue trace appears across keys with the space bar below; the delimiter is triggered by shifting gaze in or out of the keyboard. Under SwEYEpinch-Basic, a blue swipe trace is shown across the keyboard while a virtual hand performs a pinch. Under SwEYEpinch, two images appear: one showing a predicted word list during a mid-swipe and another showing a blue deletion preview box. The bottom row shows a user wearing a VR headset performing the corresponding actions: dwelling at the space bar, shifting gaze, pinching to start or end a word, and using pinch with mid-swipe prediction and cancellation.}
    \label{fig:us2_methods} 
\end{figure*}
\paragraph{SkiMR \cite{hu2024skimr}.} This method introduces a dwell-free gaze swipe where users confirm word entry by explicitly fixating on the \textit{space} key instead of relying on dwell timers or pinch gestures. Apart from the explicit space-bar confirmation and decoding algorithm, all other aspects of the method align with standard gaze-swipe approaches.

\paragraph{GlanceWriter XR} We adapt the original desktop GlanceWriter~\cite{cui2023glancewriter} to XR. Users start and stop a swipe by entering and exiting the keyboard area.

\review{We chose \textit{SkiMR} and \textit{GlanceWriter XR} because we restrict US2 to XR-native, gaze-swipe techniques that satisfy the same HWD-only sensing constraints as SwEYEpinch. This allows us to isolate the effect of the delimiter (pinch vs.\ space-key fixation vs.\ line crossing) and mid-swipe prediction without introducing confounds from additional hardware (e.g., touchscreens) or modalities (e.g., voice).}

These baselines do not involve hand motion but still couple targeting and commitment to gaze alone. US2 therefore asks whether bringing \emph{mid-swipe prediction} into a pinch-delimited design can cut verification glances and raise entry rate without harming accuracy—and how much of the gain comes from the \emph{delimiter} itself. After limited practice (three sessions), we test three hypotheses:

\begin{itemize}
  \item \textbf{H2.1.}
  \textit{SwEYEpinch} (mid-swipe prediction) is faster than \textit{SwEYEpinch-Basic} (post-swipe only) with equal or lower TER.

  \item \textbf{H2.2.}
  Pinch-delimited swipes outperform gaze-only methods (\textit{SkiMR}, \textit{GlanceWriter XR}) in WPM and preference without an accuracy penalty.

  \item \textbf{H2.3.} The two-stage pruning\,+\,spatiotemporal DTW decoder (\texttt{Gaze2Word}, Algorithm \ref{alg:g2w}) yields higher top-1/any candidate-match rates than \textit{GlanceWriter XR}’s prefix-trie decoder (under the same apparatus/keyboard), translating to higher WPM without a TER increase.

\end{itemize}

\subsection{Study Design}
The procedure is identical to that of US1, except that participants take part in three sessions instead of five.

\subsection{Participants}
Twenty-one participants (ages 19 to 34; \finalminorreview{$\bar{x}=23.6$}; \review{8 male, 13 female}) completed US2. All reported proficient English and normal or corrected-to-normal vision; one was left-handed and one reported a history of strabismus. XR familiarity was common (66.7\%), but only one participant had prior XR-typing experience.

\subsection{Results and Discussion}
\label{sec:us2_results}

\begin{figure*}[!t]
    \centering
    \begin{subfigure}[t]{0.25\textwidth}
        \centering
        \includegraphics[width=\linewidth]{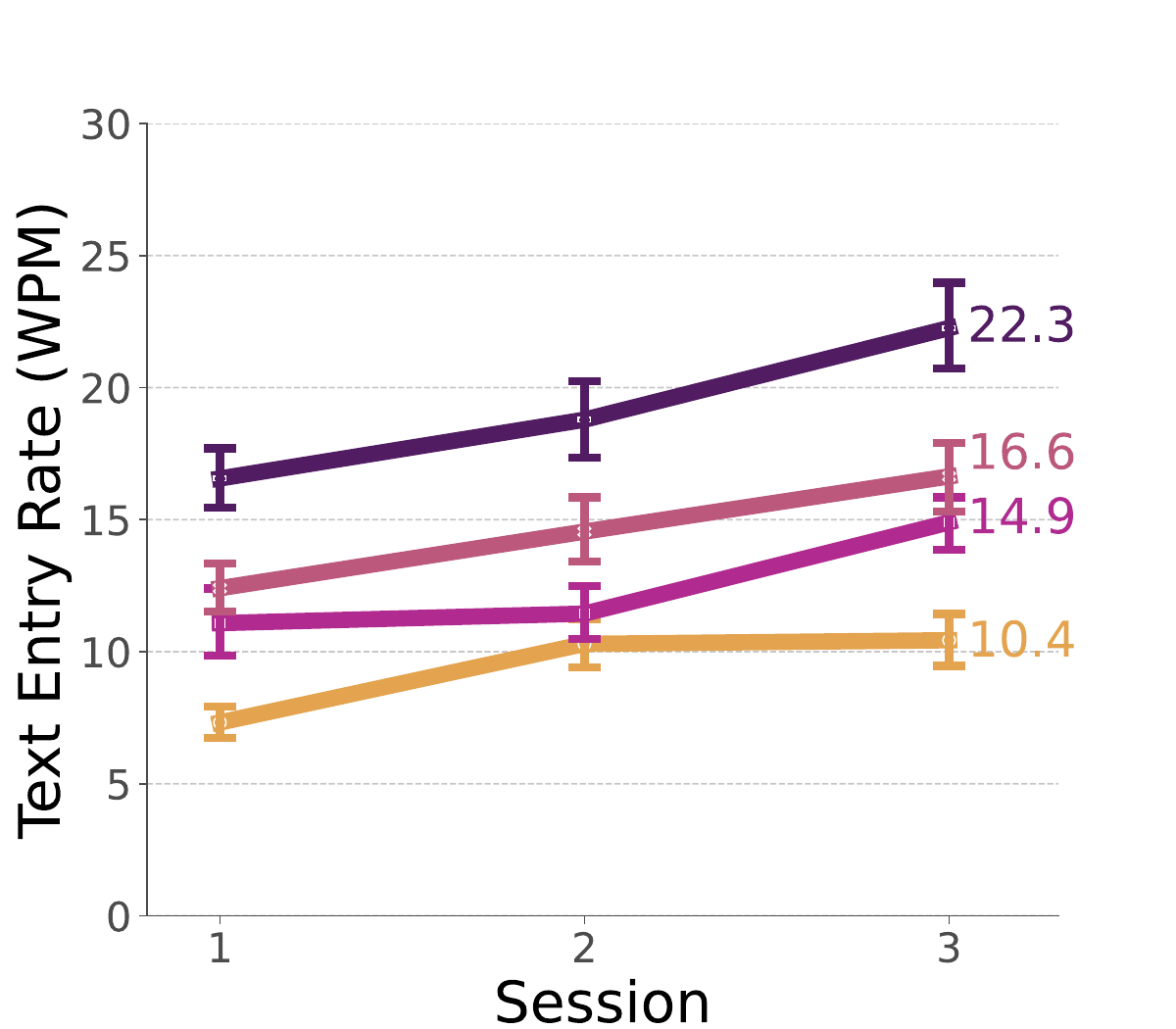}
        \phantomsubcaption\label{fig:us2_wpm}
    \end{subfigure}
    \hfill
    \begin{subfigure}[t]{0.25\textwidth}
        \centering
        \includegraphics[width=\linewidth]{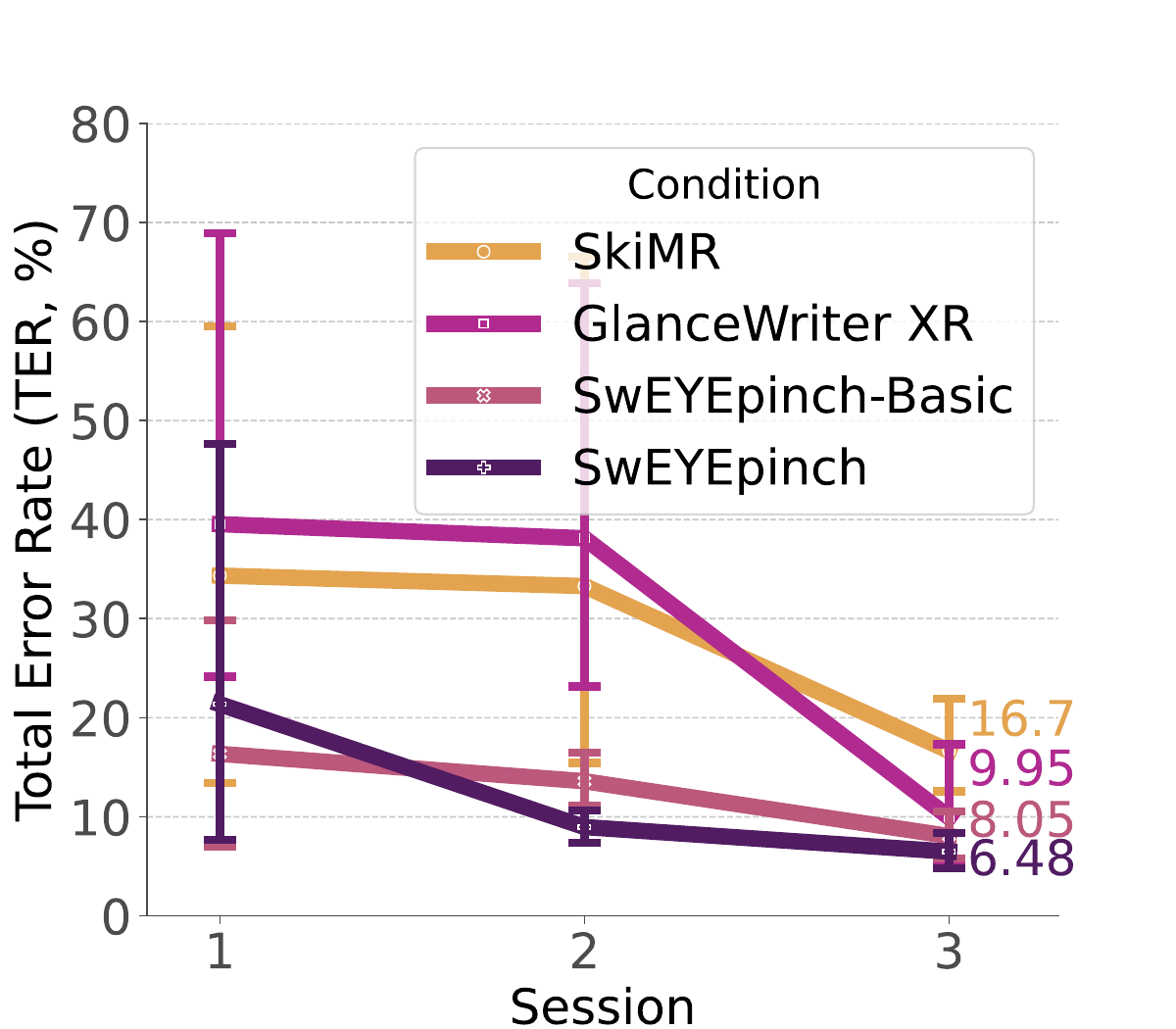}
        \phantomsubcaption\label{fig:us2_ter}
    \end{subfigure}
    \hfill
    \begin{subfigure}[t]{0.48\textwidth}
        \centering
        \includegraphics[width=\linewidth]{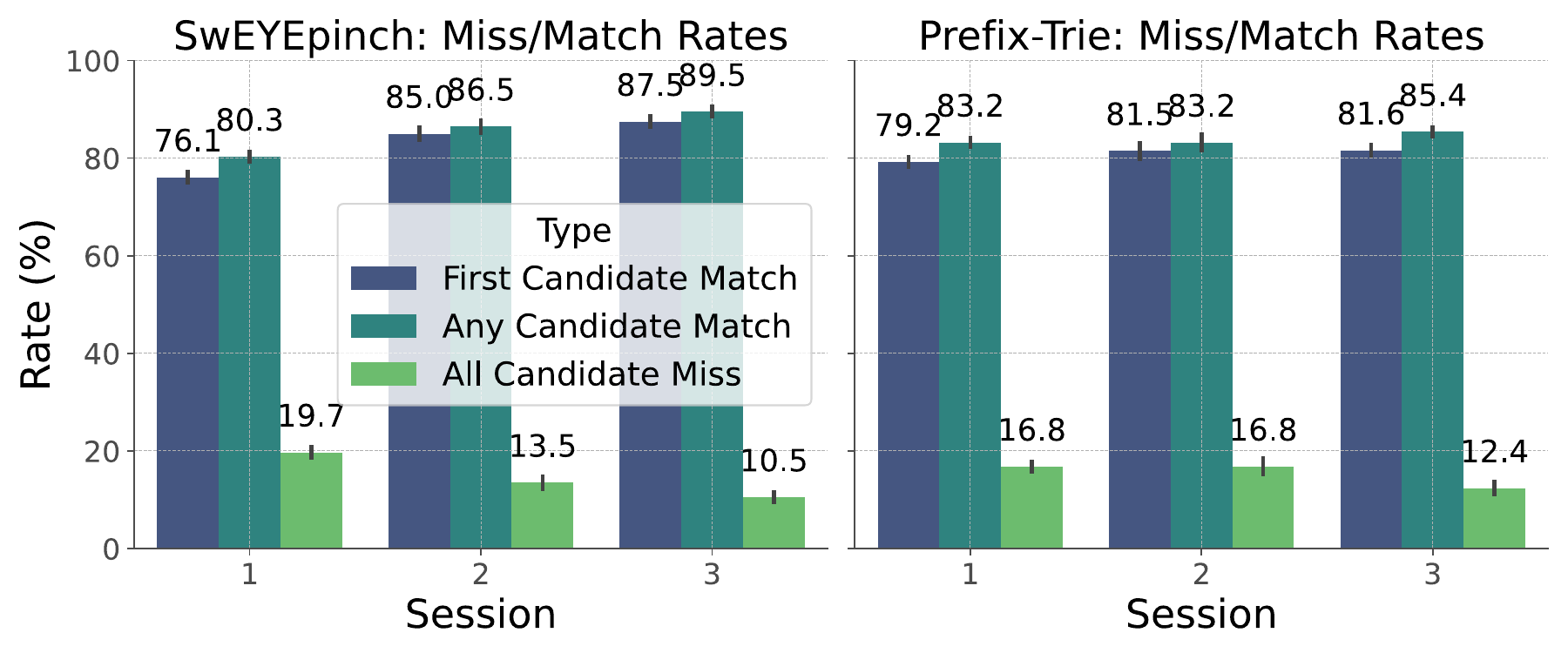}
        \phantomsubcaption\label{fig:us2_miss_rates}
    \end{subfigure}
    \caption{Performance results from US2. 
    From left to right: WPM for the three techniques across sessions, TER across conditions. Error bars show standard errors; the match and miss rates are explained in Figure \ref{fig:us1_sweyepe_miss_rate}. Comparing the decoding algorithm: \emph{SwEYEpinch} vs GlanceWriter \cite{cui2023glancewriter}.}
    \Description{The figure contains four performance charts from US2. The left graph shows text entry rate in words per minute as a line chart across Sessions~1 to~3 for SkiMR, GlanceWriter~XR, SwEYEpinch-Basic, and SwEYEpinch. All techniques show increasing text entry rate across sessions, with error bars at each point. Only Session~3 endpoints are labeled: SwEYEpinch reaches 22.3~WPM, GlanceWriter~XR reaches 16.6~WPM, SwEYEpinch-Basic reaches 14.9~WPM, and SkiMR reaches 10.4~WPM. The middle graph shows total error rate as a line chart decreasing across Sessions~1 to~3 for all four techniques. Error bars appear at each point, with larger bars in earlier sessions. At Session~3, SkiMR has a total error rate of 16.7\%, GlanceWriter~XR has 9.95\%, SwEYEpinch-Basic has 8.05\%, and SwEYEpinch has 6.48\%. The right side of the figure contains two grouped bar charts showing match and miss rates across Sessions~1 to~3 for two decoders: SwEYEpinch and Prefix-Trie. Each bar includes a numeric label and an error bar. For SwEYEpinch, Session~1 shows First Candidate Match of 76.1\%, Any Candidate Match of 80.3\%, and All Candidate Miss of 19.7\%. Session~2 shows 85.0\%, 86.5\%, and 13.5\%, respectively. Session~3 shows 87.8\%, 89.5\%, and 10.5\%. For Prefix-Trie, Session~1 shows 79.2\%, 83.2\%, and 16.8\%. Session~2 shows 81.5\%, 83.2\%, and 16.8\%. Session~3 shows 81.6\%, 85.4\%, and 12.4\%. Overall, the bar charts show increasing match rates and decreasing miss rates across sessions for both decoders.}
    \label{fig:us2_results}
\end{figure*}

\noindent\textbf{Mid-swipe preview boosts speed without hurting accuracy, supporting H2.1.} 
\finalreview{\textit{SwEYEpinch} was faster than \textit{SwEYEpinch-Basic} in every session (S1: $16.56{\pm}8.44$ vs.\ $12.40{\pm}7.05$~WPM; S2: $18.78{\pm}8.86$ vs.\ $14.55{\pm}7.09$; S3: $22.25{\pm}8.83$ vs.\ $16.64{\pm}7.30$; all \review{$p{<}0.001$}; Fig.~\ref{fig:us2_results}\subref{fig:us2_wpm}). Learning rates echoed this advantage ($2.85$ vs.\ $2.10$~WPM/session; Table~\ref{tab:us2_wpm_learning_rate}). TER did not worsen with previews and was lower in S2 (\review{$8.96\%$} vs.\ \review{$13.59\%$}, \review{$p{=}0.013$}), becoming indistinguishable by S3 (n.s.; Fig.~\ref{fig:us2_results}\subref{fig:us2_ter}). Participants attributed the gain to fewer “finish--inspect--fix” cycles: \texttt{P204} (S2) “\emph{I could stop as soon as my word showed up}.” Together, these results \textbf{support H2.1}.}

\noindent\textbf{Pinch beats gaze-only delimiters on rate with no accuracy penalty, supporting H2.2.}
\finalreview{Pinch-delimited swipes (\textit{SwEYEpinch}/\textit{SwEYEpinch-Basic}) outpaced both gaze-only baselines across sessions (vs.\ \textit{GlanceWriter XR} and \textit{SkiMR}: all \review{$p{<}0.001$} except \textit{SwEYEpinch-Basic} vs.\ \textit{GlanceWriter XR} in S1 \review{$p{=}0.014$} and S3 \review{$p{=}0.008$}; Fig.~\ref{fig:us2_results}\subref{fig:us2_wpm}), consistent with avoiding the extra saccades to gaze-based delimiter targets. TER showed no pinch penalty; by S3 both pinch conditions were lower than \textit{SkiMR} (\review{$6.48\%$} and \review{$8.05\%$} vs.\ \review{$16.65\%$}, both \review{$p{<}0.001$}), with other pairwise TERs not significant (Fig.~\ref{fig:us2_results}\subref{fig:us2_ter}). Participants emphasized the control of an explicit, low-effort delimiter: \texttt{P221} “\emph{It was easier to decide exactly when the word ended}.” These results \textbf{support H2.2}.}

\noindent\textbf{Decoder-level evidence matches the speedup: by S3, we yield more top-1/any matches, supporting H2.3.}
\finalreview{By S3, the \textit{SwEYEpinch} decoder surpassed \textit{GlanceWriter XR}'s prefix-trie on key outcomes (Fig.~\ref{fig:us2_miss_rates}): \emph{First Candidate Match} $87.5\%$ vs.\ $81.6\%$ (\review{$p{=}1.1{\times}10^{-4}$}) and \emph{Any Candidate Match} $89.5\%$ vs.\ $85.4\%$ (\review{$p{=}0.003$}); \emph{All Candidate Miss} trended lower (10.5\% vs.\ 12.4\%, n.s.). Although prefix-trie led slightly in S1 (top-1 $79.2\%$ vs.\ $76.1\%$, \review{$p{=}0.028$}), this reversed as users adapted to pinch-delimited swipes and mid-swipe decoding, reducing verification and correction overhead. One participant captured this learnability: \texttt{P215} (S3) “\emph{Auto-suggestions came much quicker as I got used to moving with my eyes}.” These results \textbf{support H2.3}.}

\paragraph{Learning trajectories}
\begin{table}[ht]
\centering
\begin{tabular}{l c}
\hline
\textbf{Condition} & \textbf{Learning Rate (WPM/Session)} \\
\hline
SkiMR & 1.49 \\
GlanceWriter XR & 2.05 \\
SwEYEpinch-Basic  & 2.10 \\
SwEYEpinch  & 2.85 \\
\hline
\end{tabular}
\caption{Average learning rate (WPM per session).}
\Description{A table listing average learning rates, measured in words per minute per session, for four text-entry conditions. SkiMR has a learning rate of 1.49 WPM per session, GlanceWriter XR has 2.05 WPM per session, SwEYEpinch-Basic has 2.10 WPM per session, and SwEYEpinch has the highest learning rate at 2.85 WPM per session.}
\label{tab:us2_wpm_learning_rate}
\end{table}


\begin{figure}[h]
    \centering
    \begin{subfigure}[t]{\columnwidth}
        \centering
        \includegraphics[width=\linewidth]{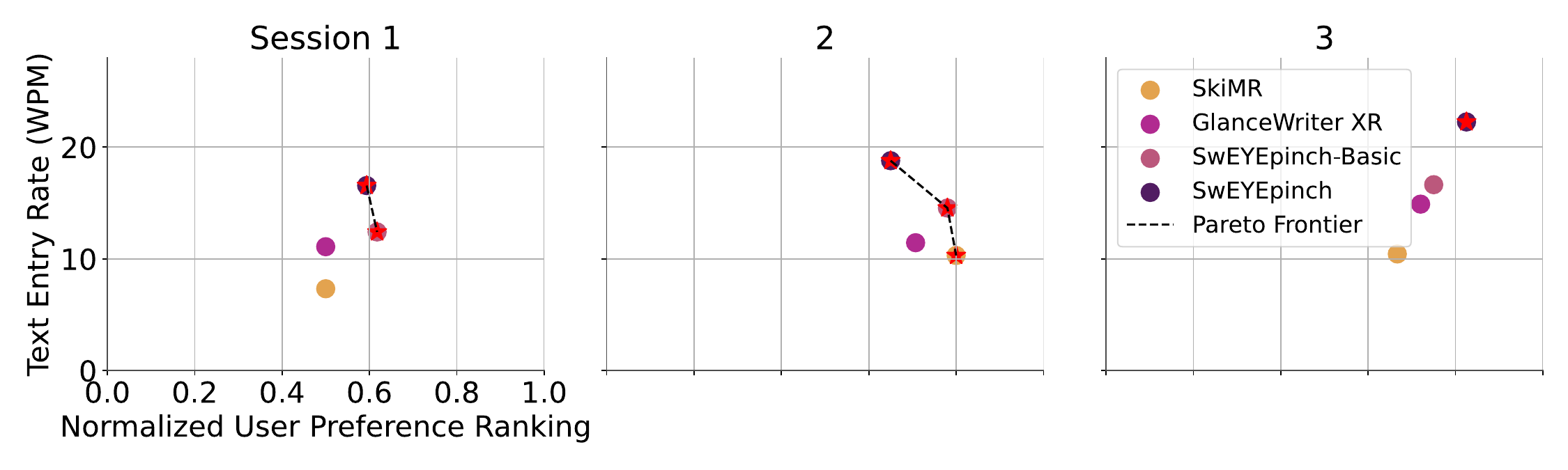}
    \end{subfigure}

    \begin{subfigure}[t]{\columnwidth}
        \centering
        \includegraphics[width=\linewidth]{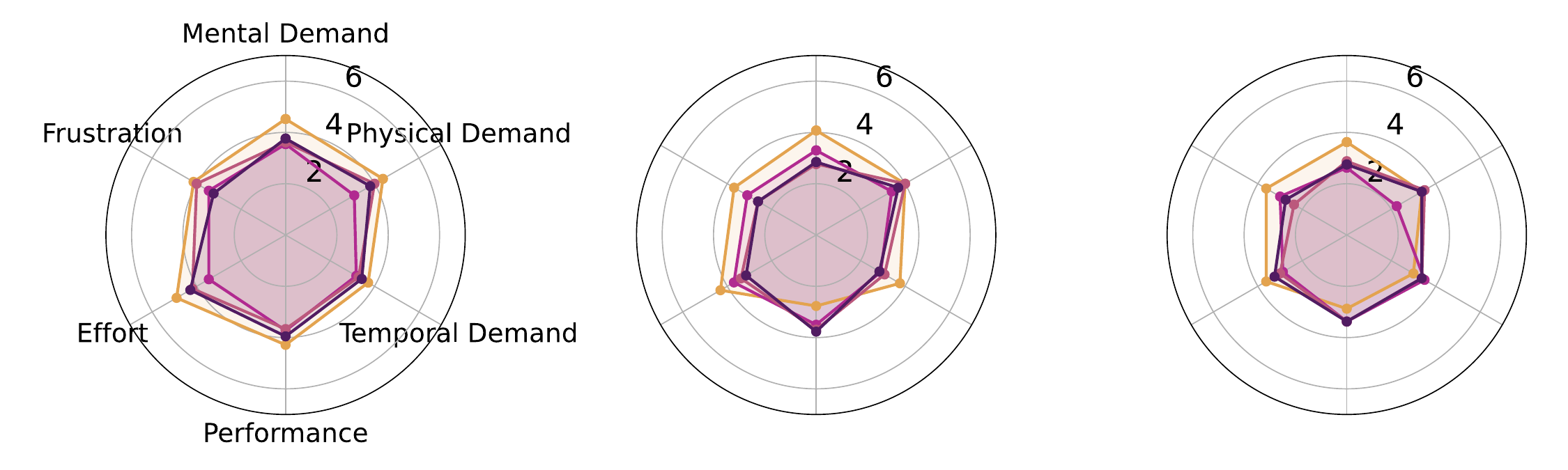}
    \end{subfigure}
    \caption{Top: Pareto frontiers showing normalized user preference vs. WPM. Bottom: Raw NASA TLX scores, across the three sessions in US2}
    \Description{Six-panel figure summarizing speed-preference trade-offs and workload by session for four text-entry methods (SkiMR, GlanceWriter XR, SwEYEpinch-Basic, SwEYEpinch). Top row—three scatter plots (Sessions 1-3): x-axis = Normalized Ranking (0 = least favorite, 1 = favorite), y-axis = WPM. A dashed line marks the Pareto frontier (best combinations of speed and preference). In Session 1, SwEYEpinch-Basic and SwEYEpinch lie on/near the frontier ahead of the gaze-only baselines. In Session 2, the frontier shifts toward SwEYEpinch. By Session 3, SwEYEpinch sits in the upper-right (highest WPM with high preference), while GlanceWriter XR and SkiMR cluster lower with slower speeds and lower preference. Bottom row-three radar (spider) charts for NASA-TLX subscales (Mental, Physical, Temporal Demand, Performance, Effort, Frustration; 1-7 scale): Across sessions, SwEYEpinch shows comparable or lower demands and effort/frustration than others and equal or higher perceived performance, with profiles converging and improving by Session 3.}
    \label{fig:us2_pareto_and_tlx}
\end{figure}

\noindent \textit{Design Insight.} US2 isolates two levers that matter: \emph{mid-swipe preview} (faster with equal or lower errors) and the \emph{pinch delimiter} (faster than gaze-only without an accuracy cost). Qualitative accounts match this picture: previews curb verification glances and shorten swipes, and pinch provides a precise, low-effort “when” that many participants preferred (\texttt{P413}: “\emph{No-hands methods took more concentration than the pinch methods.}”).

\section{User Study 3: Benchmarking Against Stronger Baselines}
\label{sec:user_study_3}

So far, US1 established that a minimal hybrid (\textit{SwEYEpinch-Basic})—using gaze for \emph{where} and a tiny pinch for \emph{when}—sits on the speed–preference Pareto frontier versus production-style, letter-by-letter input. US2 then showed \textit{why} the hybrid works: mid-swipe previews cut verification glances, and the pinch delimiter avoids gaze-only activation costs.

US3 tests \emph{external validity and skill transfer}. We ask whether the design principle \emph{decouple targeting from commitment}, plus mid-swipe feedback, still wins against production-realistic contenders. To conduct a thorough comparison, we selected two competitors that address the algorithmic and modality limitations of previous baselines:

\begin{figure*}[h] 
    \centering 
    \includegraphics[width=\textwidth]{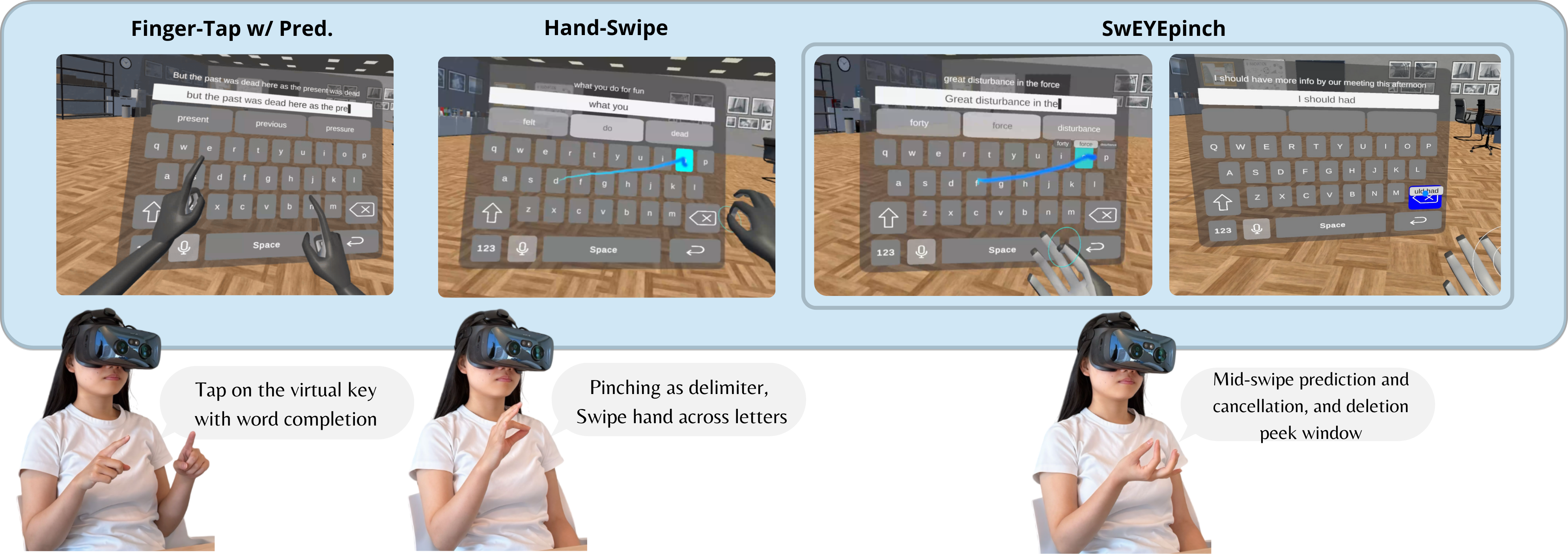} 
    \caption{Techniques evaluated in US3: the production-realistic \textit{Finger-Tap with word completion and next word prediction} (left), \textit{Hand-Swipe} that shares the same features as \textit{SwEYEpinch} but using the hand as the cursor, and \textit{SwEYEpinch} from US2.} 
    \Description{The figure compares three techniques used in US3 through VR screenshots and photos of a user demonstrating each gesture. Under Finger-Tap with Prediction, two virtual hands are shown tapping keys on a keyboard with word-completion suggestions displayed above the text box. Under Hand-Swipe, a virtual hand performs a pinch gesture while a blue swipe trace moves across the keyboard to form a word. Under SwEYEpinch, two images show mid-swipe prediction with suggested words appearing above the keyboard, and a deletion preview box shown during a cancellation gesture. Below the screenshots, photos of a user wearing a VR headset illustrate the corresponding actions: tapping forward for Finger-Tap, pinching while swiping for Hand-Swipe, and using pinch with mid-swipe prediction and deletion preview for SwEYEpinch.}
    \label{fig:us3_methods} 
\end{figure*}

\paragraph{Finger-Tap with Prediction\&Completion.}
\review{To address the algorithmic disparity in US1 where word-level swiping competed against character-level tapping, we introduce this enhanced baseline.} It extends \textit{Finger-Tap} with \textit{word completions} and \textit{next-word prediction}, similar to smartphone keyboards. While such predictive features are standard on mobile devices, they are often omitted in XR text entry studies (e.g., DwellType, Controller-Tap). \review{Including these predictive features brings comparison baselines to the same predictive power used in swipe methods.} The \textit{mid-swipe deletion} feature from \textit{SwEYEpinch} is also added here.

\paragraph{Hand-Swipe.}
\finalreview{To test the modality difference (Eye vs. Hand) using identical decoders, we implemented \textit{Hand-Swipe} as a \emph{hand-pointing} analogue of \textit{SwEYEpinch}: the user initiates a swipe by pinching, then releases the pinch to finish the swipe. \review{This technique is inspired by \textit{Vulture}~\cite {markussen2014vulture}, a seminal mid-air word-gesture keyboard that utilizes a ray-cast from the hand and a pinch delimiter. Vulture is well established in the literature \cite{Chen2019exploring,dudley2019performance,henderson2020stat} and has demonstrated a high entry rate (up to 28 WPM), ensuring that our baseline represents a high-performance standard for mid-air gestures rather than a naive implementation.} Apart from substituting gaze with hand movement, all other aspects (e.g., mid-swipe candidate previews and cancellation) match those of \textit{SwEYEpinch}.}

Finally, US3 includes two cohorts: participants new to our system and \emph{US1 alumni who \review{used only} \textit{SwEYEpinch-Basic}}. This lets us test whether skills learned with the basic hybrid transfer to \textit{SwEYEpinch}, but \emph{do not} transfer to methods that break the unique \emph{where}/\emph{when} mapping.

\begin{itemize}
  \item \textbf{H3.1.} After limited practice, \textit{SwEYEpinch} achieves higher WPM than \textit{Finger-Tap w/ Prediction\&Completion} and \textit{Hand-Swipe}.
  \item \textbf{H3.2.} \textit{SwEYEpinch} attains TER that is no worse than the contenders (and \emph{lower} than \textit{Hand-Swipe}), demonstrating that speed gains do not come at an accuracy cost.
  \item \textbf{H3.3.} Users improve in performance the fastest with \textit{SwEYEpinch}, reflecting reduced verification/correction overhead from mid-swipe previews\&cancellation, and deletion peek window.
  \item \textbf{H3.4.} \textit{SwEYEpinch} lies on (or dominates) the speed–preference Pareto frontier and yields equal or lower raw NASA TLX \review{scores} than the contenders.
  \item \textbf{H3.5.} Relative to new participants, US1 alumni more rapidly exploit mid-swipe previews in \textit{SwEYEpinch}: by Session~2 they produce more characters per swipe point (fewer swipe points per word) and shorter swipe paths, with faster time-to-confirm once the intended candidate first appears and higher top-1/any-match rates. No cohort advantage is expected for \textit{Finger-Tap w/ Prediction\&Completion} or \textit{Hand-Swipe}.
  \item \textbf{H3.6.} Prior experience influences learning: the alumni’s advantage (or slope) is larger for \textit{SwEYEpinch} than for \textit{Finger-Tap} or \textit{Hand-Swipe}, consistent with transfer specific to the decoupled targeting/commitment design.
\end{itemize}

\subsection{Study Design}
The procedure is identical to US2.

\subsection{Participants}

Forty-one volunteers participated in this study \finalminorreview{(ages 21 to 41; $\bar{x}=24.9$; 20 male, 21 female)}. All participants reported having normal or corrected-to-normal vision and no issues with hand movement. To study the effect of continued learning, among the 41 participants, 28 of them had previously participated in US1, and the rest were new. All but one participant was right-handed. All participants were familiar with typing on both computer and smartphone keyboards and had prior exposure to XR systems. However, 7 of 13 (46.7\%) US3 only participants had never used XR keyboards, while the rest reported only monthly or weekly usage. All participants were fluent in English.

\subsection{Results and Discussion}

\begin{table}[ht]
\centering
\begin{tabular}{l c c}
\hline
& \multicolumn{2}{c}{\textbf{Learning Rate (WPM/Session)}} \\
\textbf{Condition} & \textbf{US3-Only Users} & \textbf{US1 Users} \\
\hline
SwEYEpinch       & 5.09 & 2.50 \\
Finger-Tap w/ PC$^{\dagger}$   & 1.44 & 1.63 \\
Hand-Swipe            & 2.28 & 0.96 \\
\hline
\end{tabular}
\caption{Average learning rate (WPM per session) by condition for participants in US1 vs. US3. $^{\dagger}$Finger-Tap w/ PC = finger tapping with prediction\&completion.}
\Description{A table comparing average learning rates, measured in words per minute per session, between two participant groups: US3-only users and US1 users. For SwEYEpinch, the learning rate is 5.09 WPM/session for US3-only users and 2.50 WPM/session for US1 users. For Finger-Tap with Prediction, learning rates are 1.44 for US3-only users and 1.63 for US1 users. For Hand-Swipe, learning rates are 2.28 for US3-only users and 0.96 for US1 users.}
\label{tab:us3_wpm_learning_rate}
\end{table}

\begin{figure*}[!t]
    \centering

    \begin{subfigure}[t]{0.48\textwidth}
        \centering
        \includegraphics[width=\linewidth]{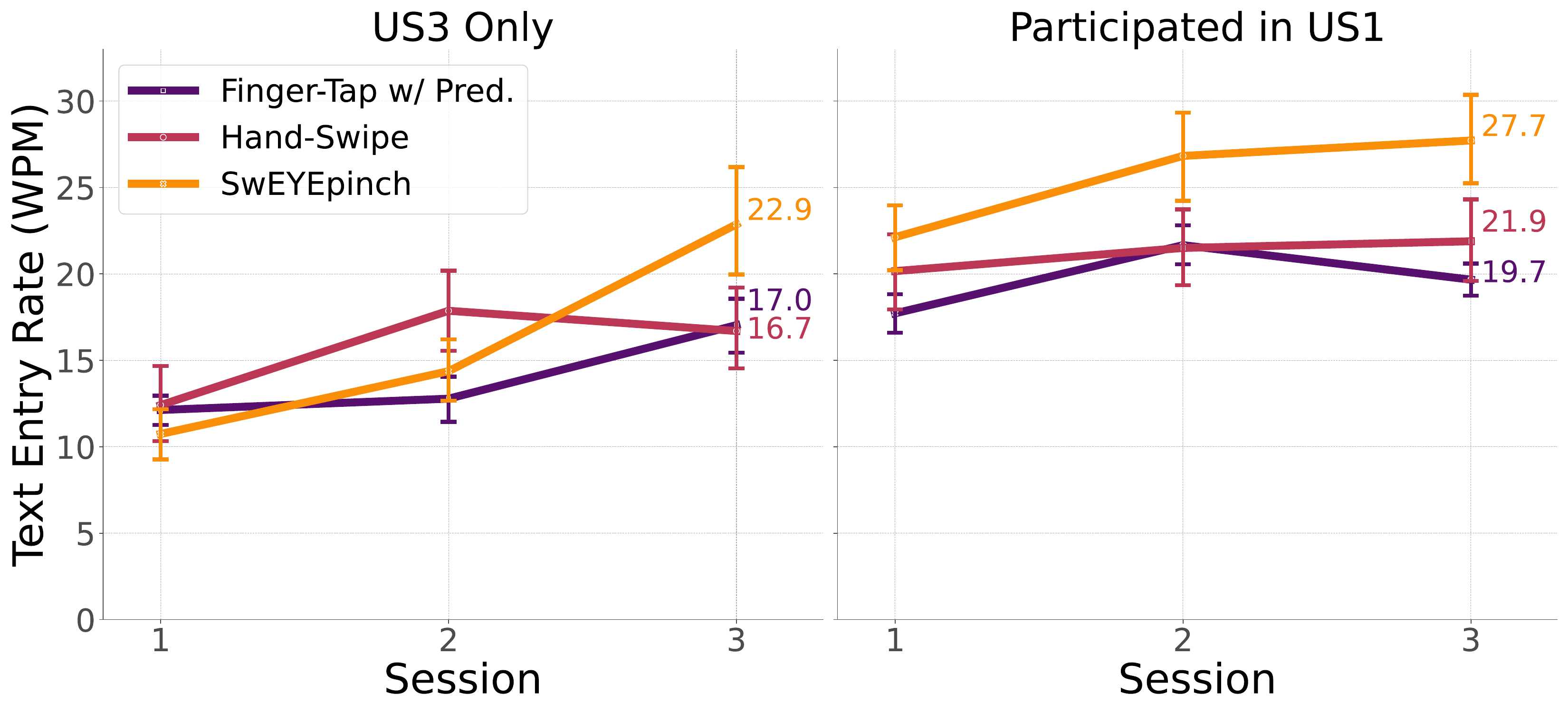}
        \phantomsubcaption\label{fig:us3_wpm}
    \end{subfigure}
    \hfill
    \begin{subfigure}[t]{0.48\textwidth}
        \centering
        \includegraphics[width=\linewidth]{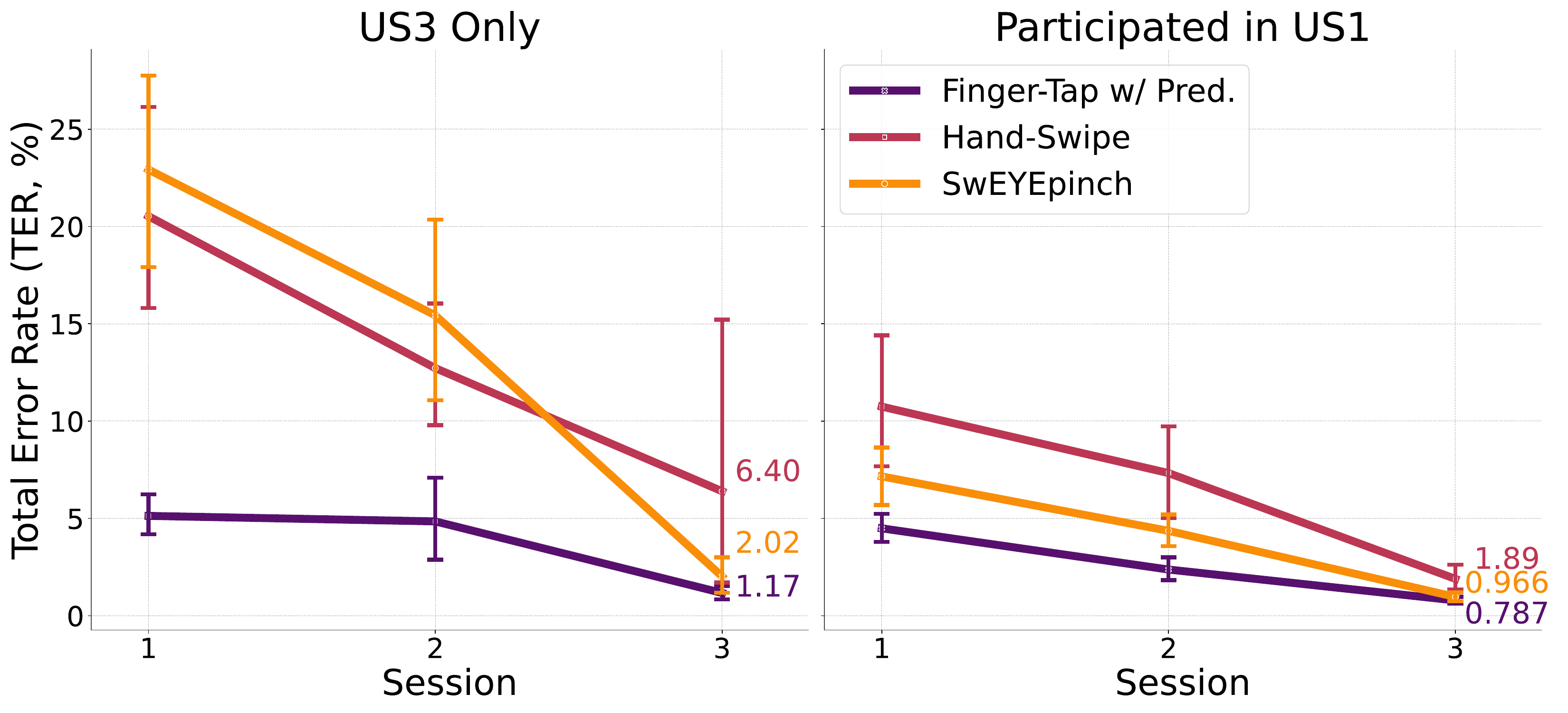}
        \phantomsubcaption\label{fig:us3_total_error_rate}
    \end{subfigure}
    \caption{Performance results from US3. 
    (Left 2) Average text entry speeds (in WPM) and (Right 2) TER by the conditions, separated for participants who only participated in US3 and those who also participated in US1. Error bars show standard errors.}
    \Description{The figure contains four line charts showing text entry rate in words per minute and total error rate from US3, separated into two participant groups: participants who only took part in US3 and participants who previously participated in US1. The left two graphs show text entry rate in words per minute across Sessions~1 to~3. For the US3-only group, three lines represent Finger-Tap with Prediction, Hand-Swipe, and SwEYEpinch, with text entry rate increasing across sessions. Only Session~3 endpoints are labeled: SwEYEpinch reaches 22.9~WPM, Finger-Tap with Prediction reaches 17.0~WPM, and Hand-Swipe reaches 16.7~WPM. For participants who previously took US1, all three techniques again increase across Sessions~1 to~3. At Session~3, SwEYEpinch reaches 27.7~WPM, Hand-Swipe reaches 21.9~WPM, and Finger-Tap with Prediction reaches 19.7~WPM. Error bars appear at each session point in both graphs. The right two graphs show total error rate decreasing across Sessions~1 to~3 for all techniques. For the US3-only group, Session~3 endpoints are labeled: Hand-Swipe has a total error rate of 6.40\%, SwEYEpinch has 2.02\%, and Finger-Tap with Prediction has 1.17\%. For participants who previously took US1, total error rate similarly declines across sessions. At Session~3, Hand-Swipe has 1.89\%, SwEYEpinch has 0.966\%, and Finger-Tap with Prediction has 0.787\%. Error bars are shown for each plotted value.}
    \label{fig:us3_results}
\end{figure*}

\begin{figure*}[!t]
    \centering
    \includegraphics[width=\textwidth]{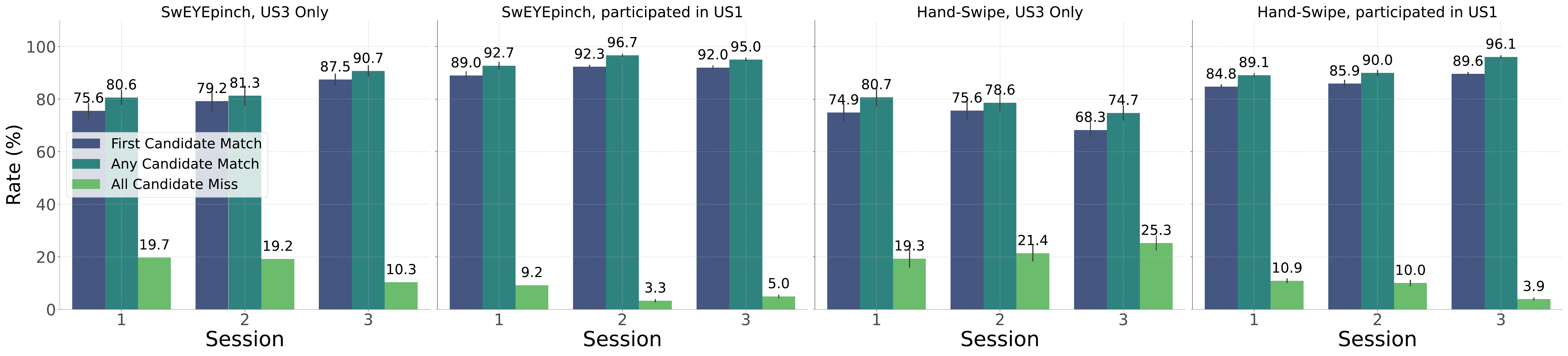}
    \caption{The match and miss rates are explained in Figure \ref{fig:us1_sweyepe_miss_rate}. US3-only participants' miss rates do not improve in Hand-Swipe.}
    \Description{The figure contains four grouped bar charts showing match and miss rates across Sessions~1 to~3 for SwEYEpinch and Hand-Swipe, separated by whether participants only took US3 or also participated in US1. Each chart displays three bar types: First Candidate Match, Any Candidate Match, and All Candidate Miss, with numeric labels on each bar. For SwEYEpinch with the US3-only group, Session~1 shows First Candidate Match of 75.6\%, Any Candidate Match of 80.6\%, and All Candidate Miss of 19.7\%. Session~2 shows 79.2\%, 81.3\%, and 19.2\%, respectively. Session~3 shows 87.5\%, 90.7\%, and 10.3\%. For SwEYEpinch with participants who also took part in US1, Session~1 shows First Candidate Match of 89.0\%, Any Candidate Match of 92.7\%, and All Candidate Miss of 9.2\%. Session~2 shows 92.3\%, 96.7\%, and 3.3\%. Session~3 shows 92.0\%, 95.0\%, and 5.0\%. For Hand-Swipe with the US3-only group, Session~1 shows First Candidate Match of 74.9\%, Any Candidate Match of 80.7\%, and All Candidate Miss of 19.3\%. Session~2 shows 75.6\%, 78.6\%, and 21.4\%. Session~3 shows 68.3\%, 74.7\%, and 25.3\%. For Hand-Swipe with participants who also took part in US1, Session~1 shows First Candidate Match of 84.8\%, Any Candidate Match of 89.1\%, and All Candidate Miss of 10.9\%. Session~2 shows 85.9\%, 90.0\%, and 10.0\%. Session~3 shows 89.6\%, 96.1\%, and 3.9\%. Across all charts, match rates increase or remain high for SwEYEpinch, while Hand-Swipe exhibits higher and less consistent miss rates, particularly for participants who only took part in US3.}
    \label{fig:us3_swipe_miss_2x2}
\end{figure*}

\begin{figure*}[!t]
    \centering
    \includegraphics[width=\textwidth]{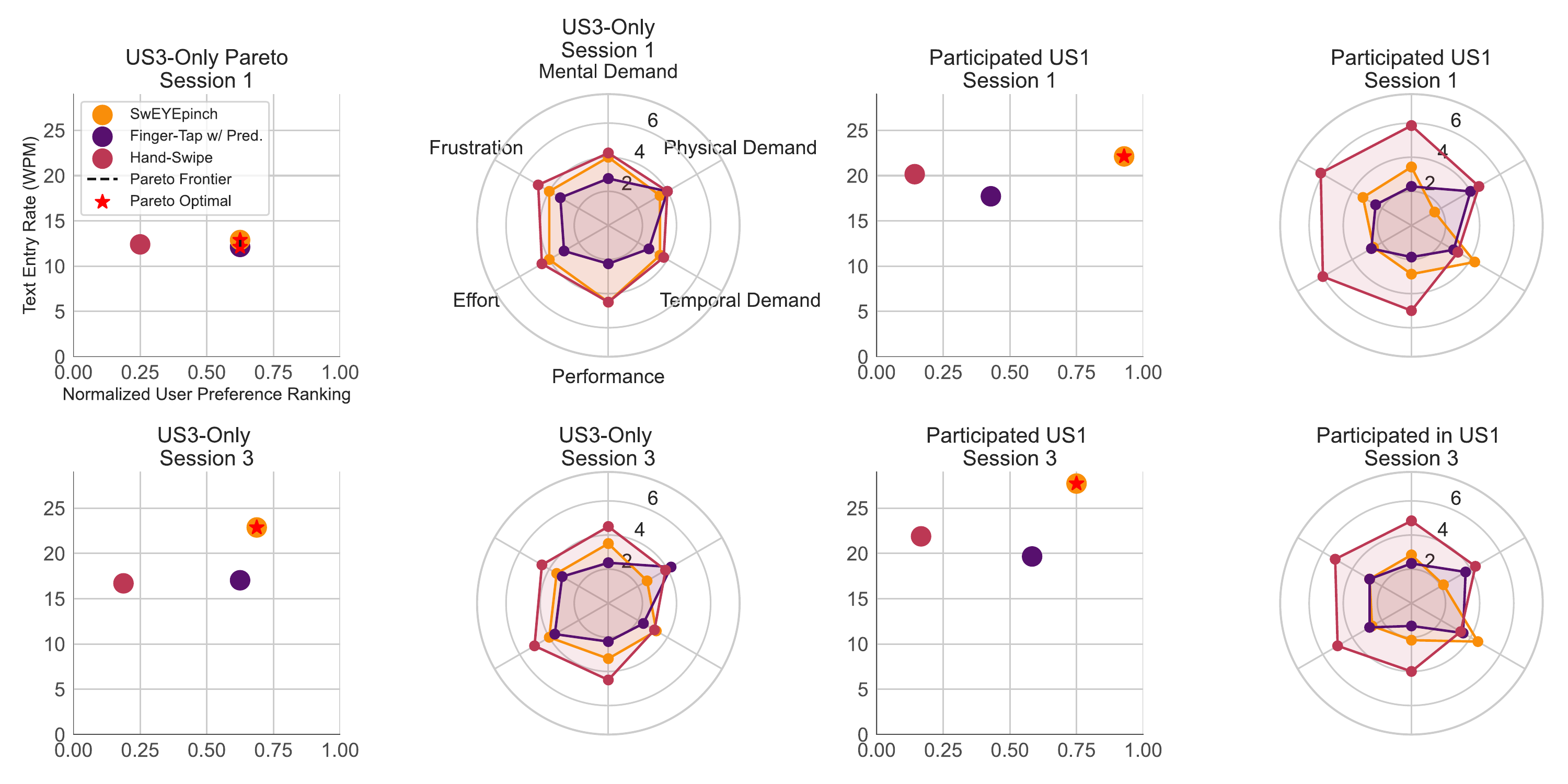}
    \caption{Pareto frontiers (normalized preference vs.\ WPM) and raw NASA TLX scores at Sessions 1 and 3 for two participant groups: those who only participated in US3 (left four figures) and those who also participated in US1 (right four figures).}
    \Description{The figure contains eight subplots showing Pareto frontiers (preference vs. WPM) and NASA-TLX workload radar charts for Sessions 1 and 3, separated into two participant groups: US3-only participants (left four plots) and those who also participated in US1 (right four plots). US3-Only Pareto Plots (top-left and bottom-left): Two scatterplots show normalized user preference on the x-axis (0–1) and text entry rate on the y-axis. Each plot includes three points—SwEYEpinch (orange), Finger-Tap with Prediction (purple), and Hand-Swipe (red). A dashed line marks the Pareto frontier, and a red star identifies the Pareto-optimal technique. In Session 1 and Session 3, SwEYEpinch appears at the highest preference values and highest WPM among the three techniques. US3-Only NASA-TLX Radar Charts (top-middle and bottom-middle): Two radar charts show six workload dimensions (Mental Demand, Physical Demand, Temporal Demand, Performance, Effort, and Frustration). Lines for the three techniques form polygon shapes with visible numeric rings labeled 2, 4, and 6. SwEYEpinch generally shows lower or similar workload ratings compared with Hand-Swipe and Finger-Tap with Prediction. Participated-in-US1 Pareto Plots (top-right scatterplot and bottom-right scatterplot): Two scatterplots similar to those on the left show higher WPM values for participants with prior experience. SwEYEpinch again appears at the upper-right region of each plot with the red star marking it as the Pareto-optimal point. Participated-in-US1 NASA-TLX Radar Charts (top-far right and bottom-far right): Two radar charts show six workload dimensions with polygons representing the three techniques. Hand-Swipe has the largest overall workload polygon, Finger-Tap with Prediction shows moderate values, and SwEYEpinch shows lower or mid-range values across most axes.}
    \label{fig:us3_session13_pareto_and_tlx}
\end{figure*}

\begin{figure*}[!t]
    \centering
    \begin{subfigure}[t]{0.32\textwidth}
        \centering
        \includegraphics[width=\linewidth]{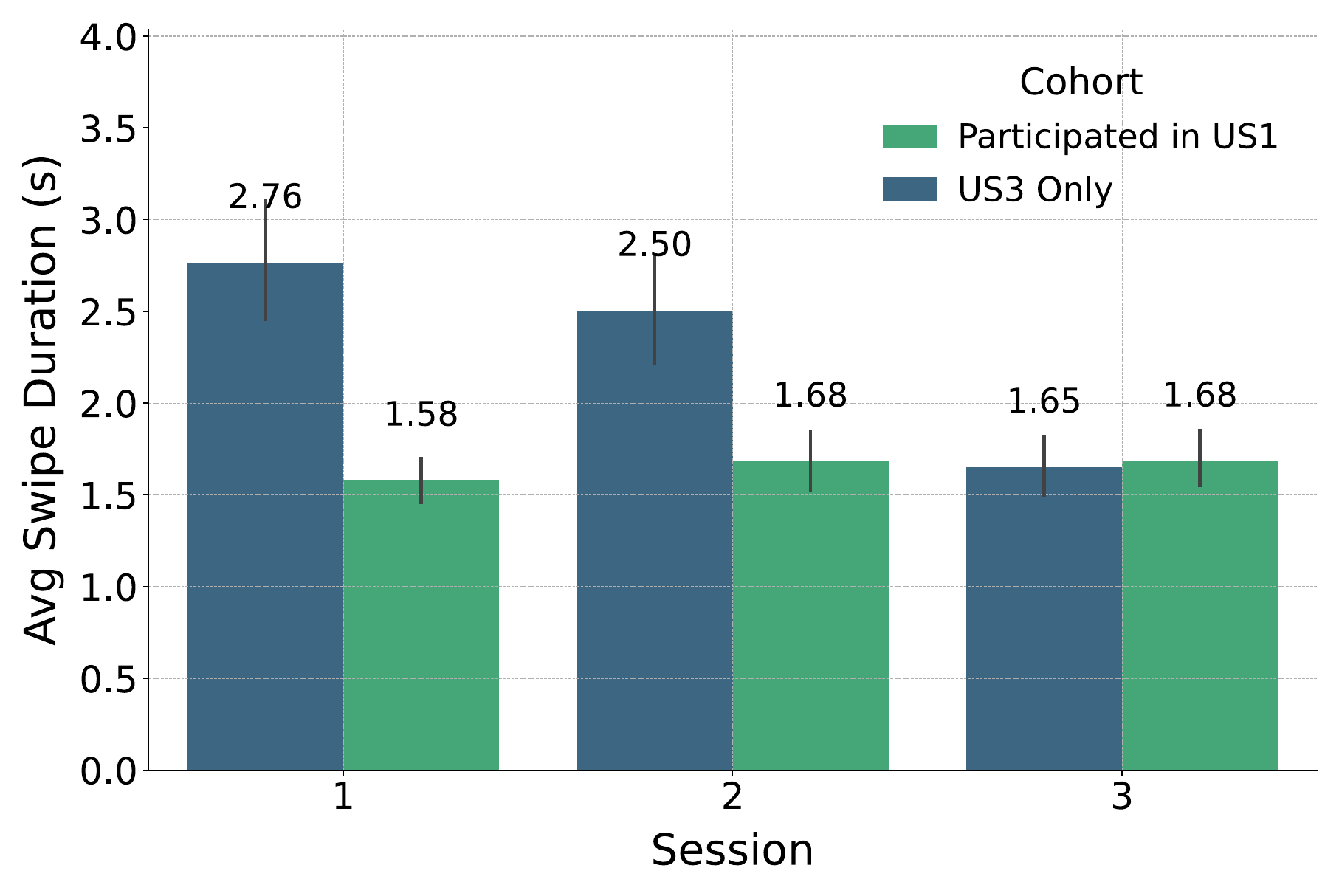}
        \subcaption*{(a)}
        \phantomsubcaption\label{fig:us3_avg_sweyepe_duration}
    \end{subfigure}
    \hfill
    \begin{subfigure}[t]{0.32\textwidth}
        \centering
        \includegraphics[width=\linewidth]{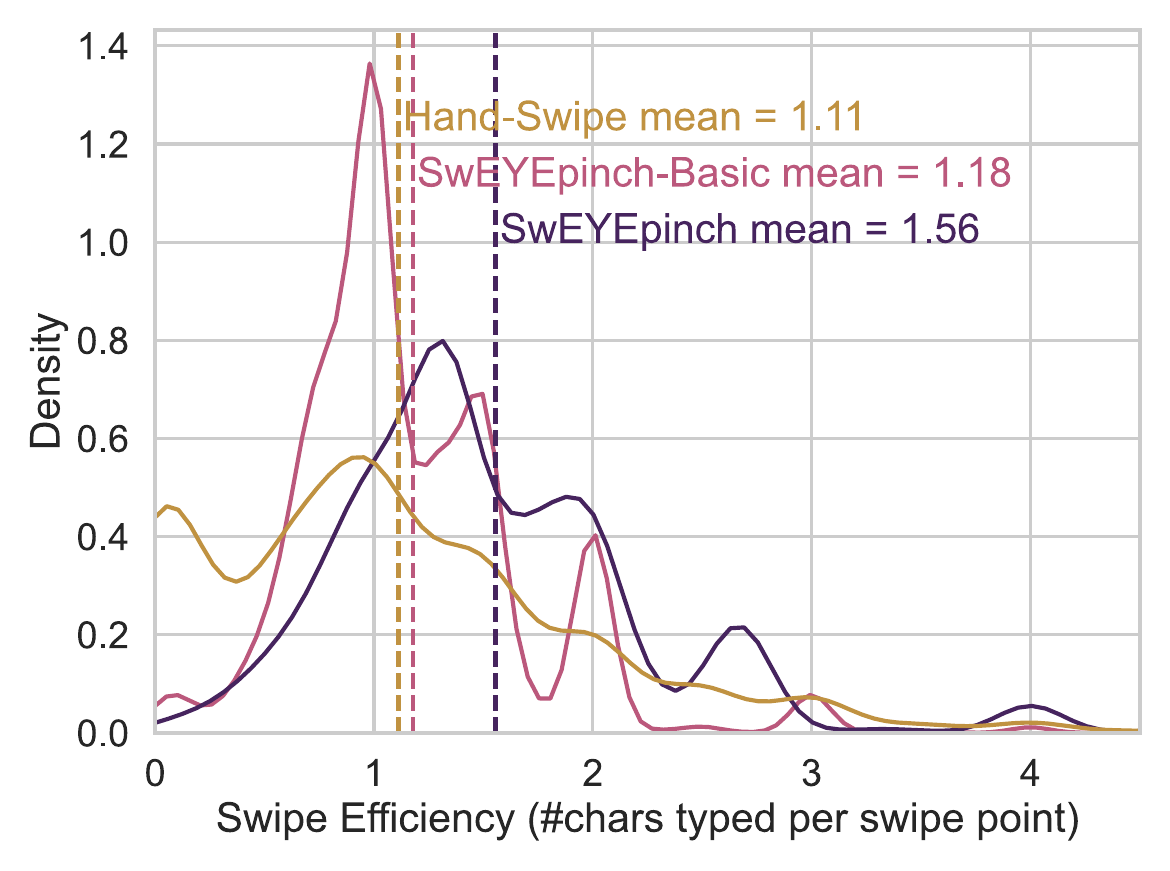}
        \subcaption*{(b)}
        \phantomsubcaption\label{fig:n_chars_per_swipe_point_kde}
    \end{subfigure}
    \hfill
    \begin{subfigure}[t]{0.32\textwidth}
        \centering
        \includegraphics[width=\linewidth]{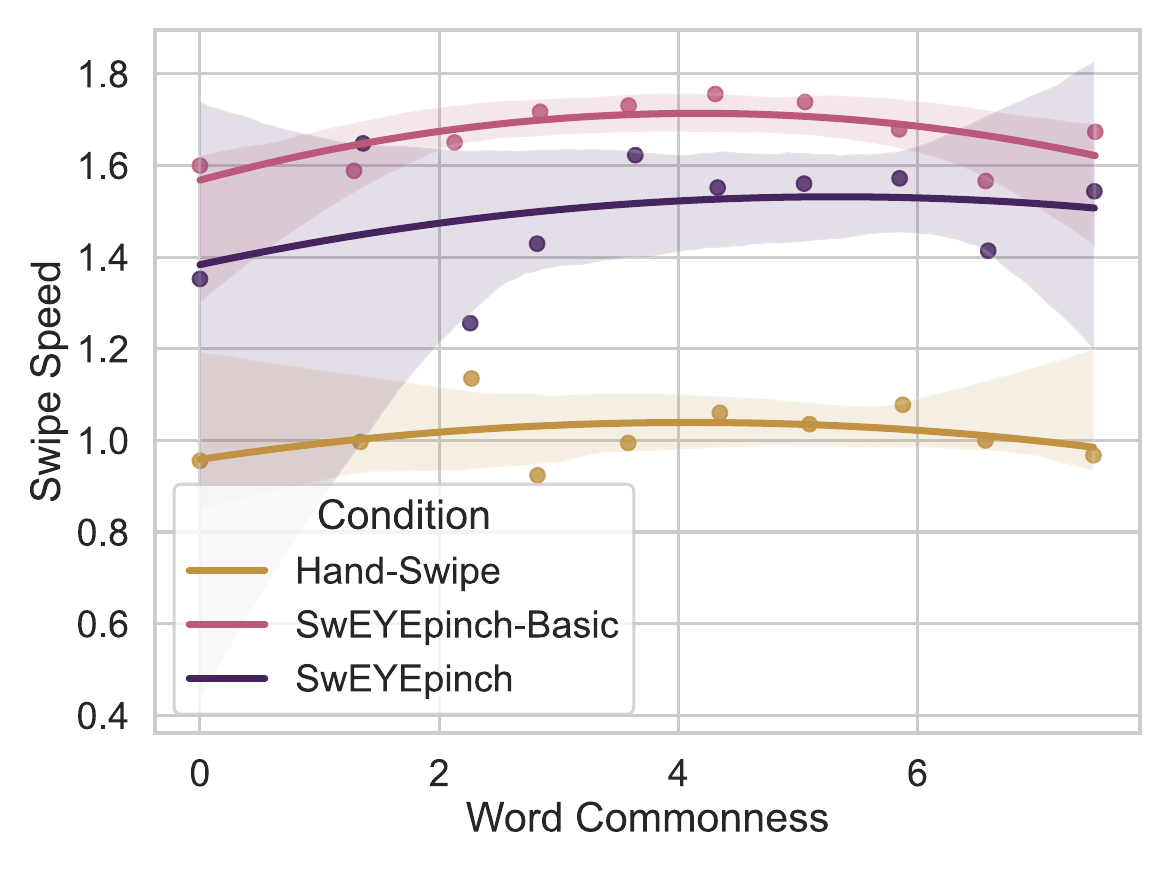}
        \subcaption*{(c)}
        \phantomsubcaption\label{fig:word_commonness_swipe_spd_fits}
    \end{subfigure}

    \caption{Mechanism uptake and efficiency in US3.
    (a) Average \textit{SwEYEpinch} swipe duration.
    (b) Swipe Efficiency distribution—plotted as a kernel density estimate (KDE) over the per-word effort metric. Higher values indicate fewer waypoints per letter.
    (c) Word commonness (Zipf frequency from \cite{robyn_speer_2022_7199437}) versus swipe speed with quadratic fits (higher Zipf = more common words).}
    \Description{The figure contains three panels illustrating mechanism uptake and swipe efficiency in US3. Panel~(a) shows average swipe duration as a grouped bar chart, reporting SwEYEpinch swipe duration in seconds for two cohorts: participants who only took part in US3 and participants who previously participated in US1, across Sessions~1 to~3. Numeric labels appear at the top of each bar, and error bars are shown for each condition. In Session~1, the US3-only group has an average swipe duration of 2.76~s, while participants who previously took US1 have 1.58~s. In Session~2, the values are 2.50~s and 1.68~s, respectively. In Session~3, the values are 1.65~s for the US3-only group and 1.68~s for participants who previously took US1. Panel~(b) shows the distribution of swipe efficiency as a density plot, measured as characters typed per swipe point, for Hand-Swipe, SwEYEpinch-Basic, and SwEYEpinch. Each technique is represented by a smooth kernel density estimate curve, with a vertical dashed line marking the mean value. The mean swipe efficiency is 1.11 for Hand-Swipe, 1.18 for SwEYEpinch-Basic, and 1.56 for SwEYEpinch. Panel~(c) shows the relationship between word commonness and swipe speed as a scatterplot with quadratic fit curves. Swipe speed is shown on the y-axis and word commonness, measured as Zipf frequency, is shown on the x-axis for Hand-Swipe, SwEYEpinch-Basic, and SwEYEpinch. Colored points and shaded regions indicate the data and confidence intervals. SwEYEpinch-Basic and SwEYEpinch exhibit higher swipe speeds across the range of word commonness, while Hand-Swipe shows lower and relatively flatter swipe speeds.}
    \label{fig:us3_three_panel_mechanism}
\end{figure*}

\noindent\textbf{SwEYEpinch is the fastest against production-realistic baselines, supporting H3.1.}
\finalreview{By Session~3, \textit{SwEYEpinch} reached $26.2{\pm}13.1$~WPM and significantly outperformed \textit{Finger-Tap w/ Pred.} (\review{$p{<}0.001$}) and \textit{Hand-Swipe} (\review{$p{<}0.001$}); see Fig.~\ref{fig:us3_wpm}.}

\noindent\textbf{Absolute TER non-inferiority to Finger-Tap did not hold, but SwEYEpinch reduced errors the fastest, not supporting H3.2.}
\finalreview{By Session~3, \textit{Finger-Tap w/ Pred.} had lower TER than \textit{SwEYEpinch} (\review{$p{=}0.018$}), while the \textit{SwEYEpinch} vs.\ \textit{Hand-Swipe} TER difference was not significant (\review{$p{=}0.31$}; Fig.~\ref{fig:us3_total_error_rate}); thus, \textbf{H3.2} is not supported on an absolute basis. However, \textit{SwEYEpinch} showed the steepest TER decline and higher first-candidate match rates than \textit{Hand-Swipe} across sessions ($p{<}10^{-4}$; Fig.~\ref{fig:us3_swipe_miss_2x2}), consistent with reduced correction overhead.}

\noindent\textbf{SwEYEpinch shows the largest WPM gains per session, supporting H3.3.}
\finalreview{Learning-rate estimates (Table~\ref{tab:us3_wpm_learning_rate}) show \textit{SwEYEpinch} improving most rapidly within each cohort (US3-only: $5.09$~WPM/session; US1-alumni: $2.50$~WPM/session), exceeding both \textit{Finger-Tap w/ Pred.} and \textit{Hand-Swipe}. This aligns with mid-swipe previews and low-effort cancellation reducing verification/correction time.}

\noindent\textbf{SwEYEpinch remains Pareto-optimal with competitive workload, supporting H3.4.}
\finalreview{By Session~3, \textit{SwEYEpinch} lies on the speed–preference Pareto frontier for both participant groups (Fig.~\ref{fig:us3_session13_pareto_and_tlx}, top). NASA TLX decreases across sessions for all techniques, with \textit{SwEYEpinch} ending at equal or lower workload than alternatives (bottom).}

\noindent\textbf{Prior exposure yields an early advantage; novices reach parity by Session~3, supporting H3.5.}
\finalreview{Alumni produced shorter \textit{SwEYEpinch} swipes in S1–S2 (\review{$p{\le}0.031$}) but converged with novices by S3 (\review{$p{=}0.90$}; Fig.~\ref{fig:us3_three_panel_mechanism},(a)). Across conditions, \textit{SwEYEpinch} shows higher swipe efficiency than \textit{SwEYEpinch-Basic} and \textit{Hand-Swipe} (Fig.~\ref{fig:us3_three_panel_mechanism},(b)), consistent with preview-and-confirm behavior rather than full-word tracing.}

\noindent\textbf{Cohort$\times$Technique effects reflect initial performance more than learning rate: alumni start faster on SwEYEpinch, but novices learn it faster, supporting H3.6.}
\finalreview{At Session~1 the cohort gap is largest for \textit{SwEYEpinch}, smaller for \textit{Hand-Swipe}, and smallest for \textit{Finger-Tap with Prediction\&Completion} (Fig.~\ref{fig:us3_wpm}), indicating technique-specific transfer. Slopes invert: novices ramp faster on \textit{SwEYEpinch} (5.09 vs.\ 2.50~WPM/session), while differences are smaller for the baselines (Table~\ref{tab:us3_wpm_learning_rate}). Thus, H3.6 is supported for \emph{initial performance} but not for a larger alumni \emph{slope}.}

\noindent\textbf{Design Insight.}
Mid-swipe previews plus a pinch delimiter promote efficient \emph{partial} swipes and early commit, yielding top entry rate and fast learning while keeping workload competitive. Transfer benefits initial performance when the where/when mapping is preserved (\textit{SwEYEpinch} vs.\ hand/tap baselines), and novices quickly acquire the same preview-and-confirm strategy. In products, surface live suggestions early, make early-commit cheap and visible, and keep low-effort correction (e.g., peek/delete) to accelerate convergence.

\section{User Study 4: Towards Everyday Typing Speeds—Seven Days of SwEYEpinch}

US1–US3 established \emph{why} the hybrid works (mid-swipe previews + pinch delimiter), \emph{that} it beats production-style baselines, and \emph{where} transfer appears. As a natural extension, \emph{does \textit{SwEYEpinch} keep improving with daily use and can it reach ordinary keyboard speeds?} To our knowledge, XR text-entry papers rarely put an interface through a longitudinal “test of time”. US4 accomplishes a week-long, multiple-times-a-day study over 30 consecutive sessions that tracks individual learning curves against normal keyboard performance. 

Guided by classic longitudinal typing work \cite{matias1996one} and by the learning patterns observed in US1–US3, we test the following:

\begin{itemize}
  \item \textbf{H4.1 (Sustained learning).} Across a week of daily use, \textit{SwEYEpinch} shows a positive per-user trend in WPM and does not plateau mid-week.
  \item \textbf{H4.2 (Everyday-speed attainability).} With routine practice, \textit{SwEYEpinch} approaches or crosses the lower band of everyday keyboard speeds for non-developer users.
  \item \textbf{H4.3 (Experience moderates, but everyone improves).} Prior exposure shapes the \emph{shape} of the curve (refinement vs.\ rapid gains), yet all users realize meaningful improvement over time.
\end{itemize}

\subsection{Study Design}
Each participant was required to complete a total of 30 sessions, distributed across seven consecutive days, to balance exposure with rest and consolidation.

Each session consisted of 14 transcription tasks, where participants typed full sentences using \textit{SwEYEpinch}. All participants were given the same sentences at the same session index. No phrases were repeated. This design resulted in 420 total transcription trials per participant (14 phrases × 30 sessions).

\subsection{Participants}
Recruitment in US4 was more limited than in other studies due to the substantial time commitment; nine participants completed all sessions. We sampled three experience cohorts \finalminorreview{(ages 20 to 31; $\bar{x}$ = 25.8 years), 7 male, 2 female}:

\review{\textbf{(1) SwEYEpinch- and XR-novices.} \texttt{P403}, \texttt{P406}, and \texttt{P407} had no prior exposure to our techniques, had never typed in XR, and reported only minimal XR use.}

\review{\textbf{(2) XR-experienced, SwEYEpinch-naïve.} \texttt{P405}, \texttt{P408}, \texttt{P409} had substantial prior XR experience (as regular users or developers) but had not used \textit{SwEYEpinch} before US4.}

\review{\textbf{(3) SwEYEpinch-experienced / experts.} \texttt{P401}, \texttt{P402}, and \texttt{P404} had prior exposure to \textit{SwEYEpinch} through earlier studies or development work, and thus entered US4 with intermediate-to-expert proficiency.}

\subsection{Results and Discussion}

\begin{figure*}[!t]
    \centering
    \includegraphics[width=\textwidth]{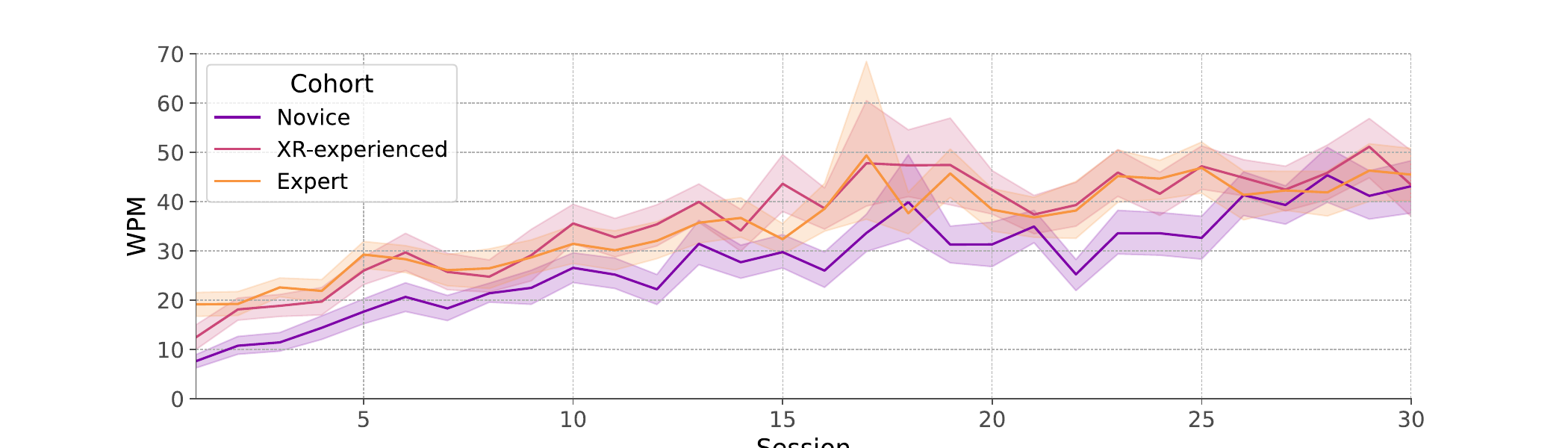}
    \caption{\review{\textit{SwEYEpinch} performance with extended daily sessions. Average WPM by participant cohort.}}
    \Description{A line chart shows SwEYEpinch typing performance over 30 sessions for three participant cohorts: Novice (purple), XR-experienced (red), and Expert (orange). The y-axis shows words per minute (WPM), and the x-axis shows session number from 1 to 30. All three lines rise visibly over time. Novices start at the lowest WPM, around single-digit values, and increase steadily into the 30–40 WPM range by the final sessions. XR-experienced participants begin higher and reach the mid-40s to low-50s WPM range. Experts start the highest and reach peak values approaching the upper 50s and near 70 WPM at one point. Shaded regions around each line represent variability within each cohort.}
    \label{fig:us4_wpm}
\end{figure*}

\noindent\textbf{SwEYEpinch keeps getting faster with daily use and does not plateau mid-week, supporting H4.1.}
All nine participants show positive overall learning slopes (0.77–1.65\,WPM/session) and rising trajectories across 30 sessions (Fig.~\ref{fig:us4_wpm}). Learning remains active well beyond Session~10 for both experts and non-experts: for example, P402 and P404 maintain strong gains in the 11–20 block (1.77 and 1.82\,WPM/session), while novices such as P406 and P407 continue to accelerate late in the week (e.g., P407: 2.98\,WPM/session in Sessions 21–30). These trends contradict a mid-week plateau. As one participant put it, “\emph{By midweek I wasn’t tracing whole words anymore—once my word appeared, I just pinched and moved on.}”

\noindent\textbf{Everyday typing speeds are attainable with routine practice, partially supporting H4.2.}
Five participants reached or exceeded the 54\,WPM band typical of everyday touch typing (P401: 60.6; P402: 64.7; P405: 56.5; P407: 57.0; P409: 60.9 median best), and the remaining four converged on high-40s (45.4, 45.9, 46.2, 48.9\,WPM for P403, P406, P404, P408). These speeds emerged from repeated daily use rather than single-session tuning (Fig.~\ref{fig:us4_wpm}). “\emph{It went from ‘cool demo’ to ‘I can actually type like this,’}” noted P402.

\noindent\textbf{Experience shapes the curve’s \emph{shape}, but everyone improves meaningfully, supporting H4.3.}
Experts who entered with prior SwEYEpinch experience (P401, P402, P404) started high and then refined: P401’s learning rate tapered over the three 10-session blocks (1.93\,$\to$\,1.14\,$\to$\,0.263\,WPM/session), while P402 and P404 showed mid-study accelerations (peaks of 1.77 and 1.82\,WPM/session). XR-experienced but SwEYEpinch-naïve users (e.g., P405) exhibited solid gains early on (1.53\,WPM/session in Sessions 1–10), then consolidated. Novices (P403, P406, P407) began at much lower speeds but posted some of the largest overall gains, including late-week acceleration in some cases (e.g., P407’s 2.98\,WPM/session in Sessions 21–30). Across all participants, performance improvements were substantial ($\Delta$WPM: 24.1–49.1), echoing US2–US3: once users come to trust mid-swipe previews, they shorten traces, commit earlier, and steadily push into everyday typing ranges.

\medskip
\noindent\textit{Design Insights.}
US4 shows that \textit{SwEYEpinch} \finalminorreview{has the potential for} everyday viability through compounding efficiency: with repeated use, mid-swipe previews shift behavior from tracing to “see–confirm–move,” producing shorter paths and steady WPM gains that persist beyond mid-week. Because the \emph{where}/\emph{when} mapping stays stable, skill transfers cleanly across days and experience levels, keeping progress monotonic. Reaching 60{+}~WPM for three participants and high-40s for the rest places \textit{SwEYEpinch} inside the everyday keyboard envelope—the technique itself (previews + early commit) has the potential to be an interface people can live with.

\section{Discussion}

This section synthesizes evidence from US1--US4 into design guidance for XR text entry. Our central result is that \emph{decoupling targeting from commitment}—using gaze for \emph{where} and a tiny explicit gesture for \emph{when}—yields a better speed–effort trade-off than production-style baselines, especially when paired with \emph{mid-swipe previews} and low-effort correction.

\paragraph{Principle: Decouple \emph{where} from \emph{when}.}
US1 showed that a minimal hybrid (\textit{SwEYEpinch-Basic}) sits on the speed–preference Pareto frontier versus \textit{Finger-Tap} and \textit{Gaze\&Pinch} (\autoref{fig:us1_results}, \autoref{fig:us1_session15_pareto_and_tlx}). US2 isolated \emph{why}: the pinch delimiter outperformed gaze-only delimiters without an accuracy cost, and \emph{mid-swipe previews} cut verification glances and raised WPM (\autoref{fig:us2_results}). US3 extended this against production-realistic contenders: \textit{SwEYEpinch} was fastest while remaining Pareto-optimal on preference and competitive on workload (\autoref{fig:us3_results}, \autoref{fig:us3_session13_pareto_and_tlx}). US4 demonstrated sustained gains, with three participants exceeding 60~WPM (\autoref{fig:us4_wpm}).

\begin{figure*}[!t]
    \centering
    \includegraphics[width=\textwidth]{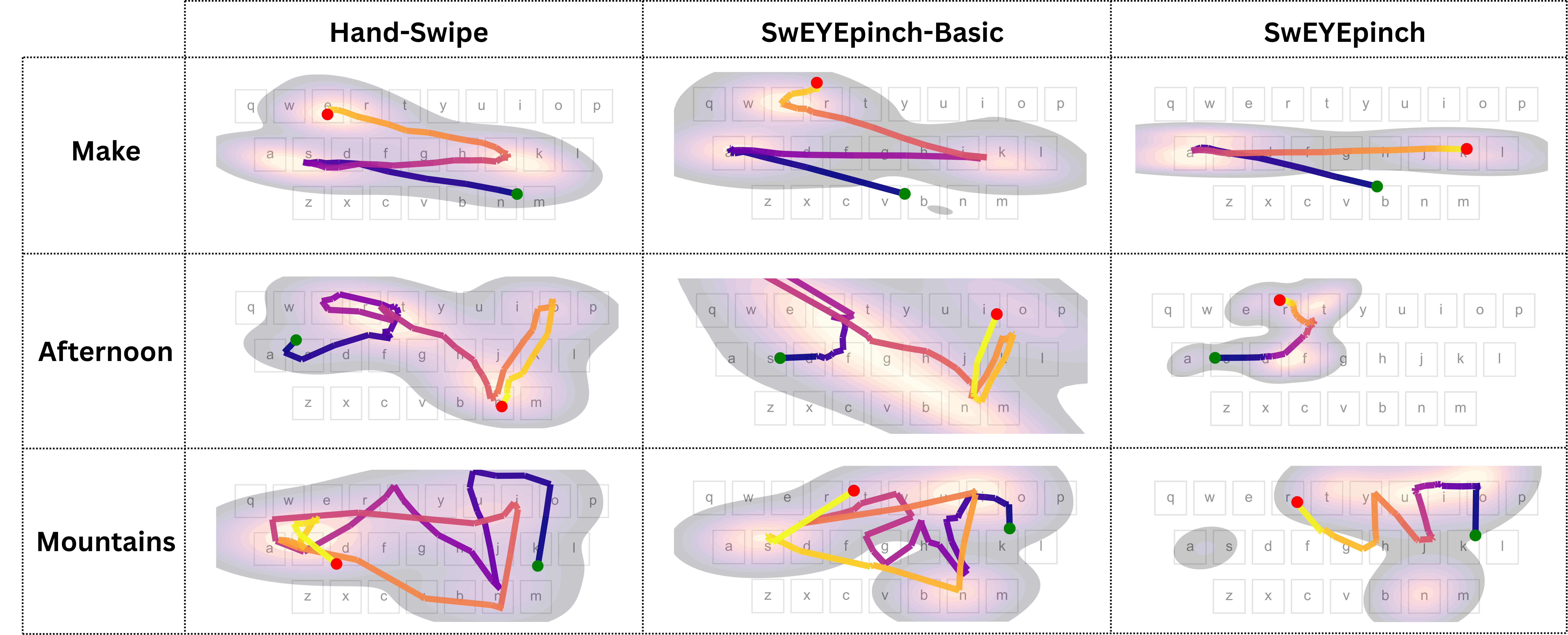}
    \caption{\finalreview{Average swipe paths for words ("make", "afternoon", "mountains") from three swipe-based techniques: \textit{Hand-Swipe}, \textit{SwEYEpinch-Basic}, and \textit{SwEYEpinch}. The red and green dots indicate the average start and end position of the swipe traces. Compared to \textit{Hand-Swipe} and \textit{SwEYEpinch-Basic}, \textit{SwEYEpinch} produces shorter paths with fewer turns and tends to end mid-word thanks to predictive completion.}}
    \Description{The figure shows a three-by-three grid of visualizations comparing average swipe paths for three swipe-based text-entry techniques: Hand-Swipe, SwEYEpinch-Basic, and SwEYEpinch. Columns correspond to techniques, labeled from left to right as Hand-Swipe, SwEYEpinch-Basic, and SwEYEpinch. Rows correspond to example words, labeled from top to bottom as ``make'', ``afternoon'', and ``mountains''.Each panel displays a QWERTY keyboard layout overlaid with a semi-transparent heatmap indicating the spatial distribution of swipe traces, along with a colored polyline representing the average swipe path for that word and technique. A red dot marks the average starting position of the swipe, and a green dot marks the average ending position. For all three words, Hand-Swipe produces relatively long swipe paths with multiple turns that closely follow the full letter sequence across the keyboard. SwEYEpinch-Basic generally shortens the swipe paths compared to Hand-Swipe but still includes several directional changes and covers most of the word length. SwEYEpinch produces the shortest and smoothest swipe paths, with fewer turns and more direct trajectories. In many cases, SwEYEpinch swipe paths terminate before reaching the final letters of the word, indicating early termination enabled by predictive completion. Overall, the figure illustrates that SwEYEpinch yields more efficient swipe trajectories than Hand-Swipe and SwEYEpinch-Basic across all three example words.}
    \label{fig:swipe_example_traces}
\end{figure*}

\paragraph{Mechanism: Previews enable partial swipes and early commit.}
Across studies, we see a consistent behavioral shift: users stop tracing once the intended word appears and commit with a small pinch. This shows up as shorter paths, fewer turns, and higher characters-per-swipe-point, alongside higher top-1/any candidate matches with \textit{SwEYEpinch}’s decoder (\autoref{fig:us2_miss_rates}, \autoref{fig:us3_swipe_miss_2x2}). The qualitative geometry is visible in \autoref{fig:swipe_example_traces}: \textit{SwEYEpinch} traces are more compact and often end mid-word, reflecting preview-and-confirm behavior rather than full-word tracing.

\paragraph{Design recommendations for XR text entry.}
\textbf{(R1) Keep the delimiter explicit and tiny.} A low-effort pinch provides precise \emph{when} without the latency/ambiguity of dwell or line-crossing delimiters, improving entry rate with no accuracy penalty (US2; \autoref{fig:us2_results}).  
\textbf{(R2) Display live candidates near the gaze locus.} Mid-swipe previews reduce “finish–inspect–fix” loops and enable early commit, driving WPM gains (US2/US3; \autoref{fig:us2_results}, \autoref{fig:us3_results}).  
\textbf{(R3) Optimize the correction loop.} Make cancellation and deletion low-friction—retain in-gesture cancel to discard a bad swipe, whole-word rollback immediately after commit, and a \emph{peek window} to minimize off-keyboard glances.  
\textbf{(R4) Bias decoding toward early partials.} Fuse a lightweight language model with a spatiotemporal matcher and ramp LM influence when evidence is sparse; this stabilizes early previews without harming later accuracy (US2; \autoref{fig:us2_miss_rates}).  
\textbf{(R5) Minimize eye travel.} Keep candidates and critical controls on or just above the keyboard plane to avoid costly verification glances and preserve an “eyes-on-keyboard” loop.  
\textbf{(R6) Provide short-string safeguards.} Very brief traces are easy to over-commit; lightweight disambiguation cues (e.g., transient alternatives or stricter commit thresholds for very short paths) preserve speed without adding dwell.

\paragraph{Learning and transfer.}
When the \emph{where}/\emph{when} mapping is preserved, skills transfer: alumni started \textit{SwEYEpinch} faster in US3, while novices quickly caught up by adopting preview-and-confirm strategies (shorter durations, higher efficiency; \autoref{fig:us3_three_panel_mechanism}). Product implication: onboarding can focus on the single mental model “trace until your word appears, then pinch,” rather than mechanics training.

\paragraph{Everyday viability.}
US4’s week-long regimen shows compounding efficiency: as users transition from tracing to “see–confirm–move,” WPM continues to rise beyond mid-week, with two users surpassing 60~WPM and others approaching high 40s (\autoref{fig:us4_wpm}). For XR platforms, this suggests that a hybrid gaze–pinch design with live previews and low-effort correction is not just learnable—it can reach everyday speeds with routine use. \review{These speeds are comparable to, or higher than, those reported for prior multimodal techniques that rely on touch surfaces or speech (\cite{kumar2020tagswipe, hedeshy2021hummer, yu2017tap}), while SwEYEpinch remains HWD-only and silent. This underscores why our in-study baselines focus on commercial XR techniques and HWD-only gaze/gesture methods: they reflect the practical comparison set for the everyday scenarios we target.}

\medskip
\noindent Overall, \textit{SwEYEpinch} reframes gaze from a \emph{click surrogate} to a \emph{continuous intent signal}, with a tiny explicit delimiter for commitment. This pairing reshapes the speed–effort frontier relative to current XR baselines, and our studies outline the concrete UI/decoder choices that make it work in practice.

\section{Limitations and Future Work}
Our investigation brings out several interactions, hardware, and ergonomic constraints. First, deletion is coarse: whole-word rollback is only available immediately after a commit; at other times, users must backspace character by character. More flexible mechanisms—e.g., gesture hotkeys \cite{10.1109/TVCG.2023.3320257}, auto-repeat on hold, or layouts beyond QWERTY \cite{10.1145/507072.507076} with dedicated keys for word/character delete—could reduce correction burden. Second, tracking and actuation are not uniformly inclusive: infrared eye tracking and hand tracking can underperform for some users, and prolonged use can lead to eye fatigue. Future work should pursue setups that are robust to diverse user groups.

\review{In line with prior gaze–swipe text-entry work \cite{kumar2020tagswipe, cui2023glancewriter, yu2017tap}, our evaluation focused on speed, accuracy, and subjective workload rather than cybersickness or spatial-awareness outcomes. Our sessions were seated, and participants did not spontaneously report nausea or disorientation. Nonetheless, longer-term and mobile use (e.g., walking or commuting) may induce different responses. \finalreview{As our studies primarily report performance (WPM, TER) and preference, we cannot conclude how fatigue accumulates over prolonged use beyond the session durations studied.} Future work should pair SwEYEpinch with standardized cybersickness measures (e.g., SSQ) and dual-task paradigms that stress navigation or spatial awareness, to more fully characterize how gaze–pinch text entry integrates into everyday XR scenarios.}

\finalreview{Methodologically, our studies used transcription tasks that may under-represent real composition and dialogue. Naturalistic scenarios such as open-ended writing, and stratification by language proficiency would better capture cognitive load and vocabulary effects. Moreover, our in-study comparisons do not include techniques that use an external button/controller (e.g., GestureType~\cite{yu2017tap}) or a touch surface (e.g., TagSwipe~\cite{kumar2020tagswipe}). As a result, our conclusions are strongest for \emph{HWD-only, silent, surface-free} XR typing, and we treat controller/surface-based techniques as external reference points rather than direct in-study baselines. Prior work suggests that sustained, targeted saccades and continuous visual focus can be fatiguing, and susceptibility to discomfort may increase with age. Our participant pools skewed young, so an important next step is to evaluate \textit{SwEYEpinch} with older adults and to explicitly quantify fatigue/comfort alongside objective proxies (gaze travel distance, saccade counts, blink rate) during longer free-form typing.}

On the systems side, our Python decoder leaves performance headroom; moving critical paths to accelerated kernels and on-device inference could further cut latency. Finally, decoding should adapt to behavior: early in a swipe, weight language priors more heavily; detect hesitation or ambiguous short traces to raise commit thresholds or surface targeted disambiguation—balancing speed and accuracy in situ.

\section{Conclusions}
This paper tested a simple idea for XR text entry: separate \emph{targeting} from \emph{commitment}—use gaze for the fast, low-effort \emph{where} and a small explicit gesture for the \emph{when}. Across four studies, this principle—instantiated as \textit{SwEYEpinch}—improved the speed–effort balance without sacrificing accuracy. US1 placed a minimal hybrid on the Pareto frontier against \textit{Finger-Tap} and \textit{Gaze\&Pinch}; US2 showed that a pinch delimiter outperforms gaze-only delimiters and that \emph{mid-swipe previews} raise text entry rate without an accuracy cost; US3 confirmed wins over production-realistic contenders; and US4 demonstrated sustained learning to everyday speeds, with three participants surpassing 60~WPM.

Mechanistically, mid-swipe candidates, low-friction correction (in-gesture cancel, immediate rollback, deletion peek), and a low-latency spatiotemporal decoder with an $n$-gram prior enable \emph{partial} swipes and \emph{early commit}: shorter paths, fewer turns, higher first-candidate matches. For designers: keep the delimiter explicit and tiny; surface candidates in place while swiping; make correction effortless; and weight language priors early, handing off to spatial matching as evidence accrues. With code and data released (Section~\ref{sec:code and data availability}), we offer a replicable pattern for XR text entry that is fast, learnable, and ready to \finalminorreview{use} beyond the lab.

\section*{Acknowledgment of AI Use}
We acknowledge the use of generative AI tools to improve the grammar, style, and readability of this manuscript; these tools played no role in the data analysis, interpretation, or generation of the core findings presented.

\begin{acks}
We thank our colleagues at Columbia University, especially Haoyan Chen, for assistance with figure preparation. We also thank Dr. Brian Smith for thoughtful discussions that informed this work. This research was supported in part by the Air Force Office of Scientific Research (FA9550-22-10337), the Army Research Laboratory (W911NF-19-2-0139, W911NF-19-2-0135, W911NF-21-2-0125), and the U.S. Department of Defense (N00014-20-1-2027). 
\end{acks}

\balance{}

\bibliographystyle{SIGCHI-Reference-Format}
\bibliography{sample}

\clearpage
\appendix
\FloatBarrier

\section{Optimization in SwEYEpinch Decoding}
\label{appendix: Optimization in SwEYEpinch Decoding}
\noindent
\paragraph{Optimization: Fixation Detection (I-VT)}
Our first optimization applies a fixation-detection algorithm to the raw gaze sequence, leveraging the extensive previous research on fixation-detection methods~\cite{salvucci2000identifying, andersson2017one}. We use a velocity-thresholding approach (I-VT) because it is efficient and does not require per-user calibration. Pseudocode is presented in Algorithm~\ref{alg:Velocity-Threshold Identification (IVT)}. Conceptually, I-VT labels each gaze point based on its angular velocity relative to the previous valid point. In our experiments, a typical swipe lasts $1.60 \pm 0.694$\,s, corresponding to $320 \pm 139$ raw gaze samples. After applying I-VT, the number of gaze points (i.e., ``fixation'' points) decreases to $14.74 \pm 2.90$.

\begin{algorithm*}
\caption{Velocity-Thresholding Identification (I-VT)} \label{alg:Velocity-Threshold Identification (IVT)}
\begin{algorithmic}[1]
\Function{I-VT}{$gaze\_points$}
    \For{$i \leftarrow 1$ \textbf{to} $N$}
        \If{$gaze\_points[i-1].valid$ \textbf{and} $gaze\_points[i].valid$}
            \State $V1 \gets gaze\_points[i-1].gazeVector$
            \State $V2 \gets gaze\_points[i].gazeVector$
            \If{$\Call{AngularVelocity}{V1, V2} < 100 \text{ deg/s}$}
                \State $gaze\_points[i].type \gets \text{"Fixation"}$
            \Else
                \State $gaze\_points[i].type \gets \text{"Saccade"}$
            \EndIf
        \Else
            \State $gaze\_points[i].type \gets \text{"Undefined"}$
        \EndIf
    \EndFor
    \State \Return $gaze\_points$
\EndFunction
\end{algorithmic}
\end{algorithm*}

\noindent
\paragraph{Optimization: Clustering (DBSCAN)}
Even after removing saccades, 50+ points can still be expensive for real-time DTW across thousands of candidate templates. Consequently, we further cluster and condense these fixation points via DBSCAN. Let us denote the 3D position (2D spatial coordinates plus 1D timestamp) of a fixation point by $\mathbf{p}_i$. DBSCAN requires two parameters, $\varepsilon_{\text{dbscan}}$ (the neighborhood radius) and $\mathit{minPts}$ (the minimum cluster size). Informally, $\mathbf{p}_i$ is a \textit{core point} if it has at least $\mathit{minPts}$ neighbors within distance $\varepsilon$. Formally, for each point $\mathbf{p}_i$,

\[
\text{neighbors}(\mathbf{p}_i) \;=\;
\{\mathbf{p}_j : \|\mathbf{p}_j - \mathbf{p}_i\| \,\le\, \varepsilon_{\text{dbscan}} \}.
\]

\noindent
If 
$\bigl|\text{neighbors}(\mathbf{p}_i)\bigr| \,\ge\, \mathit{minPts},$
then $\mathbf{p}_i$ is deemed a core point, and DBSCAN links $\mathbf{p}_i$ to other core points that share neighbors, thus forming a cluster. Any point not belonging to any cluster is labeled as noise. We then replace each cluster of fixation points with its centroid (for example, its mean position), further trimming the size of the gaze sequence. In our tests, DBSCAN reduces the number of points by about an order of magnitude, from $14.74 \pm 2.90$ down to $2.81 \pm 1.28$.

\noindent
\paragraph{Filtering by Starting Letter.}
Although the above steps significantly reduce the gaze sequence, we still run DTW against thousands of word templates. As an additional heuristic, we \textit{filter candidate words by their initial letter}. Specifically, we take the first centroid $\mathbf{p_1}$ from the DBSCAN output and measure its distance to the center $\ell \in \mathbb{R}^2$ of each letter on the keyboard. If $\|\mathbf{p_1} - \ell\|$ is below a predefined threshold $R$, we retain only those words whose first letter is $\ell$. Formally,
\[
\mathcal{V}_{\text{filtered}}
\;=\;
\Bigl\{
w \in \mathcal{V}\;\Big|\;\bigl\|\mathbf{p_1} - L(w_1)\bigr\| \le R
\Bigr\},
\]
where $\mathcal{V}$ is the entire vocabulary of candidate words, $w_1$ is the first letter of word $w$, and $L(\cdot)$ maps a letter to its canonical keyboard coordinates. This basic geometric filter sharply narrows down $\mathcal{V}_{\text{filtered}}$ before we proceed with DTW.

\begin{figure}[!t]
    \centering
    \includegraphics[width=\columnwidth]{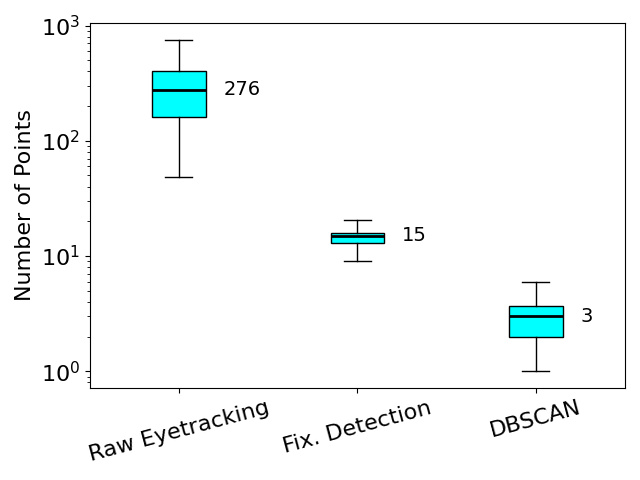}
    \caption{Number of gaze points is reduced significantly through Fixation Detection and then DBSCAN. Data from user study 1; numbers on the side are medians.}
    \Description{A box plot shows the number of gaze points at three processing stages: Raw Eyetracking, Fixation Detection, and DBSCAN. The y-axis is on a logarithmic scale labeled “Number of Points.” Three cyan box plots are displayed with median values printed beside each: 276 for Raw Eyetracking, 15 for Fixation Detection, and 3 for DBSCAN. The Raw Eyetracking distribution spans the highest range, Fixation Detection shows a substantially smaller range, and DBSCAN shows the smallest range of values. Whiskers extend vertically for each box to indicate variability.}
    \label{fig:npoints}
\end{figure}

\noindent
\paragraph{Cardinality reduction with I-VT and Clustering}
The core of \textit{Gaze2Word} is based on DTW, because it is able to align two time series of different lengths and speeds. For two time series $X = (x_1, \dots, x_n)$ and $Y = (y_1, \dots, y_m)$, DTW runs in $O(nm)$ time. In our case, $n$ corresponds to the number of points in the user’s gaze trace, while $m$ corresponds to the number of points in the candidate word’s trace (i.e., its length in letters). Because gaze is typically sampled at a high rate (200\,Hz in our experiments), $n$ can be large, making \textit{optimizations} that reduce the effective length of the gaze trace crucial. We introduce two main optimizations: \textit{Fixation Detection} and \textit{Clustering}. As shown in \autoref{fig:npoints}, these optimizations can reduce the number of gaze points by about two orders of magnitude.

\noindent
\paragraph{Spatiotemporal DTW: Detailed Derivation.}
Given a preprocessed gaze trace $\mathbf{g} = \{(x_i, y_i, t_i)\}_{i=1}^n$ and a template trace 
$\mathbf{q} = \{(x'_j, y'_j, t'_j)\}_{j=1}^m$, DTW computes the optimal alignment cost via dynamic programming:
\[
\begin{aligned}
D(i,j) \;&=\; 
dist\bigl(\mathbf{g}_i, \mathbf{q}_j\bigr)
+ \min\left\{
\begin{aligned}
&D(i-1,j), \\
&D(i,j-1), \\
&D(i-1,j-1)
\end{aligned}
\right\}, \\
\mathrm{DTW}(\mathbf{X}, \mathbf{Y}) \;&=\; D(n,m),
\end{aligned}
\]

\noindent
with boundary conditions $D(0,0)=0$, $D(i,0) = \infty$, and $D(0,j)=\infty$ for $i,j>0$. Here, $n$ and $m$ denote the lengths of sequences $\mathbf{X}$ and $\mathbf{Y}$, respectively.

Adding time and positional index as an additional dimension prevents late samples from aligning to early template points, maintaining temporal monotonicity. Doing so also preserves speed invariance—because $t'_j$ encodes only order, not absolute time. In practice, this reduces pathological alignments that Fréchet- or purely spatial DTW can admit when a single noisy fixation dominates. Our post hoc evaluation shows that including timestamps improved the top candidate match rate from 72.1\% to 77.9\% (more details in table \autoref{tab:decoder_comp}). We convert the raw DTW distance $\mathrm{DTW}(\mathbf{g}, \mathbf{t}_w)$ into probabilities through inverting and min-max normalization, as described in Algorithm \ref{alg:g2w}, line 13. Hence, a smaller DTW distance yields a higher value of $p_{\mathrm{dist}}(w; \mathbf{g})$.

\section{Fat-Finger Avoidance via 2D Gaussian and Character-Level n-gram}
\label{appendix: Fat-Finger Avoidance via 2D Gaussian and Character-Level n-gram}

To mitigate these “fat-finger” errors, we combine a 2D Gaussian-based model with a language model to infer the most likely intended letter. Let $\mathbf{z} = (z_x, z_y)$ be the user’s raw input (the tap or the gaze intersection point), and let $\mathbf{c}_\ell = (c_{\ell x}, c_{\ell y})$ be the center of letter~$\ell$ on the virtual keyboard. We define a 2D Gaussian likelihood as:
\[
p_{\mathrm{gauss}}(\ell \,\mid\, \mathbf{z}) 
\;=\;
\frac{1}{2\pi \,\sigma_x \sigma_y}
\exp\!\Bigl(-\tfrac{1}{2}\bigl(\mathbf{z} - \mathbf{c}_\ell\bigr)^\top 
\Sigma^{-1}
\bigl(\mathbf{z} - \mathbf{c}_\ell\bigr)\Bigr),
\]
where $\Sigma = \mathrm{diag}(\sigma_x^2, \sigma_y^2)$ and $(\sigma_x, \sigma_y)$ approximate “key size” or tapping variance. This gives higher probability to letters near $\mathbf{z}$, yet still accommodates nearby keys.

We then blend this likelihood with a character-level $n$-gram model: $p_{\mathrm{ngram}}(\ell \mid \text{prefix})$, where $\ell$ is the next character and \textit{prefix} is the typed text thus far (we use a bigram in practice). Let $\alpha \in [0,1]$ control the relative contribution of the $n$-gram. The final probability is
\[
p(\ell \,\mid\, \mathbf{z}, \text{prefix})
\;=\;
\alpha \,p_{\mathrm{ngram}}(\ell \,\mid\, \text{prefix})
\;+\;
(1 - \alpha)\,p_{\mathrm{gauss}}(\ell \,\mid\, \mathbf{z}),
\]
which yields a higher score for letters favored by both the geometric and language models.

\section{Late-Trigger Drift}
Following reports on \textit{Gaze\&Pinch} temporal mis-synchrony and late-trigger selection \cite{park2024impact,pfeuffer2024design}, we quantified drift between the intended key center and the actual gaze location at pinch. We found the mean drift magnitude was 21.2\,$\pm$\,30.6\,mm. We mitigate with a \emph{fat‑finger} Bayesian letter inference (see \autoref{appendix: Fat-Finger Avoidance via 2D Gaussian and Character-Level n-gram}); $2\sigma$ covers $\sim95\%$ of observed drifts. We observe that the drift remained constant across sessions while Gaze\&Pinch WPM rose, indicating the model absorbs temporal mis‑sync without added user burden.

\section{Aggregated Results from Prior Text-Entry Studies}

This section provides a detailed overview of existing XR text-input techniques evaluated in prior research. \autoref{tab:aggregated_prior_works} compiles reported performance metrics across diverse modalities (e.g., controller-based, hand-tracking, gaze-based, other eye swipes), listing average speeds (WPM), error rates, and study details such as participant counts and publication venues. Although this table is by no means complete, we hope it offers the reader a broader context of how recent methods compare with one another and with our proposed techniques.

\begin{table*}
  \centering
  \resizebox{\textwidth}{!}{
  \begin{tabular}{c c c c c c c c}
    \rowcolor{gray!30}
    Technique  & Modality & \begin{tabular}[c]{@{}c@{}}Speed\\(WPM)\end{tabular} & \begin{tabular}[c]{@{}c@{}}Error\\Evaluation\\Metrics\end{tabular} & \begin{tabular}[c]{@{}c@{}}Error\\(\%)\end{tabular} & \begin{tabular}[c]{@{}c@{}}Number of\\Participants\end{tabular} & Publication & \begin{tabular}[c]{@{}c@{}}Year of\\Publication\end{tabular}\\
    \toprule
    \midrule
    Raycasting\cite{boletsis2019controller}           
      & Controller & \textasciitilde16.65 & Total Error Rate (TER) & \textasciitilde11.05\% & 22 & IJVR & 2019 \\
    Drum-like Keyboard\cite{boletsis2019controller}   
      & Controller & \textasciitilde21.01 & Total Error Rate (TER) & \textasciitilde12.11\% & 22 & IJVR & 2019 \\
    Head-directed Input\cite{boletsis2019controller}  
      & Controller & \textasciitilde10.83 & Total Error Rate (TER) & \textasciitilde10.15\% & 22 & IJVR & 2019 \\
    Split Keyboard\cite{boletsis2019controller}       
      & Controller & \textasciitilde10.17 & Total Error Rate (TER) & \textasciitilde8.11\%  & 22 & IJVR & 2019 \\
    Controller + Tap\cite{xu2019pointing}
      & Controller & 14.60 & Total Error Rate (TER) & 1.65\% & 12 & ISMAR & 2019 \\

    Touch + Controller\cite{luong2023controllers}     
      & Controller & N/A (faster)   & Selection Error Rate & \textasciitilde7.3\% & 24 & TVCG & 2023 \\
    Raycast + Controller\cite{luong2023controllers}   
      & Controller & 11.8\% faster  & Selection Error Rate & \textasciitilde8.1\% & 24 & TVCG & 2023 \\

    Standard Qwerty Keyboard (STD)\cite{wan2024design} 
      & Controller & \textasciitilde13.23 & Total Error Rate (TER) & \textasciitilde8.27\% & 24 & TVCG & 2024 \\
    Layer-PointSwitch (LPS)\cite{wan2024design}        
      & Controller & \textasciitilde12.04 & Total Error Rate (TER) & \textasciitilde8.64\% & 24 & TVCG & 2024 \\
    Layer-ButtonSwitch (LBS)\cite{wan2024design}       
      & Controller & \textasciitilde15.94 & Total Error Rate (TER) & \textasciitilde4.65\% & 24 & TVCG & 2024 \\
    Key-ButtonSwitch (KBS)\cite{wan2024design}         
      & Controller & \textasciitilde15.08 & Total Error Rate (TER) & \textasciitilde4.46\% & 24 & TVCG & 2024 \\
    Controller + 2D Keys\cite{akhoroz2024poke}         
      & Controller & 14.08  & Character Error Rate (CER) & 1.91\%  & 28 & ICMI & 2024 \\
    Controller + 3D Keys\cite{akhoroz2024poke}         
      & Controller & 16.52  & Character Error Rate (CER) & 1.17\%  & 28 & ICMI & 2024 \\

    \midrule
    DesktopKeyboard + NoReposition\cite{grubert2018text}   
      & Desktop Keyboard & 26.3  & Character Error Rate (CER) & 2.1\%  & 24 & IEEE VR & 2018 \\
    DesktopKeyboard + Reposition\cite{grubert2018text}     
      & Desktop Keyboard & 25.5  & Character Error Rate (CER) & 2.4\%  & 24 & IEEE VR & 2018 \\

    \midrule
    BlinkType\cite{lu2020handsfree}   
      & Gaze + Blink & 13.47 & Total Error Rate (TER) & 10.44\% & 36 & arXiv & 2020 \\

    \midrule
    Click\cite{mutasim2021pinch}      
      & Gaze + Controller & N/A & Error Rate & \textasciitilde11\% & 12 & ETRA & 2021 \\

    \midrule
    Pinch\cite{mutasim2021pinch}      
      & Gaze + Hand & N/A & Error Rate & \textasciitilde13\% & 12 & ETRA & 2021 \\

    \midrule
    Gaze + Keyboard Button\cite{gizatdinova2012comparison}  
      & Gaze & 10.98 & Relative Error Rate & 8.0\%  & 15 & AVI & 2012 \\

    EyeSwipe (Dwell-free)\cite{kurauchi2016eyeswipe}    
      & Gaze & 11.7 & Minimum String Distance (MSD) Error & 1.31\% & 10 & CHI & 2016 \\
    Dwell-time Gaze Typing\cite{kurauchi2016eyeswipe}   
      & Gaze & 9.5  & Minimum String Distance (MSD) Error & 1.01\% & 10 & CHI & 2016 \\

    Static Dwell Gaze Typing\cite{mott2017improving}   
      & Gaze & 10.62 & Total Error Rate (TER) & 16.83\% & 17 & CHI & 2017 \\
    Cascading Dwell Gaze Typing\cite{mott2017improving}
      & Gaze & 12.39 & Total Error Rate (TER) & 11.08\% & 17 & CHI & 2017 \\

    Flat Keyboard + Sitting + Dwell\cite{rajanna2018gaze} 
      & Gaze & 9.36 & Rate of Backspace Activation (RBA) & 2.0\% & 16 & ETRA & 2018 \\
    Flat Keyboard + Sitting + Click\cite{rajanna2018gaze}
      & Gaze & 10.15 & RBA & 7.0\% & 16 & ETRA & 2018 \\
    Flat Keyboard + Biking + Dwell\cite{rajanna2018gaze}
      & Gaze & 8.07 & RBA & 4.0\% & 16 & ETRA & 2018 \\
    Flat Keyboard + Biking + Click\cite{rajanna2018gaze}
      & Gaze & 8.58 & RBA & 8.0\% & 16 & ETRA & 2018 \\
    Curved Keyboard + Sitting + Dwell\cite{rajanna2018gaze}
      & Gaze & 7.48 & RBA & 6.0\% & 16 & ETRA & 2018 \\
    Curved Keyboard + Sitting + Click\cite{rajanna2018gaze}
      & Gaze & 9.15 & RBA & 3.0\% & 16 & ETRA & 2018 \\
    Curved Keyboard + Biking + Dwell\cite{rajanna2018gaze}
      & Gaze & 6.77 & RBA & 4.0\% & 16 & ETRA & 2018 \\
    Curved Keyboard + Biking + Click\cite{rajanna2018gaze}
      & Gaze & 8.29 & RBA & 3.0\% & 16 & ETRA & 2018 \\

    Dwell\cite{mutasim2021pinch}
      & Gaze & N/A & Error Rate & \textasciitilde7\%  & 12 & ETRA & 2021 \\
    Eye-typing with Dwell (OptiKey)\cite{bafna2021mental}
      & Gaze & 13.0 & Uncorrected Error Rate & 5.7\% & 18 & PLOS ONE & 2021 \\
    Dwell-Typing\cite{lystbaek2022gaze}
      & Gaze & 9.50  & Total Error Rate (TER) & 4.06\% & 16 & ETRA & 2022 \\

    GlanceWriter (Exp I)\cite{cui2023glancewriter}       
      & Gaze & 10.89 & Word Error Rate (WER) & 2.71\% & 14 & CHI & 2023 \\
    EyeSwipe\cite{cui2023glancewriter}                   
      & Gaze & 6.49  & Word Error Rate (WER) & 6.85\% & 14 & CHI & 2023 \\
    GlanceWriter (Exp II)\cite{cui2023glancewriter}      
      & Gaze & 9.54  & Word Error Rate (WER) & 12.89\% & 12 & CHI & 2023 \\
    Tobii Communicator 5\cite{cui2023glancewriter}       
      & Gaze & 7.41  & Word Error Rate (WER) & 16.32\% & 12 & CHI & 2023 \\

    \midrule
    S-Gaze\&Finger\cite{lystbaek2022gaze}   
      & Hand + Gaze & 10.66 & Total Error Rate (TER) & 2.9\%  & 16 & ETRA & 2022 \\
    S-Gaze\&Hand\cite{lystbaek2022gaze}     
      & Hand + Gaze & 9.49  & Total Error Rate (TER) & 2.98\% & 16 & ETRA & 2022 \\
    Speedup (Gaze-Assisted)\cite{zhao2023gaze}        
      & Hand + Gaze & 17.6  & Word Error Rate (WER) & 1.01\% & 12 & IUI & 2023 \\
    Gaussian Speedup (Gaze-Assisted)\cite{zhao2023gaze}
      & Hand + Gaze & 17.1  & Word Error Rate (WER) & 0.71\% & 12 & IUI & 2023 \\

    \midrule
    Hand + Tap\cite{xu2019pointing}       
      & Hand & 6.74  & Total Error Rate (TER) & 6.48\% & 12 & ISMAR & 2019 \\
    AirTap\cite{lystbaek2022gaze}          
      & Hand & 11.37 & Total Error Rate (TER) & 3.54\% & 16 & ETRA & 2022 \\
    Touch + Hand\cite{luong2023controllers}
      & Hand & N/A            & Selection Error Rate & \textasciitilde7.3\% & 24 & TVCG & 2023 \\
    Raycast + Hand\cite{luong2023controllers}
      & Hand & slower         & Selection Error Rate & \textasciitilde8.1\% & 24 & TVCG & 2023 \\
    Wrist-Only Gesture Typing\cite{zhao2023gaze} 
      & Hand & 16.4  & Word Error Rate (WER) & 0.98\% & 12 & IUI & 2023 \\

    \midrule
    Midair Keyboard Only (NoSpeech)\cite{adhikary2021text} 
      & Hand-tracking & 11.1 & Character Error Rate (CER) & 1.2\% & 18 & TVCG & 2021 \\
    Hand-tracking + 2D Keys\cite{akhoroz2024poke} 
      & Hand-tracking & 12.78  & Character Error Rate (CER) & 0.98\%  & 28 & ICMI & 2024 \\
    Hand-tracking + 3D Keys\cite{akhoroz2024poke} 
      & Hand-tracking & 14.73  & Character Error Rate (CER) & 0.85\%  & 28 & ICMI & 2024 \\

    \midrule
    HGaze Typing (Head + Gaze)\cite{feng2021hgaze} 
      & Head + Gaze & 11.22 & Minimum String Distance (MSD) Error & \textasciitilde0.37\% & 10 & ETRA & 2021 \\

    \midrule
    Hybrid + Tap\cite{xu2019pointing}     
      & Head + Hand & 8.53  & Total Error Rate (TER) & 3.85\% & 12 & ISMAR & 2019 \\

    \midrule
    Head + Keyboard Button\cite{gizatdinova2012comparison}         
      & Head & 4.42  & Relative Error Rate & 3.8\%  & 15 & AVI & 2012 \\
    Head + Mouth Open Gesture\cite{gizatdinova2012comparison}      
      & Head & 3.07  & Relative Error Rate & 6.0\%  & 13 & AVI & 2012 \\
    Head + Brows Up Gesture\cite{gizatdinova2012comparison}        
      & Head & 2.85  & Relative Error Rate & 21.0\% & 13 & AVI & 2012 \\

    TapType\cite{yu2017tap}       
      & Head & 15.58 & Total Error Rate (TER) & 2.02\% & 6 & CHI & 2017 \\
    DwellType\cite{yu2017tap}     
      & Head & 10.59 & Total Error Rate (TER) & 3.69\% & 6 & CHI & 2017 \\
    GestureType\cite{yu2017tap}   
      & Head & 19.04 & Total Error Rate (TER) & 4.21\% & 6 & CHI & 2017 \\
    GestureType (trained)\cite{yu2017tap} 
      & Head & 24.73 & Total Error Rate (TER) & 5.82\% & 12 & CHI & 2017 \\

    Head + Tap\cite{xu2019pointing}       
      & Head & 5.62  & Total Error Rate (TER) & 1.06\% & 12 & ISMAR & 2019 \\

    DwellType\cite{lu2020handsfree}   
      & Head & 11.65 & Total Error Rate (TER) & 9.64\% & 36 & arXiv & 2020 \\
    NeckType\cite{lu2020handsfree}    
      & Head & 11.18 & Total Error Rate (TER) & 9.04\% & 36 & arXiv & 2020 \\

    \midrule
    Speech + Midair Keyboard\cite{adhikary2021text}        
      & Speech + Hand-tracking & 27.9 & Character Error Rate (CER) & 0.5\% & 18 & TVCG & 2020 \\

    \midrule
    Touchscreen Keyboard + NoReposition\cite{grubert2018text}
      & Touchscreen Keyboard & 11.6  & Character Error Rate (CER) & 2.7\%  & 24 & IEEE VR & 2018 \\
    TouchscreenKeyboard + Reposition\cite{grubert2018text}
      & Touchscreen Keyboard & 8.8   & Character Error Rate (CER) & 3.6\%  & 24 & IEEE VR & 2018 \\

    \bottomrule
  \end{tabular}
  }
  \caption{XR text-input techniques proposed over the years, sorted first by \textbf{Modality} and then by \textbf{Year of Publication}.}
  \Description{A comprehensive reference table listing XR text-input techniques proposed in prior work, organized first by input modality (e.g., gaze, hand, controller, hybrid, touchscreen) and then by year of publication. For each technique, the table reports its name, input modality, performance metric (such as WPM, TER, CER, WER, MSD, or RBA), the reported value, the number of participants, and the publication venue and year.}
  \label{tab:aggregated_prior_works}
\end{table*}

\section{Participant Demographics}
\begin{figure*}[h]
    \centering
    \includegraphics[width=\textwidth]{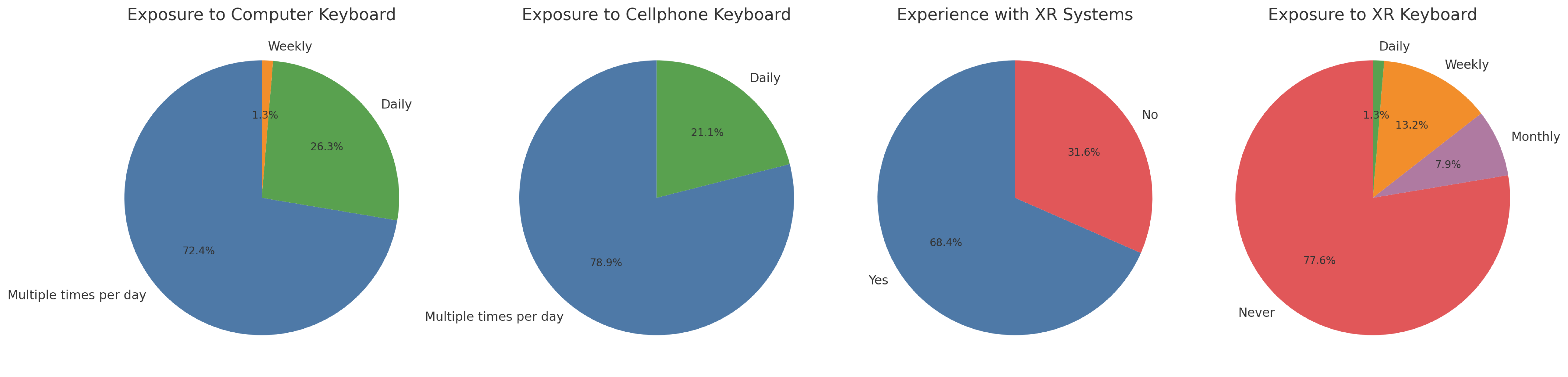}
    \caption{Participant demographics through US1, US2 and US3, showing participants' reported exposure to computer keyboards, cellphone keyboards, participants’ experience with XR systems and exposure to XR keyboard.}
    \Description{The figure contains four pie charts summarizing participant demographics across US1, US2, and US3. The first pie chart shows exposure to computer keyboards, with 72.4\% of participants reporting use multiple times per day, 26.3\% reporting daily use, and 1.3\% reporting weekly use. The second pie chart shows exposure to cellphone keyboards, with 78.9\% of participants reporting use multiple times per day and 21.1\% reporting daily use. The third pie chart shows experience with XR systems, with 68.4\% of participants reporting prior XR experience and 31.6\% reporting no prior XR experience. The fourth pie chart shows exposure to XR keyboards, with 77.6\% of participants reporting never using an XR keyboard, 13.2\% reporting weekly use, 7.9\% reporting monthly use, and 1.3\% reporting daily use.}
    \label{fig:demographics}
\end{figure*}

We provide additional detail on the prior experience of participants in US1, US2, and US3. Figure~\ref{fig:demographics} displays the distribution of self-reported usage frequencies for computer and cellphone keyboards, along with participants’ XR exposure and XR keyboard experience. We hope this demographic information helps readers contextualize subsequent performance.

Across US1, US2, and US3 survey responses, participants who reported prior experience with XR systems indicated that their exposure primarily involved gaming, academic projects, software development, public demos, or research, frequently citing devices such as Meta Quest and Apple Vision Pro.

Among those who have used XR keyboards, many described the interaction as unintuitive or inefficient, often attributing challenges to tracking inaccuracies, the absence of haptic feedback, and physical discomfort. However, some participants noted improved usability over time or with the use of more advanced systems such as the Apple Vision Pro.

\section{Additional Results}
We present supplementary user study results that were omitted from the primary text. While data presented here did not significantly alter our primary conclusions, they are included here to provide a more complete picture of each user study and to address secondary questions readers may have.

\begin{figure*}[h]
    \centering
    \begin{subfigure}[t]{\columnwidth}
        \centering
        \includegraphics[width=\linewidth]{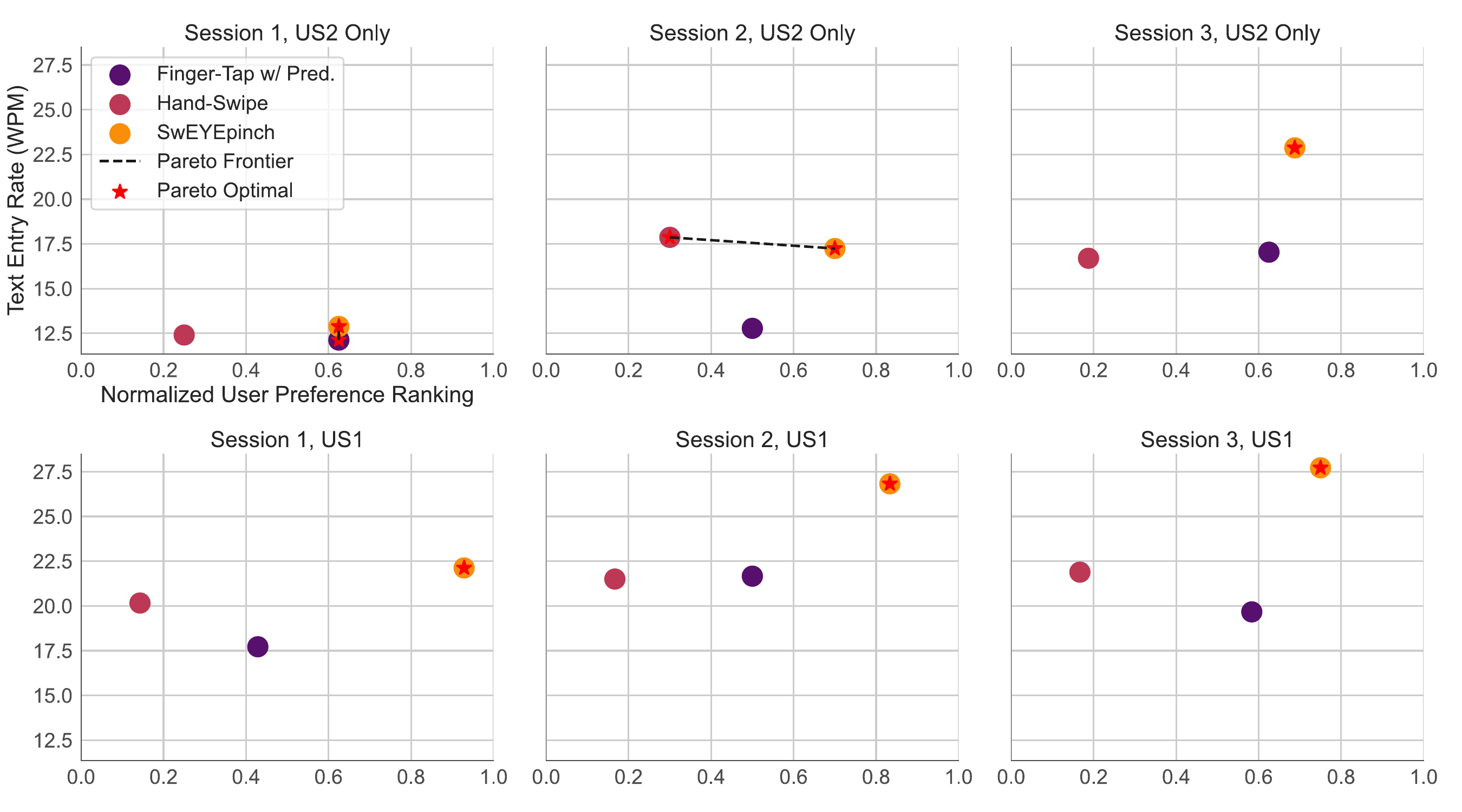}
        \label{fig:study3_pareto_sessions_1to3_2x3}
    \end{subfigure}\hfill
    \begin{subfigure}[t]{\columnwidth}
        \centering
        \includegraphics[width=\linewidth]{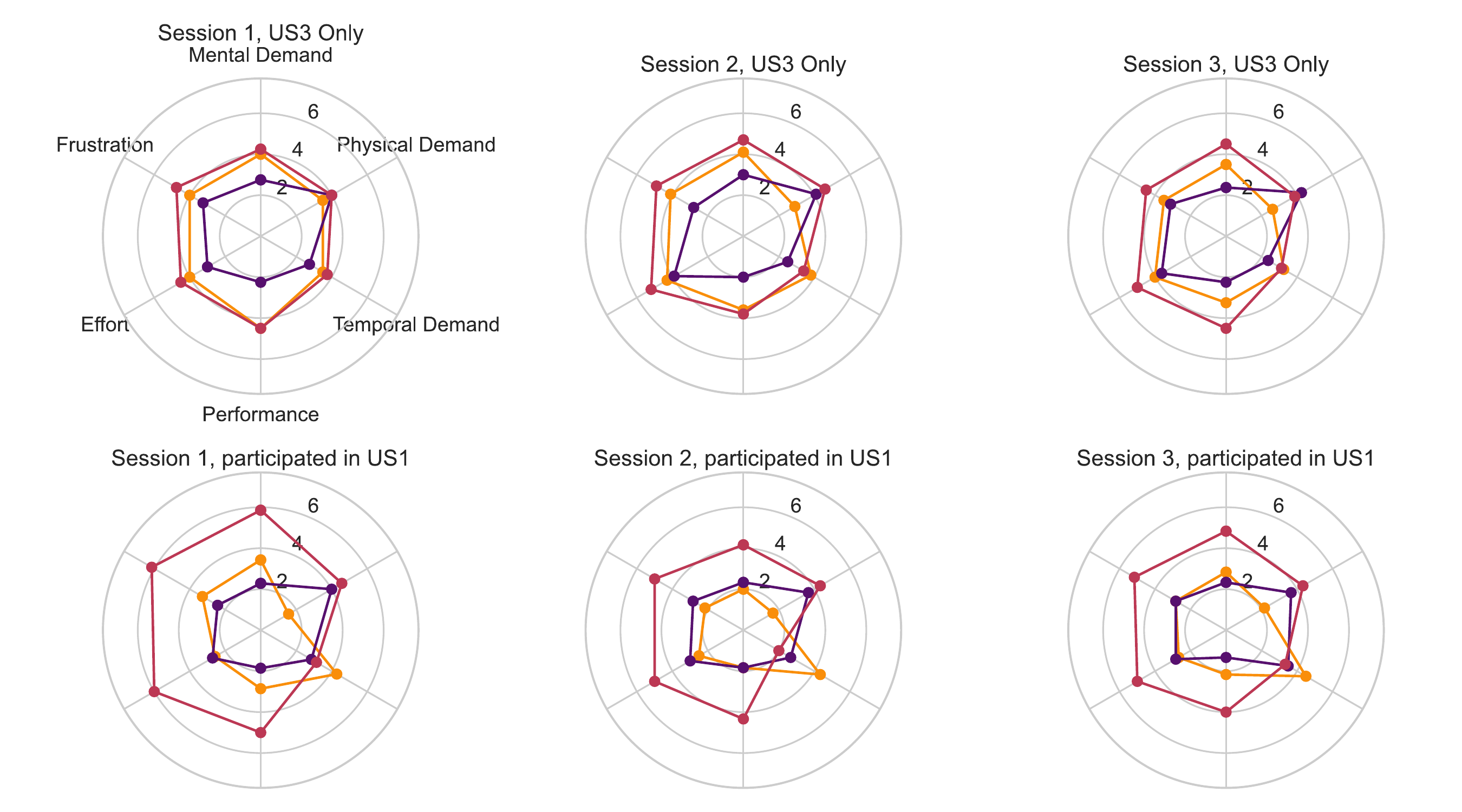}
        \label{fig:study3_tlx_sessions_1to3_2x3}
    \end{subfigure}
    \caption{Left: Pareto Frontier. Right: NASA TLX. Data are shown across all \textit{three} sessions from \textit{US3}. ``US3 Only'' includes participants who only participated in US3. Bottom row ``US1'' includes participants who have also participated in US1.}
    \Description{The figure contains two sets of visualizations. The top six panels show Pareto frontier plots comparing normalized user preference ranking (x-axis) and text entry rate in WPM (y-axis) for three techniques—Finger-Tap with Prediction (purple), Hand-Swipe (red), and SwEYEpinch (orange)—across Sessions 1–3 for two participant groups: US3 Only (top row) and US1 participants (middle row). In each scatterplot, individual techniques appear as colored circular markers, and orange star markers indicate Pareto-optimal points. In Session 1 for US3 Only, points cluster between 0.2 and 0.7 on preference ranking and between roughly 12.5 and 17.5 WPM. In Session 2, the Pareto frontier is shown as a dashed black line leading to the Pareto-optimal SwEYEpinch point around 17–18 WPM and preference near 0.7. In Session 3, SwEYEpinch remains the highest-performing technique with a Pareto-optimal point around 22 WPM and preference near 0.75. Similar patterns appear in the US1 group plots, where SwEYEpinch continues to occupy the upper-right region of each chart. The lower six panels show NASA TLX radar charts for mental demand, physical demand, temporal demand, performance, effort, and frustration across Sessions 1–3. The left column is for US3 Only participants and the right column is for US1 participants. Each radar chart contains three colored traces: purple for Finger-Tap with Prediction, red for Hand-Swipe, and orange for SwEYEpinch. The axes are numbered from 0 to 6. Across sessions, SwEYEpinch (orange) generally shows lower values on several workload dimensions—particularly mental demand and frustration—while Hand-Swipe (red) consistently shows higher ratings on physical demand and effort. Finger-Tap with Prediction (purple) typically lies between the two. Shapes remain similar across Sessions 1, 2, and 3, with minor reductions in workload for most techniques over time.}
    \label{fig:us3_pareto_tlx_sessions_1to3}
\end{figure*}

\autoref{fig:us3_pareto_tlx_sessions_1to3} shows the Pareto and NASA TLX results from \textit{US3}—stratifying participants into those who only took part in US3 versus those who also participated in US1.

\begin{figure*}[h]
    \includegraphics[width=\textwidth]{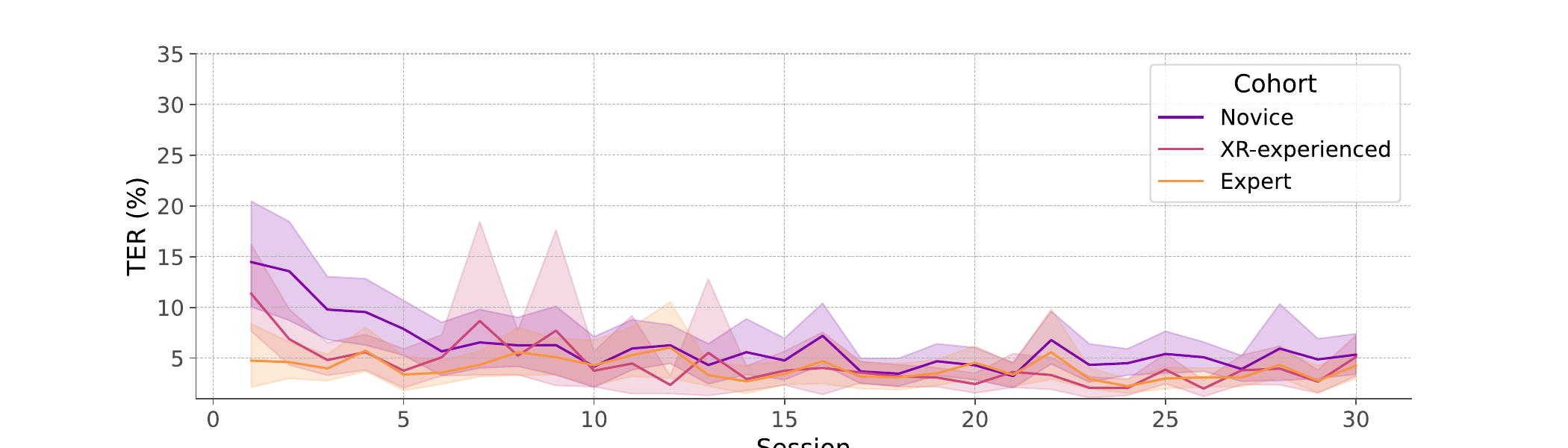}
    \caption{TER for the three user cohorts in \textit{US4}, across the 30 sessions.}
    \Description{A line chart shows typing performance across 30 sessions for three individual users in US4, grouped by cohort: Novice (purple), XR-experienced (pink), and Expert (orange). The y-axis is labeled WPM, ranging from 0 to 35, and the x-axis shows sessions from 1 to 30. All three lines begin with higher variability in the early sessions and gradually stabilize toward lower WPM in later sessions. The Novice user begins around the high teens to low 20s WPM, then declines steadily into the 4-6 WPM range. The XR-experienced user shows large fluctuations early on, including several peaks above 15 WPM, then converges to around 4-6 WPM. The Expert user begins near 8 WPM, briefly dips to around 3 WPM, and stabilizes around 3-5 WPM for the remaining sessions. Shaded regions around each line indicate variability across trials for each user.}
    \label{fig:us4_ter}
\end{figure*}

In our paper, we show WPM and TER from US3 with stratified participants (based on their participation in just US3 or both). Here in the supplementary material, we present \autoref{fig:us3_wpm_ter}, showing WPM and TER without stratification. Finally, \autoref{fig:us4_ter} plots the TER observed over the 30-session period in \textit{User Study 4}. We hope this can provide additional insight into the long-term performance trajectories of individual participants.

\begin{figure}[t]
    \centering
    \begin{subfigure}[b]{0.49\columnwidth}
        \centering
        \includegraphics[width=\linewidth]{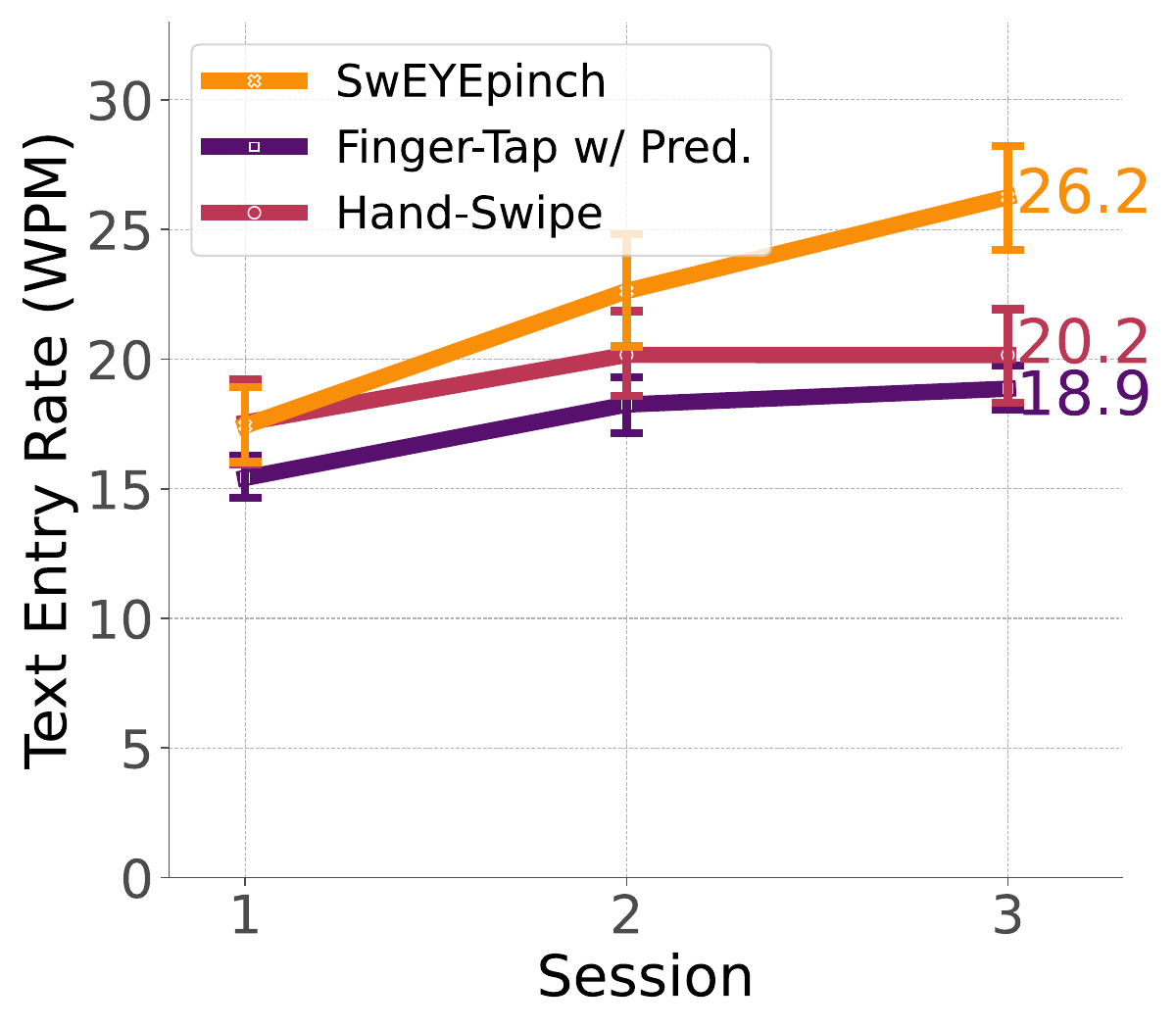}
        \label{fig:us3_wpm_wo_is_in_study_1}
    \end{subfigure}\hfill
    \begin{subfigure}[b]{0.49\columnwidth}
        \centering
        \includegraphics[width=\linewidth]{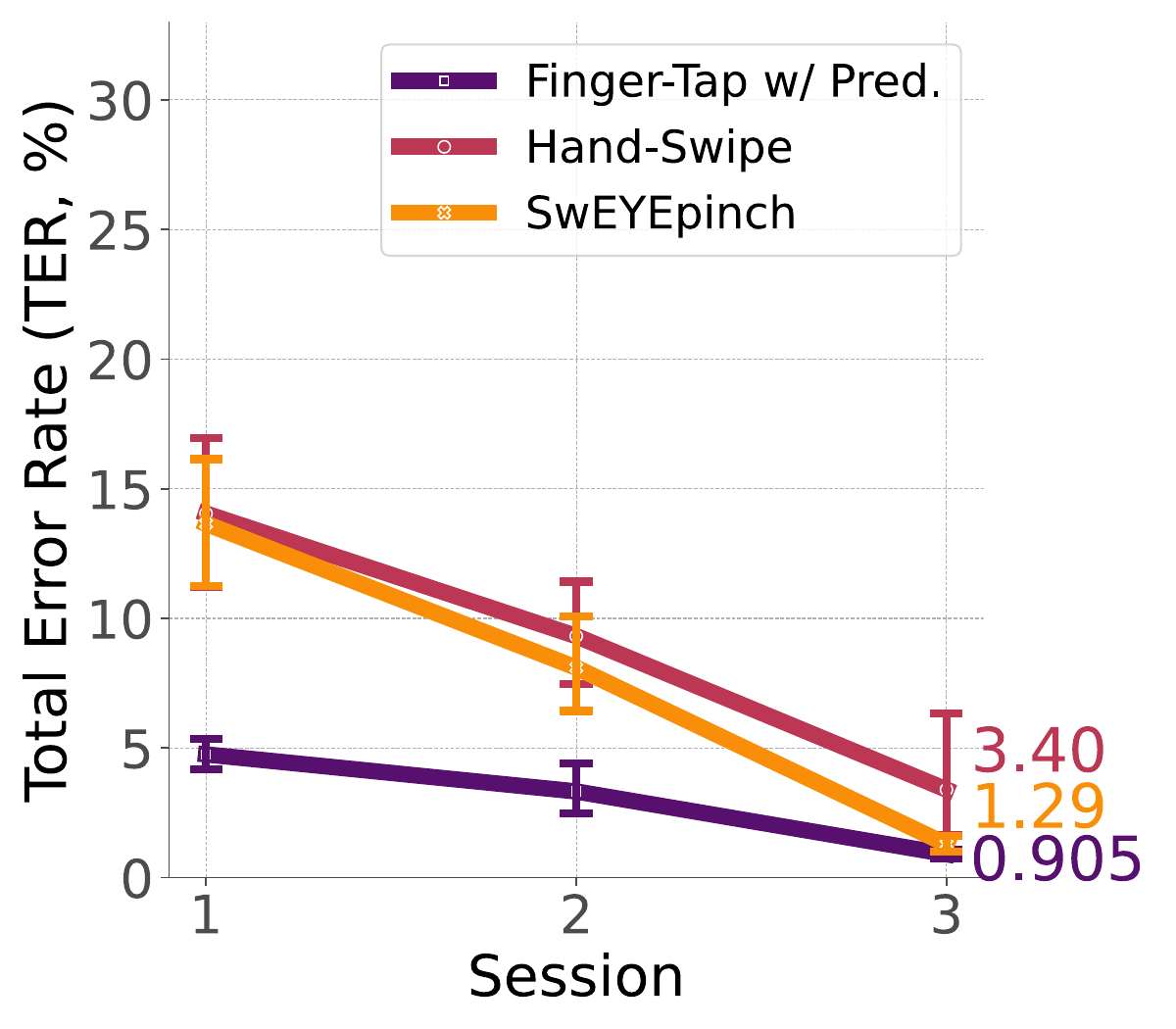}
        \label{fig:us3_ter_wo_is_in_study_1}
    \end{subfigure}
    \caption{\textit{left}: Average text entry speeds (WPM). \textit{right}: TER. Data are shown across all \textit{three} sessions from \textit{US3}.}
    \Description{The figure contains two line charts showing results from US3 across three sessions. The left chart shows text entry rate in words per minute for three techniques: SwEYEpinch, Finger-Tap with Prediction, and Hand-Swipe. Error bars are shown for all conditions. In Session~1, SwEYEpinch achieves approximately 18~WPM, Finger-Tap with Prediction achieves approximately 15~WPM, and Hand-Swipe achieves approximately 17~WPM. In Session~2, SwEYEpinch increases to above 22~WPM, Finger-Tap with Prediction reaches approximately 18~WPM, and Hand-Swipe reaches approximately 20~WPM. In Session~3, SwEYEpinch reaches 26.2~WPM, Hand-Swipe reaches 20.2~WPM, and Finger-Tap with Prediction reaches 18.9~WPM. The right chart shows total error rate for the same three techniques, with error decreasing across sessions and error bars shown for all values. In Session~1, Finger-Tap with Prediction shows the lowest error at approximately 5\%, while Hand-Swipe and SwEYEpinch show higher error rates of approximately 16\% and 14\%, respectively. In Session~2, Finger-Tap with Prediction decreases to approximately 3\%, while Hand-Swipe and SwEYEpinch drop to approximately 10\% and 8\%, respectively. In Session~3, Finger-Tap with Prediction reaches 0.905\%, SwEYEpinch reaches 1.29\%, and Hand-Swipe reaches 3.40\%.}
    \label{fig:us3_wpm_ter}
\end{figure}

\begin{table}[!t]
\centering
\begin{tabular}{l
                S[table-format=1.2]
                S[table-format=1.2]
                S[table-format=1.3]
                S[table-format=1.3]
                S[table-format=2.1]
                S[table-format=2.1]
                S[table-format=2.1]}
\toprule
\textbf{User} &
\multicolumn{4}{c}{\textbf{Learning rate (WPM/session)}} &
\multicolumn{3}{c}{\textbf{Median WPM}} \\
\cmidrule(lr){2-5} \cmidrule(lr){6-8}
 & \textbf{1--10} & \textbf{11--20} & \textbf{21--30} & \textbf{All} &
   \textbf{First} & \textbf{Best} & \(\Delta\) \textbf{WPM} \\
\midrule
P401 & 1.93 & 1.14 & 0.263 & 1.02 & 26.7 & 60.6 & 33.9 \\
P402 & 1.33 & 1.77 & 1.11 & 1.29 & 19.8 & 64.7 & 44.9 \\
P403 & 1.92 & 1.67 & 0.524 & 0.943 & 8.35 & 45.4 & 35.8 \\
P404 & 0.95 & 1.82 & 0.961 & 0.807 & 22.0 & 46.2 & 24.1 \\
P405 & 1.53 & 1.12 & 0.774 & 1.14 &  9.8 & 56.5 & 46.8 \\
P406 & 2.43 & 1.08 & 1.43 & 1.65 &  5.5 & 45.9 & 40.4 \\
P407 & 1.51 & 0.38 & 2.98 & 1.62 &  8.0 & 57.0 & 49.1 \\
P408 & 2.35 & 1.45 & 1.65 & 1.09 &  6.4 & 48.9 & 42.5 \\
P409 & 1.16 & 1.02 & 0.118 & 0.767 & 15.1 & 60.9 & 45.8 \\
\bottomrule
\end{tabular}
\caption{Per-user learning rates (WPM/session) in 10-session blocks and overall, and median WPM at first vs.\ best session with gains.}
\Description{The table reports per-user learning rates, measured in words per minute per session, across three 10-session blocks (Sessions~1--10, 11--20, and 21--30), along with an overall learning rate and median words per minute at the user’s first session, best session, and the resulting improvement in words per minute. Each row corresponds to one participant, labeled P401 through P409. For participant P401, the learning rate is 1.93 for Sessions~1--10, 1.14 for Sessions~11--20, 0.263 for Sessions~21--30, and 1.02 overall. Median text entry rate is 26.7~WPM at the first session and 60.6~WPM at the best session, yielding an improvement of 33.9~WPM. For participant P402, the learning rate is 1.33 for Sessions~1--10, 1.77 for Sessions~11--20, 1.11 for Sessions~21--30, and 1.29 overall. Median text entry rate is 19.8~WPM at the first session and 64.7~WPM at the best session, yielding an improvement of 44.9~WPM. For participant P403, the learning rate is 1.92 for Sessions~1--10, 1.67 for Sessions~11--20, 0.524 for Sessions~21--30, and 0.943 overall. Median text entry rate is 8.35~WPM at the first session and 45.4~WPM at the best session, yielding an improvement of 35.8~WPM. For participant P404, the learning rate is 0.95 for Sessions~1--10, 1.82 for Sessions~11--20, 0.961 for Sessions~21--30, and 0.807 overall. Median text entry rate is 22.0~WPM at the first session and 46.2~WPM at the best session, yielding an improvement of 24.1~WPM. For participant P405, the learning rate is 1.53 for Sessions~1--10, 1.12 for Sessions~11--20, 0.774 for Sessions~21--30, and 1.14 overall. Median text entry rate is 9.8~WPM at the first session and 56.5~WPM at the best session, yielding an improvement of 46.8~WPM. For participant P406, the learning rate is 2.43 for Sessions~1--10, 1.08 for Sessions~11--20, 1.43 for Sessions~21--30, and 1.65 overall. Median text entry rate is 5.5~WPM at the first session and 45.9~WPM at the best session, yielding an improvement of 40.4~WPM. For participant P407, the learning rate is 1.51 for Sessions~1--10, 0.38 for Sessions~11--20, 2.98 for Sessions~21--30, and 1.62 overall. Median text entry rate is 8.0~WPM at the first session and 57.0~WPM at the best session, yielding an improvement of 49.1~WPM. For participant P408, the learning rate is 2.35 for Sessions~1--10, 1.45 for Sessions~11--20, 1.65 for Sessions~21--30, and 1.09 overall. Median text entry rate is 6.4~WPM at the first session and 48.9~WPM at the best session, yielding an improvement of 42.5~WPM. For participant P409, the learning rate is 1.16 for Sessions~1--10, 1.02 for Sessions~11--20, 0.118 for Sessions~21--30, and 0.767 overall. Median text entry rate is 15.1~WPM at the first session and 60.9~WPM at the best session, yielding an improvement of 45.8~WPM.}
\label{tab:us3_learning_rate_and_wpm}
\end{table}

\subsection{Summary of Qualitative User Feedback}
In this section, we present the excerpts from participant comments and a thematic breakdown of user impressions across different typing methods, drawn from both \textit{US1} and \textit{US3}. \autoref{tab:typing-feedback} summarizes key feedback themes (e.g., ease of use, speed, error handling) for Finger-Tap, Gaze\&Pinch, and SwEYEpinch-Basic in \textit{US1}. Representative user quotes appear in \autoref{tab:quotes}.

For \textit{US3}, \autoref{tab:typing-feedback-us3} and \autoref{tab:quotes-us3} provides a similar thematic summary contrasting Finger-Tap with Prediction\&Completion, Hand-Swipe, and SwEYEpinch. \autoref{tab:quotes-us3} contains additional direct quotes illustrating how participants perceived each method’s advantages and drawbacks. We hope these summaries and excerpts can help contextualize our quantitative findings.

\begin{table*}[ht]
\resizebox{\textwidth}{!}{
\centering
\begin{tabular}{|p{3.5cm}|p{4.2cm}|p{4.2cm}|p{4.2cm}|}
\hline
\textbf{Theme} & \textbf{Finger-Tap} & \textbf{Gaze\&Pinch} & \textbf{SwEYEpinch-Basic} \\
\hline
\textbf{Ease of Use / Intuitiveness} & Familiar, intuitive like mobile typing; easy to correct errors & Requires learning curve; hard to control gaze and pinch coordination & Easy after practice; auto-suggestion helps users improve speed \\
\hline
\textbf{Physical Effort} & Most physically demanding; arm fatigue common due to hand position & Less physically demanding than Finger-Tap, but tiring over long use & Least physically demanding; only eye and light hand gestures needed \\
\hline
\textbf{Mental Demand / Fatigue} & Low cognitive load; users could monitor text while typing & High mental load; gaze focus and hand coordination stressful & Moderate mental demand; stressful for long or complex words \\
\hline
\textbf{Accuracy / Reliability} & High accuracy; hand tracking occasionally caused unintended double taps & Least accurate; heavily dependent on calibration and tracking quality & Moderate accuracy; autocorrect can help but also cause unintended words \\
\hline
\textbf{Calibration Sensitivity} & Less sensitive; works even with some hand tracking drift & Highly sensitive; even minor tracking drift ruins input & Somewhat sensitive; tracking drift affects swipe path recognition \\
\hline
\textbf{Error Recovery} & Easy to fix errors manually & Error correction slow and frustrating; often required full re-typing & Autocorrect helpful, but hard to fix partial mistakes \\
\hline
\textbf{Speed / Efficiency} & Slowest due to physical input speed & Slow due to gaze/pinch coordination & Fastest once mastered; boosted by predictive text \\
\hline
\textbf{Learning Curve} & Very low; resembles real-life typing & Steep; most users struggled to master it in short time & Moderate; improved notably with repeated use \\
\hline
\textbf{User Preference Trend} & Most reliable but tiring; good fallback method & Least preferred due to frustration and tracking issues & Most preferred overall for speed and comfort \\
\hline
\end{tabular}
}
\caption{Summary of user feedback for Finger-Tap, Gaze\&Pinch, and SwEYEpinch-Basic (US1).}
\Description{Table~7 summarizes user feedback for three text-entry techniques: Finger-Tap with Prediction, Hand-Swipe, and SwEYEpinch, organized across multiple qualitative themes. Regarding ease of use and intuitiveness, Finger-Tap is described as familiar and similar to phone typing, resulting in a low learning curve. Hand-Swipe is reported as not intuitive for many users and requires difficult gesture control. SwEYEpinch is initially unfamiliar but becomes intuitive after learning and is described by some users as futuristic. In terms of physical effort, Finger-Tap is reported to be physically taxing and to cause arm fatigue. Hand-Swipe is described as the most exhausting technique because holding the hand up for extended periods is tiring. SwEYEpinch is perceived as the least physically demanding technique, requiring primarily eye movement and light pinch gestures. With respect to mental demand and fatigue, Finger-Tap is associated with low mental effort. Hand-Swipe imposes high mental demand due to the need to coordinate gestures.}
\label{tab:typing-feedback}
\end{table*}

\begin{table*}[ht]
\resizebox{\textwidth}{!}{
\centering
\renewcommand{\arraystretch}{1.4}
\begin{tabular}{|p{3.2cm}|p{3.2cm}|p{7.5cm}|}
\hline
\textbf{Typing Method} & \textbf{Theme} & \textbf{Representative Quote} \\
\hline

\multirow{3}{*}{Finger-Tap} 
& Physical Effort & ``Finger-Tap was easy but physically demanding.'' \\
&  & ``Finger tap required a lot of upper arm strain in order for the computer to pick up my hand movements.'' \\
\cline{2-3}
& Accuracy / Control & ``Finger tap is the easiest because it resembles the keyboard most.'' \\
&  & ``Easy to fix the mistake, unlike the other two methods.'' \\
\cline{2-3}
& Familiarity & ``Finger-Tap is still the most intuitive way to type... muscle memory typing on a keyboard.'' \\
\hline

\multirow{4}{*}{Gaze\&Pinch} 
& Calibration Issues & ``Gaze\&pinch was hardest due to eye tracking not being accurate.'' \\
&  & ``If the calibration was even slightly off, it made the task extremely frustrating.'' \\
\cline{2-3}
& Physical / Mental Demand & ``Gaze\&Pinch was the most mentally and physically exhausting.'' 
\\
\cline{2-3}
& Improvement with Familiarity & ``I would rank Gaze\&Pinch over Finger-Tap because I am familiar with the distance to trigger the pinch action and thus typing speed is improved.'' \\
\hline

\multirow{4}{*}{SwEYEpinch-Basic} 
& Speed / Ease & ``Swipe can be the fastest and easiest because you don't have to focus on the letters.'' \\
&  & ``Once I mastered the coordination, swipe was the simplest to use.'' \\
\cline{2-3}
& Autocorrect / Error Handling & ``Swipe was fun and quite easy to fix when it went wrong.'' \\
&  & ``Swipe had a hard time recognizing some common words like 'the' and 'strawberries'.'' \\
\cline{2-3}
& Learning Curve / Fatigue & ``Swipe was initially hard to get used to, but it became a lot less demanding once I understood how it works.'' \\
&  & ``Swipe required a lot of effort because often if I would misread the word... it required effort to retype it.'' \\
\hline

\end{tabular}
}
\caption{Representative user quotes for US1}
\Description{A table presenting representative user quotes for the three typing methods—Finger-Tap, Gaze & Pinch, and SwEYEPinch-Basic—organized by themes. For Finger-Tap, under Physical Effort, users describe it as easy but physically demanding, causing upper-arm strain. Under Accuracy / Control, users say it resembles keyboard typing, is easiest to control, and mistakes are easy to fix. Under Familiarity, users note it feels intuitive due to keyboard muscle memory. For Gaze & Pinch, under Calibration Issues, users report inaccuracy and frustration when calibration is even slightly off. Under Physical / Mental Demand, users describe it as the most mentally and physically exhausting. Under Improvement with Familiarity, one user says familiarity with pinch distance improves performance. For SwEYEPinch-Basic, under Speed / Ease, users state swipe can be the fastest and easiest once mastered. Under Autocorrect / Error Handling, users say swipe is fun and mistakes are easy to fix, though autocorrect sometimes misrecognizes words like “the” or “strawberries.” Under Learning Curve / Fatigue, users report initial difficulty, but reduced effort with understanding; some note the need to retype words when mis-swiped.}
\label{tab:quotes}
\end{table*}

\begin{table*}[ht]
\resizebox{\textwidth}{!}{
\centering
\begin{tabular}{|p{3.5cm}|p{4.2cm}|p{4.2cm}|p{4.2cm}|}
\hline
\textbf{Theme} & \textbf{Finger-Tap w/ Pred.} & \textbf{Hand-Swipe} & \textbf{SwEYEpinch} \\
\hline
\textbf{Ease of Use / Intuitiveness} & Familiar and similar to phone typing; low learning curve & Not intuitive for many; difficult gesture control & Intuitive after learning; feels futuristic for some \\
\hline
\textbf{Physical Effort} & Physically taxing; arm fatigue common & Most physically exhausting; holding hand up is tiring & Least physically demanding; only requires eye movement \\
\hline
\textbf{Mental Demand / Fatigue} & Low mental effort; simple concept & High mental demand from coordinating swipes and gestures & Medium; some strain from focusing gaze \\
\hline
\textbf{Accuracy / Reliability} & Reliable; occasional finger tracking issues & Inconsistent detection; trouble with long/complex words & Good when calibrated; errors at periphery of vision \\
\hline
\textbf{Calibration Sensitivity} & Less sensitive to misalignment & Sensitive to hand angle and distance & Requires precise calibration for gaze to work well \\
\hline
\textbf{Error Recovery} & Easy to correct manually & Difficult to delete or redo words; no easy back gesture & Better with autocomplete, but hard to recover from mistakes \\
\hline
\textbf{Speed / Efficiency} & Slower but reliable & Slower due to gesture complexity & Fastest when functioning correctly; improved by predictive text \\
\hline
\textbf{Learning Curve} & Very low; familiar typing behavior & Steep; requires practice and good calibration & Moderate; intuitive after learning curve \\
\hline
\textbf{User Preference Trend} & Preferred for familiarity and control & Least preferred due to frustration and physical effort & Frequently preferred for long-term use if calibrated \\
\hline
\end{tabular}
}
\caption{Summary of user feedback for \textit{Finger-Tap}, \textit{Hand-Swipe}, and \textit{SwEYEpinch} (US3).}
\Description{Table 9 summarizes qualitative user feedback for three text-entry methods in US3: Finger-Tap with Prediction, Hand-Swipe, and SwEYEPinch, organized by interaction theme. Finger-Tap is described as familiar and intuitive with a low learning curve, but physically taxing with common arm fatigue; it has low mental demand, reliable accuracy with occasional finger-tracking issues, low calibration sensitivity, easy error recovery, slower but reliable speed, and is preferred for familiarity and control. Hand-Swipe is reported as not intuitive for many users, physically exhausting, and mentally demanding due to gesture coordination; it shows inconsistent accuracy for long or complex words, high sensitivity to hand angle and distance, difficult error recovery, slower speed due to gesture complexity, a steep learning curve, and is the least preferred due to frustration and physical effort. SwEYEPinch is described as intuitive after learning and futuristic for some users; it requires the least physical effort but moderate mental demand from sustained gaze, provides good accuracy when well calibrated with errors at the visual periphery, requires precise gaze calibration, supports better error recovery with autocomplete, achieves the fastest speed when functioning correctly, has a moderate learning curve, and is frequently preferred for long-term use when calibrated.}
\label{tab:typing-feedback-us3}
\end{table*}

\begin{table*}[ht]
\resizebox{\textwidth}{!}{
\centering
\renewcommand{\arraystretch}{1.4}
\begin{tabular}{|p{3.2cm}|p{3.2cm}|p{7.5cm}|}
\hline
\textbf{Typing Method} & \textbf{Theme} & \textbf{Representative Quote} \\
\hline

\multirow{3}{*}{Finger-Tap w/ Pred.} 
& Physical Effort & ``Finger tap hurts my arms.'' \\
&  & ``Finger tap required a lot of upper arm strain...'' \\
\cline{2-3}
& Familiarity & ``Finger tap was most familiar and intuitive.'' \\
&  & ``Finger tap mimics real-life typing and is easiest to use.'' \\
\cline{2-3}
& Accuracy / Reliability & ``Most reliable, but sometimes missed my finger.'' \\
\hline

\multirow{4}{*}{Hand-Swipe} 
& Physical Demand & ``Hand-Swipe is physically exhausting; my arms got tired.'' \\
&  & ``Holding hand up and pinching for each word is tedious.'' \\
\cline{2-3}
& Accuracy / Control & ``Not intuitive at all.'' \\
&  & ``High learning curve and frequent gesture misfires.'' \\
\cline{2-3}
& Frustration & ``Hand-Swipe was the most frustrating method.'' \\
\hline

\multirow{4}{*}{SwEYEpinch} 
& Speed / Comfort & ``Felt like the future; quick and minimal effort.'' \\
&  & ``Eye swipe was quite intuitive after learning.'' \\
\cline{2-3}
& Calibration Issues & ``Only works well with proper calibration.'' \\
&  & ``Had issues tracking eyes at edge of keyboard.'' \\
\cline{2-3}
& Prediction Benefit & ``Autocomplete made eye swipe very fast.'' \\
&  & ``Suggestions helped speed up typing significantly.'' \\
\hline

\end{tabular}
}
\caption{Representative User Quotes from US3}
\Description{The table presents representative user quotes from US3 for three typing methods: Finger-Tap with Prediction, Hand-Swipe, and SwEYEpinch. The table is organized by typing method, theme, and corresponding participant quotes. For Finger-Tap with Prediction, user feedback highlights several themes. With respect to physical effort, users reported arm strain, with comments such as ``Finger tap hurts my arms'' and ``Finger tap required a lot of upper arm strain.'' Regarding familiarity, users described the technique as familiar and intuitive, stating that ``Finger tap was most familiar and intuitive'' and that ``Finger tap mimics real-life typing and is easiest to use.'' In terms of accuracy and reliability, users characterized Finger-Tap as generally reliable but not perfect, noting that it was the ``most reliable, but sometimes missed my finger.'' For Hand-Swipe, user feedback emphasizes physical demand, control, and frustration. Users reported that Hand-Swipe was physically tiring, with remarks such as ``Hand-Swipe is physically exhausting; my arms got tired'' and ``Holding hand up and pinching for each word is tedious.'' In terms of accuracy and control, users struggled with the technique, describing it as ``not intuitive at all'' and noting a ``high learning curve and frequent gesture misfires.'' Users also explicitly expressed frustration, stating that ``Hand-Swipe was the most frustrating method.'' For SwEYEpinch, user feedback focuses on speed, comfort, calibration, and prediction benefits. Users praised the technique for its speed and low effort, describing it as ``felt like the future; quick and minimal effort'' and noting that ``eye swipe was quite intuitive after learning.'' Users also pointed out calibration-related issues, stating that the technique ``only works well with proper calibration'' and that they ``had issues tracking eyes at edge of keyboard.'' Finally, users emphasized the benefit of prediction, reporting that ``autocomplete made eye swipe very fast'' and that ``suggestions helped speed up typing significantly.''}
\label{tab:quotes-us3}
\end{table*}

\FloatBarrier
\section{Questionnaire}
We use the following survey for our study. The exact text of each question as presented to the participants is below. The content here is for US1. The US2 and US3 surveys are identical except for the text entry techniques covered.

\begin{enumerate}
  \item Age  

  \item Gender  

  \item Right-handed or left-handed?  

  \item I have normal or corrected-to-normal vision (with or without glasses).  

  \item Strabismus (crossed-eyes)?  

  \item I use a computer keyboard...  

  \item I use a cellphone keyboard...  

  \item I have experience using Augmented Reality/Virtual Reality systems  

  \item If you answered "Yes" to the previous question, please explain your experience.  

  \item I use Augmented Reality/Virtual Reality keyboards...  

  \item If you have used an Augmented Reality/Virtual Reality keyboard, please tell us more about your experience
        (e.g., how intuitive/efficient do you find them to be)?  

  \item Mental Demand: how mentally demanding was the task?: Finger-Tap  

  \item Mental Demand: how mentally demanding was the task?: Gaze\&Pinch  

  \item Mental Demand: how mentally demanding was the task?: Swipe  

  \item Physical Demand: how physically demanding was the task?: Finger-Tap  

  \item Physical Demand: how physically demanding was the task?: Gaze\&Pinch  

  \item Physical Demand: how physically demanding was the task?: Swipe  

  \item Temporal Demand: how hurried or rushed was the pace of the task?: Finger-Tap  

  \item Temporal Demand: how hurried or rushed was the pace of the task?: Gaze\&Pinch  

  \item Temporal Demand: how hurried or rushed was the pace of the task?: Swipe  

  \item Performance: how successful were you in accomplishing what you were asked to do?: Finger-Tap  

  \item Performance: how successful were you in accomplishing what you were asked to do?: Gaze\&Pinch  

  \item Performance: how successful were you in accomplishing what you were asked to do?: Swipe  

  \item Effort: how hard did you have to work to accomplish your level of performance?: Finger-Tap  

  \item Effort: how hard did you have to work to accomplish your level of performance?: Gaze\&Pinch  

  \item Effort: how hard did you have to work to accomplish your level of performance?: Swipe  

  \item Frustration: how insecure, discouraged, irritated, stressed, and annoyed were you?: Finger-Tap  

  \item Frustration: how insecure, discouraged, irritated, stressed, and annoyed were you?: Gaze\&Pinch

  \item Frustration: how insecure, discouraged, irritated, stressed, and annoyed were you?: Swipe  

  \item Please rank the styles of typing by your overall experience with them.: Finger-Tap  

  \item Please rank the styles of typing by your overall experience with them.: Gaze\&Pinch  

  \item Please rank the styles of typing by your overall experience with them.: Swipe  

  \item Please explain your rankings.  

  \item Do you have any additional comment about the three typing methods (Finger-Tap, Gaze\&Pinch, Swipe)?  

\end{enumerate}

\end{document}